\documentclass[acmlarge]{acmart}
\AtBeginDocument{%
  }

\setcopyright{cc}
\setcctype{by-nc-nd}
\acmJournal{IMWUT}
\acmYear{2026} \acmVolume{10} \acmNumber{4} \acmArticle{225}
\acmMonth{12} \acmDOI{10.1145/3858001}

\usepackage{tabularx}
\usepackage{makecell}
\usepackage{enumitem}

\usepackage{rotating}
\usepackage{multirow}
\usepackage[table]{xcolor}
\usepackage{ragged2e}

\begin{document}

\newcommand{\sys}[0]{FlexGuard}
\title{\sys{}: A Design Space for On-Body Feedback for Safety Scaffolding in Strength Training}

\definecolor{stblue}{HTML}{0B6FA4}  
\definecolor{fdpurple}{HTML}{7A3E9D} 

\newcommand{\ST}[1]{\colorbox{stblue!12}{\textcolor{stblue}{\textsc{#1}}}}
\newcommand{\FD}[1]{\colorbox{fdpurple!12}{\textcolor{fdpurple}{\textsc{#1}}}}

\author{Panayu Keelawat}
\orcid{0000-0002-2819-1705}
\affiliation{%
  \institution{Virginia Tech}
  \city{Blacksburg}
  \state{Virginia}
  \country{USA}}
\email{panayu@vt.edu}

\author{Darshan Nere}
\orcid{0009-0006-7682-6396}
\affiliation{%
  \institution{Virginia Tech}
  \city{Blacksburg}
  \state{Virginia}
  \country{USA}}
\email{darshannere@vt.edu}

\author{Jyotshna Bali}
\orcid{0009-0009-3911-2716}
\affiliation{%
  \institution{Virginia Tech}
  \city{Blacksburg}
  \state{Virginia}
  \country{USA}}
\email{jyotshna@vt.edu}

\author{Rezky Dwisantika}
\orcid{0009-0003-3192-090X}
\affiliation{%
  \institution{Sepuluh Nopember Institute of Technology}
  \city{Surabaya}
  \country{Indonesia}}

\author{Yogesh Phalak}
\orcid{0000-0001-5279-2271}
\affiliation{%
  \institution{Virginia Tech}
  \city{Blacksburg}
  \state{Virginia}
  \country{USA}}
\email{yphalak@vt.edu}

\author{Ardalan Kahak}
\orcid{0009-0003-8367-1298}
\affiliation{%
  \institution{Virginia Tech}
  \city{Blacksburg}
  \state{Virginia}
  \country{USA}}
\email{ardalankahak@vt.edu}

\author{Anekan Naicker}
\orcid{0009-0004-9749-1519}
\affiliation{%
  \institution{Academies of Loudoun}
  \city{Ashburn}
  \state{Virginia}
  \country{USA}}

\author{Liang He}
\orcid{0000-0003-4826-629X}
\affiliation{%
  \institution{The University of Texas at Dallas}
  \city{Richardson}
  \state{Texas}
  \country{USA}}
\email{liang.he@utdallas.edu}

\author{Suyi Li}
\orcid{0000-0002-0355-1655}
\affiliation{%
  \institution{Virginia Tech}
  \city{Blacksburg}
  \state{Virginia}
  \country{USA}}
\email{suyili@vt.edu}

\author{Yan Chen}
\orcid{0000-0002-1646-6935}
\affiliation{%
  \institution{Virginia Tech}
  \city{Blacksburg}
  \state{Virginia}
  \country{USA}}
\email{ych@vt.edu}

\renewcommand{\shortauthors}{Panayu Keelawat et al.}

\begin{abstract}
    Strength training carries inherent safety risks when exercises are performed without supervision. While haptics research has advanced, there remains a gap in how to integrate on-body feedback into intelligent wearables. Developing such a design space requires experiencing feedback in context, yet obtaining functional systems is costly. By addressing these challenges, we introduce FlexGuard, a design space for on-body feedback that scaffolds safety during strength training. The design space was derived from nine co-design workshops, where novice trainees and expert trainers DIY'd low-fidelity on-body feedback systems, tried them immediately, and surfaced needs and challenges encountered in real exercising contexts. We then evaluated and refined the design space through a multi-stakeholder speed dating study, using storyboards to cover the design dimensions. The study involved novice trainees and experienced trainers, allowing us to validate the design space from both usability and safety perspectives while synthesizing design implications for future wearable systems. Our findings extend the design space for sports and fitness wearables in the context of strength training.
\end{abstract}



\begin{teaserfigure}
  \centering
  \includegraphics[width=0.82\textwidth]{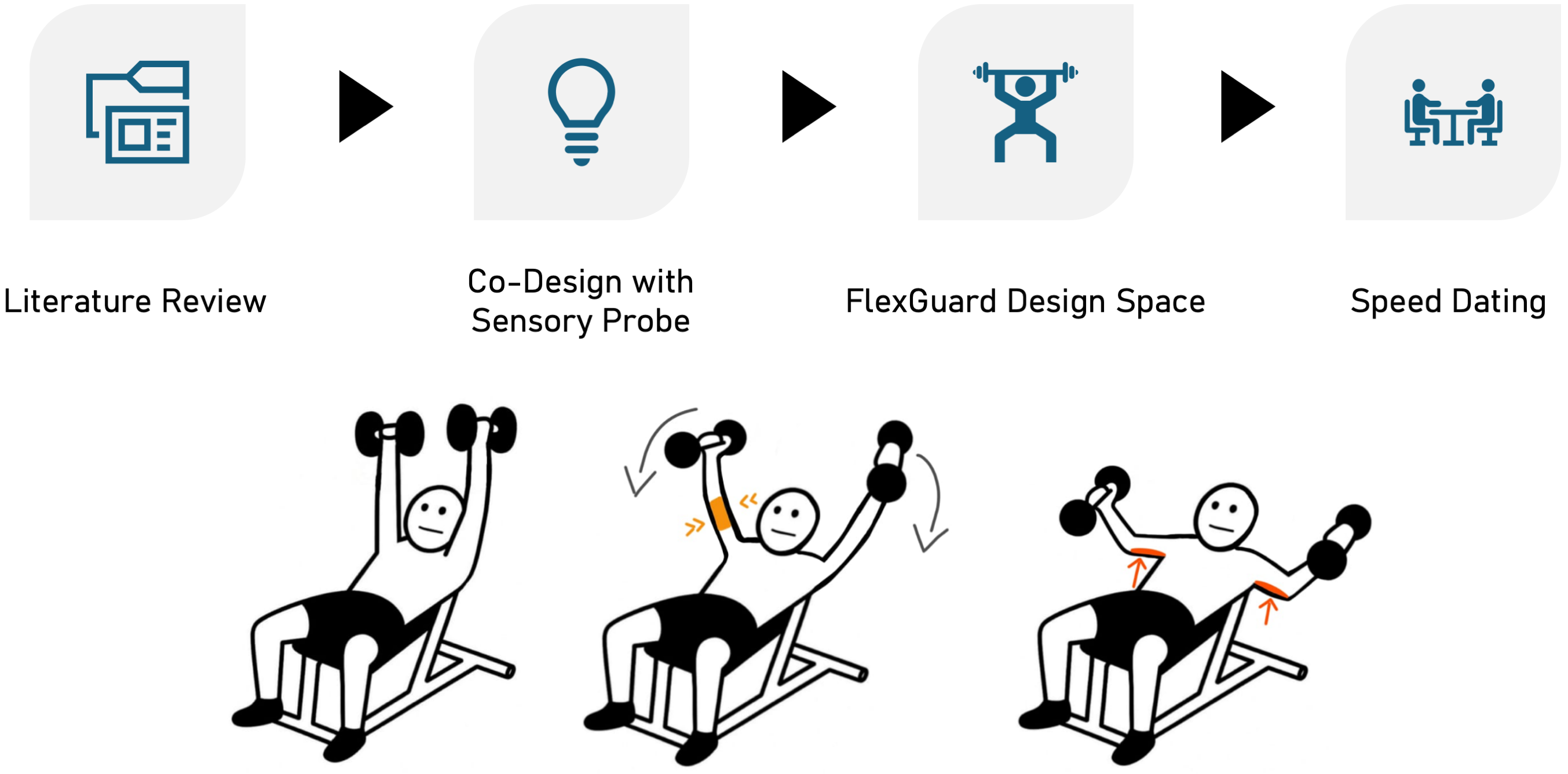}
  \caption{Research process for developing and validating \sys{}. The process includes a literature review, co-design workshops, the development of the FlexGuard design space, and its evaluation through speed-dating sessions.}
  \label{fig:teaser}
\end{teaserfigure}

\received{1 February 2026}
\received[revised]{1 August 2026}
\received[accepted]{18 September 2026}

\maketitle

\section{INTRODUCTION}
\label{sec:introduction}

Strength training is a widely practiced method for improving strength, mobility, and overall health. However, errors such as joint misalignment or overextension can increase the risk of injury, especially for novice practitioners \cite{fisher2025supervision}. Personal trainers help mitigate these risks by providing tailored feedback that cultivates trainees’ awareness of potentially unsafe movement and offers moment-to-moment physical support during execution. Such in-person supervision, however, is costly and depends on others’ availability, making it difficult to scale. As a result, there is a need for alternative forms of support that can scaffold trainees’ experience by helping them remain aware of safety-relevant movement while also providing embodied assistance during training. Advances in haptics now enable the emulation of physical sensations commonly used in in-person coaching, such as taps, pressure, and resistance. This creates an opportunity for intelligent wearable systems that can deliver on-body feedback to scaffold safety during strength training. What remains missing, however, is the design knowledge needed to integrate such on-body feedback into wearable systems.

Designing on-body feedback systems is challenging because users must physically experience feedback during strength training \cite{somaDesign, SomaTheory}, and obtaining functional systems is costly \cite{teslasuitSensors, genEMG}. Investing heavily in a prototype to serve as a probe \cite{codesIMWUT} is often impractical in early-stage exploration. Moreover, any probing system must be representative of the diverse on-body feedback mechanisms that have been studied and developed. Together, these challenges hinder systematic investigation of the design space.

By addressing these challenges, we developed \sys{}, a design space for on-body feedback to scaffold safety in strength training. We began by surveying the landscape of on-body feedback, and identified five open questions. Moreover, we grouped feedback modalities into two types: \textit{extrinsic}, which require active actuation (e.g., vibration or air pressure), and \textit{intrinsic}, which rely on material properties (e.g., orthoses or braces) to provide passive feedback once configured. Based on these types, we then prepared sensory probes, along with supporting props, and conducted nine co-design workshops in our institution's gym. Each session paired a novice trainee with an expert trainer, who collaboratively DIY’d personalized low-fidelity on-body feedback systems for five representative exercises and tested them immediately in practice. Through thematic analysis, we distilled the needs and challenges that informed the derivation of \sys{}.

\sys{} is organized around two complementary components: \textit{Sensing \& Triggering} and \textit{Feedback Delivery}. \textit{Sensing \& Triggering} specifies which indicators should be monitored (e.g., joint angle, movement dynamics, motion path, muscle activation, or inter-limb coordination), how triggering policies are defined (e.g., expert-based or personalized), and where monitoring should occur (e.g., target, stabilizing, moving, or compensation regions). \textit{Feedback Delivery} describes how feedback is provided once triggered, including the body zones where different feedback types are applied, the intensity of feedback, its directional strategies, and its timing across movement phases. Together, these components define how on-body feedback can be timed, targeted, and configured to scaffold safety during exercise.

We evaluated and refined \sys{} through a multi-stakeholder speed dating study \cite{PSD_AIED, PSD_CSCW}. We recruited seven beginner trainees and six experienced personal trainers, conducting three workshops with the beginner trainees and six one-on-one sessions with the trainers. Participants discussed a series of storyboard scenarios that varied design dimensions within the \textit{Sensing \& Triggering} and \textit{Feedback Delivery} components of \sys{}. Experienced trainers were asked to ground their responses in their coaching practices and safety considerations, while beginner trainees performed the exercises using the provided equipment and followed tutorial videos to ground their reflections.

The results validated the dimensions of \sys{} by providing empirical evidence for the perceived usefulness of its dimensions. We further reported participants' responses with respect to the open design questions identified from the literature landscape. Finally, we synthesized these findings together with prior literature to report design guidelines that clarify how different feedback modalities should be assigned, how feedback should be prioritized, how should wearables balance standardized expert knowledge with personalized calibration, how on-body feedback should be designed to support clear interpretation, and how physical assistance should preserve user agency while supporting safe and effective physical training.

Our contributions are as follows:
\begin{itemize}
    \item We introduce \sys{}, a design space for on-body feedback that scaffolds safety in strength training, grounded in literature review and nine co-design workshops with trainers and trainees.
    \item We present a multi-stakeholder speed dating results that validates and refines \sys{} and introduce design implications for future wearable systems.
\end{itemize}

This work extends prior design spaces for sports and fitness wearables \cite{sportsWearablesDesignSpace, nzWorkshopPaper} by addressing the embodied, phase-specific realities of strength training and advancing toward controllable, personalized intelligent wearable systems.

\section{RELATED WORK}
\label{sec:related-work}

Our research builds upon prior work in three key areas: Safety and Risk in Strength Training, Sensing and Feedback Systems for Physical Training, Intelligent Systems in Movement and Skill Learning.

\subsection{Safety and Risk in Strength Training}
Strength training carries inherent safety considerations, particularly when exercises involve high external loads that stress the musculoskeletal system. Body regions that both bear the load and move dynamically, such as the lower back and shoulders, are especially prone to overload or misalignment-related issues \cite{commonInjuries, barnes2021peak}. While such issues are common, they are often avoidable. Research indicates that a large portion of lower back strain incidents are linked to improper lifting form, excessive loads, and overtraining, which all are factors that can be mitigated through proper technique and training practices \cite{alqarni2019common}. Faulty alignment and compensatory movement patterns introduce abnormal stress across joints and soft tissues, contributing not only to unsafe strain but also to long-term issues. This risk is especially pronounced in novice trainees, whose motor control systems may not yet be developed enough to coordinate complex compound movements effectively \cite{aslam2025neuromuscular}. Thus, consistent attention to motor learning is fundamental for safe and effective strength training \cite{latella2019differences}.

This is why working with a personal trainer is widely recommended, particularly for beginners. Professional supervision ensures that trainees are taught correct movement techniques, are properly spotted to maximize repetitions safely, and are guided through appropriate loading progressions. When properly coached, strength training is a safe activity relative to other sports \cite{fisher2023supervision}. In contrast, unsupervised training is associated with higher risks due to inconsistent technique and overly aggressive load selection \cite{2010youngatheletes, 2020progOverload}. While technologies exist to support self-coaching \cite{workoutcool2025}, human coaches provide multifaceted guidance that extends beyond form correction and motor habit development to include enforcing safety practices such as spotting, load monitoring, and early detection of execution breakdowns \cite{fisher2023supervision}.

In addition to coaching, trainees frequently use supportive wearable gear to enhance safe training practices. Equipment such as weightlifting belts, wrist wraps, knee sleeves, and elbow sleeves is commonly used in gyms to help maintain alignment and distribute stress during heavy lifts. Weightlifting belts, for instance, are used to increase intra-abdominal pressure and reduce spinal loading during squats and deadlifts \cite{beltsOnLowerBack}. Knee sleeves improve proprioception during deep knee flexion movements \cite{kneeSleeves}. Users can also define and produce custom wearables for their specific needs \cite{xu2025smart}. For example, tailoring their own knee patch or squat suit to support heavy lifting \cite{squatSuit}.

\sys{} builds on this trajectory by extending safety considerations beyond passive support and human supervision. Leveraging haptics for on-body sensation, \sys{} advances toward adaptive systems that actively sense and respond to cues and deviations during exercise, providing intuitive feedback to support safety scaffolding during training.  

\subsection{Sensing and Feedback Systems for Physical Training}

Designing wearables for physical training requires attention to both sensing bodily states and delivering feedback that accounts for users' felt experience \cite{somaDesign, SomaBits}. This framing draws on soma-based design theory and somatic design work that treats the lived body and body perception as central to physical engagement \cite{SomaTheory, codesIMWUT}. Sensing technologies have enabled researchers and designers to monitor a wide range of body states during physical training. Existing systems monitor muscle activity \cite{MuscleRehab, FSRandMMG, mmgTei}, assess exercise performance and repetitions through wearable and magnetic sensing \cite{RecoFit, LiftSmartUbiComp, ProxiFitIMWUT}, and support posture and movement monitoring through pressure sensing \cite{vizCgDeadLifts}, computer vision \cite{rao2024formfit}, smart garments and textile-based sensors \cite{Zishi, stretchyKneeElsevier, intelligentTextiles, SeamFitIMWUT}, electronically augmented orthoses \cite{EOrthosis}, and deformable or shape-changing sensor platforms \cite{umichShoulder, FluxLab}. Commercial platforms extend these capabilities to elite sports analytics \cite{catapult}.

A parallel stream of work focuses on feedback modalities for physical training. Systems deliver guidance through wearable and projected outputs \cite{smartRingsIMWUT, BodyLights}, instructional videos and XR frameworks \cite{phyTrainingVisual, XRVisualDesSpace}, AR/VR overlays and spatial visualizations \cite{InstructAR, avaTTAR, TheYogaVR, danceVisionDesignSpace}, and combined visual--auditory cues \cite{Weight-Mate}. Vibrotactile systems support physical-activity guidance \cite{snowboardVibro, rockClimbing, Subtletee}, simulated trainer taps and scalable spatial vibration \cite{TingleTouch, TrainerTap, VibraForge}, accessible yoga practice \cite{YogaVI}, and varied sensations through different tactor designs \cite{ParametricHaptics}. More physically intervening approaches include EMS/TENS for movement correction and exercise monitoring \cite{FootStrikerIMWUT, teslasuitSensors}, wearable robotics and pneumatic garments for distributed or kinesthetic feedback \cite{HapticSnakes, PneumAct, ForceJacket, PneuMa}, and modular pneumatic, and compression interfaces \cite{PneuMod, Springlets, mitTunableSpaceSuit}. Material-driven interfaces further expand this repertoire through fluidic and origami-inspired actuation \cite{milliMorph, oriArtificialMuscle}, kirigami and dynamic surface structures \cite{KiriHapticSwatches, FlexTure}, and programmable, multistable, or snap-through mechanisms \cite{WaxpaperSeq, ConeAct, SnapInflatables}. Other mechanisms include skin stretch for upper- and lower-body proprioceptive cues \cite{skinStretching, Gaiters}, pneumatic jets and localized pressure \cite{JetUnit, kneeStraigtener}, and torque guidance \cite{torqueFeedback}. Across SportsHCI, researchers emphasize adaptation to individual contexts and goals, and the objective relevance of feedback \cite{sportsWearablesDesignSpace, grandChallengesSportsHCI, nzWorkshopPaper, mocoWorkshop}. 

Prior systems offer diverse sensing and feedback mechanisms but limited guidance for coordinating them. \sys{} organizes these decisions based on stakeholder knowledge for safe strength training.

\subsection{Intelligent Systems in Movement and Skill Learning}

Interactive Machine Learning (IML) refers to systems that allow end-users to guide, correct, or shape the learning process through iterative interaction. Unlike traditional supervised ML, IML invites user-in-the-loop contributions during model refinement \cite{powerToThePeople}. Hartmann et al.\ \cite{bjornDemo} emphasize the value of interactivity in enabling more intuitive and user-aligned learning. Related end-user authoring systems extend this interactive principle by enabling non-experts to use sensor data and demonstrations to specify adaptive behaviors and on-body feedback \cite{CollabJam, HapticPilotIMWUT, ShapeKit, ModElecIMWUT}.

In movement learning and demonstration-based systems, researchers have applied IML techniques to encode and reuse human motion, which often involves tacit knowledge that is difficult to articulate explicitly \cite{MLMovementDes}. \textit{MotionMA} \cite{MotionMA} enables users to demonstrate movement patterns that the system can model and compare in real time. \textit{Mimic} \cite{Mimic} applies similar principles in teleoperation, letting users record and re-trigger physical behaviors. Other approaches like \textit{danceON} \cite{danceON} and \textit{Rapido} \cite{Rapido} focus on embodied programming, through specification or demonstration, for creative and spatial applications. Vidal et al. propose a framework and infrastructure for applying IML in physiotherapy to support both in-clinic and out-of-clinic patients \cite{physiotherapyML}. Wearable systems have coupled movement sensing with EMS and tactile feedback to support physical skill learning in domains such as dance, piano playing, and hand gesture training \cite{hapticScores, danceInflatables, glovePiano, ReKnitCare}. In sport-specific contexts, Weng et al. decompose movement into feedback-friendly units to integrate AI-driven textual coaching for basketball shooting \cite{basketballCoach}. However, such textual feedback approaches \cite{expressiveRobotLLM, assistanceOrDisruption} may be less effective in physical tasks like strength training. 

\sys{} builds on these directions by embedding interactive intelligence directly into the wearable interface, enabling trainees to shape feedback for safer and more personalized strength training.

\section{OPEN QUESTIONS IN THE DESIGN LANDSCAPE OF ON-BODY FEEDBACK}
\label{sec:design-landscape}

We reviewed prior work to characterize the design landscape of on-body feedback, and also to identify open questions for designing on-body feedback for safety scaffolding in strength training.

\subsection{Methodology}
\label{sec:landscape-methodology}

We searched libraries including the ACM Digital Library and IEEE Xplore using combinations of keywords (Appendix~\ref{app:literature-search-terms}). We filtered the results to focus on top, relevant venues, including CHI, IMWUT, UIST, ToCHI, and TEI. To broaden coverage beyond the initial database searches, we iteratively applied backward and forward snowballing from relevant review papers, design spaces, and representative systems. We additionally examined publication records of recurring authors to identify continued lines of research relevant to on-body feedback. This process also led us to papers and systems indexed outside the initial search venues.

The first, second, fourth and seventh authors analyzed the corpus through an iterative, design-oriented thematic synthesis in a shared Miro~\cite{miro} workspace. The analysis proceeded at two connected levels. First, we used descriptive affinity diagramming \cite{affinityDiagramming} to organize articles by extracted characteristics, such as the source and location of sensing and feedback, the logic governing feedback, and the intended meaning of the cue. Papers could contribute to multiple groups because these characteristics were not mutually exclusive. 

The groups subsequently served as the basis for comparison and abductive interpretation. We used constant comparison \cite{groundedTheory} to examine variation among publications within the same group and relationships across groups. In particular, we considered how systems addressing a similar design concern differed in their technical realization, underlying assumptions, intended interactional role, and consequences for users. These comparisons were guided by three questions: (1) What capability has prior work established? (2) What assumptions, limitations, or trade-offs accompany that capability? and (3) What design decision remains unresolved when the capability is considered in the context of safety scaffolding?

We then conducted an abductive synthesis \cite{abductiveSynthesis} to interpret the design significance of these comparisons. We considered what higher-level design problem could account for the observed differences and tensions. For example, coding revealed a growing range of bodily states that wearable systems could detect. Comparing these systems, however, suggested that broader sensing coverage did not resolve which among several simultaneously detected movement issues should be communicated to the user. We therefore interpreted prioritization as the underlying unresolved design problem. We applied this form of reasoning across the descriptive groups.

The process produced five findings (F1--F5), and the analysis remained iterative as additional papers were identified through continued searching and feedback from collaborators. New publications were coded against both the descriptive groups and the candidate findings. The landscape was considered sufficiently stable when newly identified publications could be incorporated without requiring changes to the groupings. The final corpus consisted of 188 papers, including 56.92\% from ACM venues, 21.81\% from IEEE venues, and 21.27\% from others. Of the 188 papers, 135 (71.8\%) were published from 2020 onward, and 108 (57.4\%) appeared between 2022 and 2026.

\subsection{Findings}

We identified five design considerations (F1--F5), revealing trade-offs across sensing, feedback, and intervention strategies. These design choices within one dimension may influence the possibilities and constraints of others. For instance, broader sensing capabilities can identify multiple movement deviations, yet delivering simultaneous cues may introduce perceptual interference and reduce feedback interpretability (F2). For another example, expert-defined movement references provide useful standards, but appropriate movement varies with trainee capabilities and exercise contexts (F3). 

\definecolor{FGTableHeader}{RGB}{210,228,236}
\definecolor{FGTableCluster}{RGB}{232,241,245}
\definecolor{FGTableGap}{RGB}{235,235,235}
\definecolor{FGTableQuestion}{RGB}{220,232,238}

\subsubsection{F1: Different Feedback Mechanisms Serve Complementary Intervention Roles}

F1 emerged from comparing how on-body mechanisms create a sensation and how directly that sensation involves in supporting movements. We distinguish \textit{extrinsic feedback}, generated directly by an actively controlled actuator whose output can be produced independently of the trainee's movement, from \textit{intrinsic feedback}, experienced through the configured mechanical response of the worn structure as the body moves. This is an analytical distinction rather than a mutually exclusive classification: one system may actively reconfigure a material whose resulting stiffness or geometry is then experienced intrinsically (Table~\ref{tab:feedbackmodalities}).

\begin{table}[H]
\centering
\caption{F1: Prior mechanisms can be broadly organized as extrinsic or intrinsic feedback.}
\label{tab:feedbackmodalities}
\begingroup
\scriptsize
\setlength{\tabcolsep}{3pt}
\renewcommand{\arraystretch}{1.00}
\arrayrulecolor{black!55}
\begin{tabularx}
{\linewidth}{|>{\columncolor{FGTableHeader}\centering\arraybackslash}p{0.065\linewidth}|>{\columncolor{FGTableCluster}\RaggedRight\arraybackslash}p{0.185\linewidth}|>{\RaggedRight\arraybackslash}X|}
\hline
\rowcolor{FGTableHeader}
\shortstack{\textbf{Feedback}\\\textbf{origin}} & \shortstack[l]{\textbf{Modality or}\\\textbf{mechanism}} & \textbf{Prior work within the finding theme} \\
\hline
\multicolumn{3}{|>{\columncolor{FGTableQuestion}\RaggedRight\arraybackslash}p{\dimexpr\linewidth-2\tabcolsep-2\arrayrulewidth\relax}|}{\textbf{Open question:} How should designers assign complementary roles to extrinsic and intrinsic feedback?} \\
\hline
\multirow[c]{1}{*}{\textbf{Extrinsic}} & \textbf{Vibration} & \cite{rockClimbing, TrainerTap, Subtletee, VibraForge, Zishi, HapticHead, HapticPilotIMWUT, VibroMap, MagTex, snowboardVibro, YogaVI, ieeeVibration16, motionlessMovement, ieeeVibration4, CollabJam}
 \\
\cline{2-3}
& \textbf{Electrical stimulation and electronic textiles} & \cite{TingleTouch, FootStrikerIMWUT, smartwatchEms, ReKnitCare, teslasuitSensors, hapticScores, glovePiano}
 \\
\cline{2-3}
& \textbf{Pneumatic and fluidic} & \cite{Bellowband, aSpireIMWUT, ieeePneum1, ForceJacket, PneumAct, PneuMa, kneeStraigtener, JetUnit, VabricBeads, milliMorph, SnapInflatables, oriArtificialMuscle, purnendu, OmniFiber}
 \\
\cline{2-3}
& \textbf{Thermal} & \cite{PneuMod, ThermoVR, thermalBreathing, ShapeTex}
 \\
\cline{2-3}
& \textbf{Mechanical force and skin deformation} & \cite{torqueFeedback, Gaiters, HapticSnakes, ImpactVest, Springlets, skinStretching, ieeeMech4, fingerTipVRForce, FW-Touch, ieeeSkinStrechForXRFeeling, ImaginaryJoint, ieeeMech1, ieeeMech3, ConeAct, mimickMuscle, allonic, sourcePaperAllonic, humanScaleRobotic, ShapeKit, ParametricHaptics, Kinethreads, reelFeel, Mod2Hap}
 \\
\hline
\multirow[c]{1}{*}{\textbf{Intrinsic}} & \textbf{Structural supports} & \cite{BistableOrthoses, selectivelyStiffening, ThermoFitIMWUT, EOrthosis, FrozenSuit, SkinMorph, laminarJamming, mitTunableSpaceSuit, kneeSleeves, squatSuit,beltsOnLowerBack, wearableMaterialProps}
 \\
\cline{2-3}
& \textbf{Textile, cut/fold, and metamaterial mechanics} & \cite{MagKnitic, FlexTruss, ardalanPaper, KiriHapticSwatches, FlexTure, oriHapticsReview, WaxpaperSeq, Embrogami}
 \\
\hline
\end{tabularx}
\arrayrulecolor{black}
\endgroup
\end{table}

\subsubsection{F2: Broader Sensing Capabilities Create Competing Feedback Opportunities}

F2 emerged from comparing the states or metrics that wearable systems can detect with the limited opportunities to communicate during a loaded movement. Table~\ref{tab:landscape-f2} therefore groups prior work by the kinds of observations that become candidates for feedback.

\begin{table}[H]
\centering
\caption{F2: Broader sensing exposes more candidate issues than a wearable can communicate at once.}
\label{tab:landscape-f2}
\begingroup
\scriptsize
\setlength{\tabcolsep}{3pt}
\renewcommand{\arraystretch}{1.08}
\arrayrulecolor{black!55}
\begin{tabularx}{\linewidth}{|>{\columncolor{FGTableCluster}\RaggedRight\arraybackslash}p{0.205\linewidth}|>{\RaggedRight\arraybackslash}X|}
\hline
\rowcolor{FGTableHeader}
\textbf{Candidate feedback source} & \textbf{Prior work within the finding theme} \\
\hline
\multicolumn{2}{|>{\columncolor{FGTableQuestion}\RaggedRight\arraybackslash}p{\dimexpr\linewidth-2\tabcolsep-2\arrayrulewidth\relax}|}{\textbf{Open question:} When several detectable issues occur simultaneously, which issue warrants immediate feedback, which can be deferred, and which requires no intervention?} \\
\hline
\textbf{Movement identity and geometry} & \cite{RecoFit, LiftSmartUbiComp, ProxiFitIMWUT, SeamFitIMWUT, stretchyKneeElsevier, umichShoulder, Zishi, EOrthosis, rao2024formfit, mediapipeaccuracy, mediaPipeLatency, tactilePose}
 \\
\hline
\textbf{Mechanical loading and muscle state} & \cite{vizCgDeadLifts, ACLGuard, UbiPhysio, MuscleRehab, FSRandMMG, mmgTei, BIT, ieeeGlovePostureDetection, IntelliLining, teslasuitSensors, fishKiri, fiberFabric}
 \\
\hline
\textbf{Integrated and longitudinal context} & \cite{intelligentWearableSurvey, sittingReview, catapult, workoutcool2025, longTermPhyData}
 \\
\hline
\textbf{Coaching interpretation} & \cite{fisher2023supervision, fisher2025supervision, grandChallengesSportsHCI}
 \\
\hline
\end{tabularx}
\arrayrulecolor{black}
\endgroup
\end{table}

\subsubsection{F3: Movement Correctness Is Situated Rather Than Universal}

F3 emerged from comparing the sources through which prior systems operationalize ``correct'' movement. Table~\ref{tab:landscape-f3} groups encoded expert references, personal calibration, measurement practices, and demonstrated or socially negotiated knowledge. Their comparison shows that correctness is produced through a reference and a context.

\begin{table}[H]
\centering
\caption{F3: Different sources of movement correctness encode different assumptions.}
\label{tab:landscape-f3}
\begingroup
\scriptsize
\setlength{\tabcolsep}{3pt}
\renewcommand{\arraystretch}{1.08}
\arrayrulecolor{black!55}
\begin{tabularx}{\linewidth}{|>{\columncolor{FGTableCluster}\RaggedRight\arraybackslash}p{0.205\linewidth}|>{\RaggedRight\arraybackslash}X|}
\hline
\rowcolor{FGTableHeader}
\textbf{Basis for ``correct''} & \textbf{Prior work within the finding theme} \\
\hline
\multicolumn{2}{|>{\columncolor{FGTableQuestion}\RaggedRight\arraybackslash}p{\dimexpr\linewidth-2\tabcolsep-2\arrayrulewidth\relax}|}{\textbf{Open question:} How should expert knowledge and individual calibration be combined to determine when a deviation warrants feedback across users, exercises, and contexts?} \\
\hline
\textbf{Encoded expert references} & \cite{teslasuitSensors, rao2024formfit, basketballCoach, BodyLights, headVRCorrection, LegolasGolf, stretchyKneeElsevier, pilatesCorrection}
 \\
\hline
\textbf{Personal and contextual calibration} & \cite{nzWorkshopPaper, aslam2025neuromuscular, latella2019differences, physiotherapyML, genEMG, ieeeEMSTENS2, personalSensingNoti, phyActivityGoalsModeration, lowcostFitnessWearables, dias2017influence, chernozub2024correlation}
 \\
\hline
\textbf{Demonstrated knowledge from peers} & \cite{MotionMA, bjornDemo, MLMovementDes, exoRobotDemoLearning, Mimic, Rapido, danceON, ASTEROIDS, wearableWisdom, OpenDanceLab, hereAndNow}
 \\
\hline
\end{tabularx}
\arrayrulecolor{black}
\endgroup
\end{table}

\subsubsection{F4: Feedback Meaning Must Be Designed}

F4 emerged from comparing work and examining how feedback could communicate messages that are comprehensible by humans. The groups in Table~\ref{tab:landscape-f4} follow a progression from noticing and locating a cue through interpreting it.

\begin{table}[H]
\centering
\caption{F4: Perceptibility is necessary but insufficient for a learnable coaching language.}
\label{tab:landscape-f4}
\begingroup
\scriptsize
\setlength{\tabcolsep}{3pt}
\renewcommand{\arraystretch}{1.08}
\arrayrulecolor{black!55}
\begin{tabularx}{\linewidth}{|>{\columncolor{FGTableCluster}\RaggedRight\arraybackslash}p{0.205\linewidth}|>{\RaggedRight\arraybackslash}X|}
\hline
\rowcolor{FGTableHeader}
\textbf{Meaning layer} & \textbf{Prior work within the finding theme} \\
\hline
\multicolumn{2}{|>{\columncolor{FGTableQuestion}\RaggedRight\arraybackslash}p{\dimexpr\linewidth-2\tabcolsep-2\arrayrulewidth\relax}|}{\textbf{Open question:} How can on-body feedback form a perceptible, interpretable, and learnable coaching language for the trainee?} \\
\hline
\textbf{Detection and localization} & \cite{VibroMap, HapticHead, HapticPilotIMWUT, HapticSnakes, ImpactVest, ieeeVibration1, ieeeVibration7, ieeeMech2, ieeeVibration12, ieeeVibration14, 7591147}
 \\
\hline
\textbf{Temporal and multimodal interference} & \cite{ieeeSkinStretchTesting, ieeeVibration13, ieeeVibration11, ieeeVibration6, ieeeVibration9, motionlessMovement, ieeeVibration10, ieeeVibration15}
 \\
\hline
\textbf{Task-to-action mapping} & \cite{VibraForge, ieeeCripHapticJam, snowboardVibro, rockClimbing, Subtletee, YogaVI, AeroHaptix, ieeeVibration2, ieeeVibration3, ieeeVibration5, ieeeVibration8}
 \\
\hline
\end{tabularx}
\arrayrulecolor{black}
\endgroup
\end{table}

\subsubsection{F5: Physical Intervention Introduces Trade-offs in Safety, Agency, and Learning}

F5 emerged from comparing systems according to how much action authority remains with the user, is shared with a mechanism, or is transferred to a system or another person. Table~\ref{tab:landscape-f5} orders the thematic groups from informational cues to direct bodily control.

\begin{table}[H]
\centering
\caption{F5: On-body systems distribute intervention authority differently among users, mechanisms, and human coaches.}
\label{tab:landscape-f5}
\begingroup
\scriptsize
\setlength{\tabcolsep}{3pt}
\renewcommand{\arraystretch}{1.08}
\arrayrulecolor{black!55}
\begin{tabularx}{\linewidth}{|>{\columncolor{FGTableCluster}\RaggedRight\arraybackslash}p{0.205\linewidth}|>{\RaggedRight\arraybackslash}X|}
\hline
\rowcolor{FGTableHeader}
\textbf{Intervention authority} & \textbf{Prior work within the finding theme} \\
\hline
\multicolumn{2}{|>{\columncolor{FGTableQuestion}\RaggedRight\arraybackslash}p{\dimexpr\linewidth-2\tabcolsep-2\arrayrulewidth\relax}|}{\textbf{Open question:} When should a wearable intervene during movement, and when should it instead preserve user autonomy or yield to human coaching?} \\
\hline
\textbf{Inform and the user acts} & \cite{YogaVI, personalSensingNoti, expressiveRobotLLM}
 \\
\hline
\textbf{Shared control} & \cite{torqueFeedback, PneumAct, FrozenSuit, kneeStraigtener, ieeeForceFeedbackInMRI, ieeeEMGtbd, coldPack}
 \\
\hline
\textbf{The system acts on the body} & \cite{FootStrikerIMWUT, genEMG, ieeeEMSTENS1, ieeeExoStairClimbing, ieeeEMSTENS3, ieeeExoSupportMaybe, actuatingWearablesSkillLearning, ieeeEMSTENS4}
 \\
\hline
\end{tabularx}
\arrayrulecolor{black}
\endgroup
\end{table}

\section{ON-BODY FEEDBACK FOR STRENGTH TRAINING}
\label{sec:on-body-feedback-for-strength-training}

To understand the needs and challenges of on-body feedback systems for strength training, we conducted a co-design workshop study with novice trainees, supported and informed by expert trainers. To achieve that goal, we have to prepare materials for the study including sensory probe and worksheets, that will cover comprehensive space of strength training.

We focused on a set of upper-body strength exercises where errors could compromise safety, to scope the study. Although not full-body, the selected exercises encompass a broad range of movement demands in strength training, offering insights that can largely generalize to other body regions and exercises. The representative set included five exercises: dumbbell front hold, dumbbell incline fly, bench press, lateral raise, and Arnold press. Each exercise highlighted a distinct movement error risk, such as posture holding, range of motion (ROM), inter-limb balance, tempo, or complex transitions. 

\subsection{Probe Design for DIY Low-Fi Prototyping}

To enable rapid DIY prototyping and immediate exploration of feedback sensations, we developed a sensory probe \cite{codesIMWUT, mocoWorkshop} designed for quick reconfiguration. The base system included adjustable shoulder braces and elbow sleeves, equipped with customizable feedback components such as inflatable air wedges on the shoulders, and rigid and stretchy straps on the elbows. The probe supported two distinct modes of on-body feedback, according to F1:

\subsubsection{Extrinsic Feedback} 
Feedback generated by actively controlled components, simulated through the modulation of air wedges.

\subsubsection{Intrinsic Feedback} 
Feedback experienced through the wearable's own mechanical response, achieved through tension in rigid or elastic straps that respond directly to the user’s movement.

\begin{figure}
    \centering
    \includegraphics[width=\linewidth]{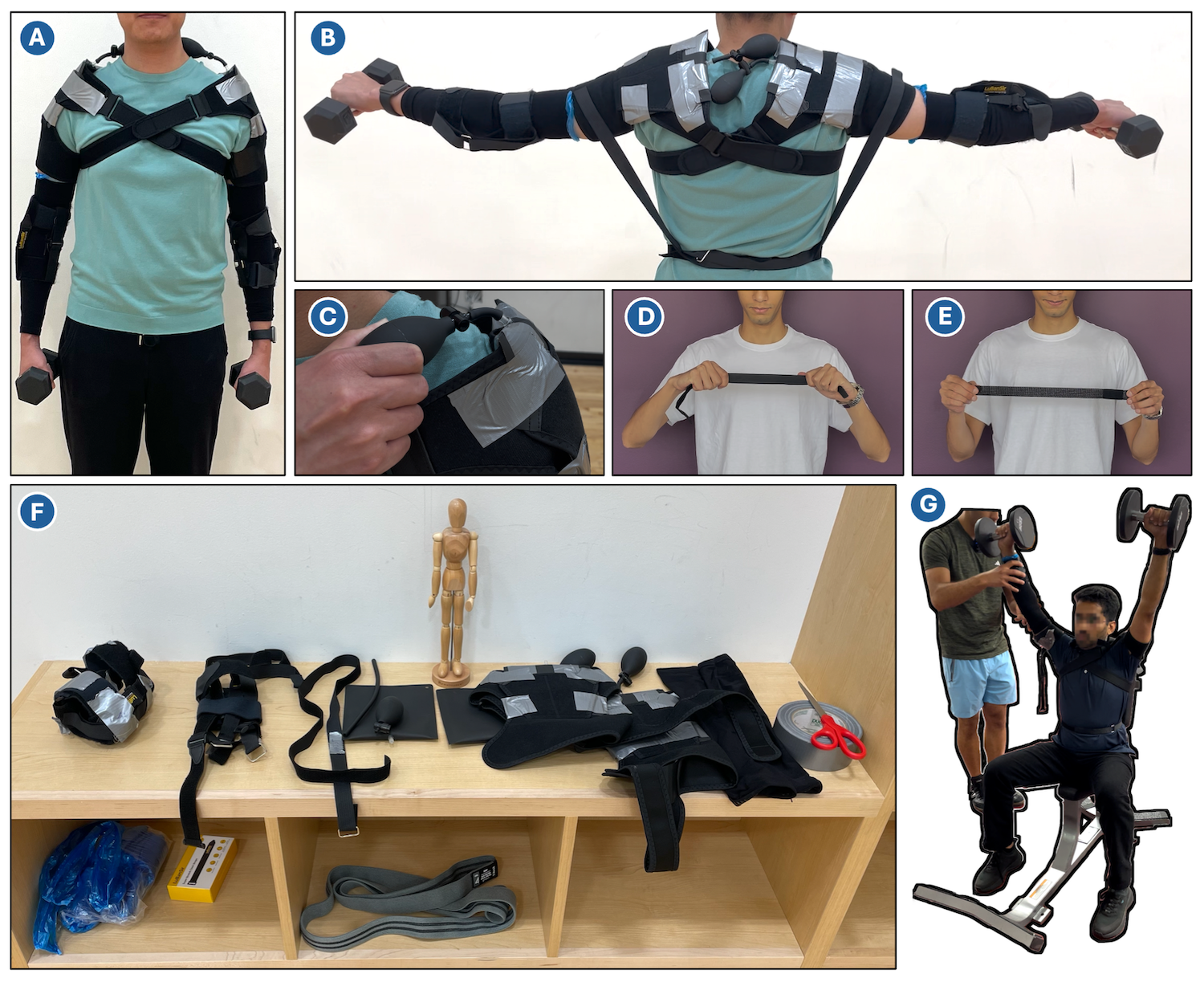}
    \caption{Sensory probe provided to participants for creating low-fidelity feedback prototypes. 
    (A) Default setup on shoulders and elbows, using braces with inflatable air wedges and elbow straps.
    (B) Alternative configurations with attachable waist straps using Velcro for customization. 
    (C) Air wedge simulating on-demand extrinsic actuation. 
    (D) Rigid strap providing non-elastic feedback. 
    (E) Stretchy strap providing elastic and variable feedback when stretched. 
    (F) Props available for prototyping, including additional air wedges, straps, elastic bands, a mini-mannequin, and duct tape. 
    (G) Example of participant N2 wearing the probe during the study with trainer E2.}
    \label{fig:wearingTechProbe}
\end{figure}

We selected these modalities for the study based on their practicality for rapid prototyping in a gym environment (e.g., without electronic wiring) and their coverage of diverse experiences. The air wedges, serving as extrinsic feedback, allowed participants to explore a range of sensations on demand by varying pumping patterns and intensity. In contrast, intrinsic feedback depended almost entirely on material properties, so we provided both rigid and elastic options to offer participants a broader range of experiences.

Velcro interfaces allowed participants to quickly attach, reposition, or remove components as needed. To further support collaborative prototyping, we provided additional materials such as duct tape, resistance bands, and a mini-mannequin to facilitate communication between trainers and trainees. This setup enabled hands-on experimentation: participants configured feedback on their own bodies, performed representative movements, and reflected on the sensations in real time (Figure~\ref{fig:wearingTechProbe}).

\subsection{Study Protocol}

We conducted nine co-design workshops, each involving a matched trainer–trainee pair. 
Appendix~\ref{app:participant-background} summarizes participant demographics, training experience, and notable athletic backgrounds.

All novice trainees (N1–N9) had less than three months of consistent strength training experience, while expert trainers (E1–E9) were recruited based on having at least one year of consistent training experience and prior coaching experience to ensure that trainees received appropriate guidance during the study.

Participants were instructed to wear shirts with sleeves, as the sensory probe was designed to be worn on the shoulders and arms. To ensure hygiene and minimize direct skin contact, we provided disposable elbow sleeves for participants to wear under the probe components.

Workshops (W1--W9) took place in a private studio room at our institution’s gym and lasted approximately 90 minutes. To ensure participant comfort, especially during on-body probe configuration, each pair was matched by gender. All participants provided informed consent for video and audio recording. When participants used public gym equipment (e.g., for the bench press), continuous audio recording was maintained, and short video clips were recorded when appropriate, focusing only on consenting individuals. Throughout the session, participants were encouraged to think aloud and take breaks as needed. Each participant received \$37.50 in compensation (\$25 per hour) for their time and effort.

Our study was approved by our institution’s IRB. Each workshop followed the structure below:

\subsubsection{Initial Interview}
Each workshop began with two brief, separate audio-recorded interviews, one with the trainee and one with the trainer. Participants discussed their individual training backgrounds, any prior injuries, experiences, or safety concerns related to strength training, and their experience coaching or being coached. These interviews provided context for interpreting their design decisions during the rest of the session.

\subsubsection{Worksheet Introduction}
We introduced the study using worksheets to guide participants through the activities. Worksheet details are in Appendix~\ref{app:worksheet-design}. We explained the project and highlighted three design goals to optimize:  

\begin{itemize}
    \item \textbf{DG1:} Promote safety or prevent injuries by providing feedback.  
    \item \textbf{DG2:} Deliver on-body sensations.  
    \item \textbf{DG3:} Personalize feedback to the individual needs of each trainee.  
\end{itemize}

\subsubsection{Probe Demonstration}
Next, the research team introduced the sensory probe. We first demonstrated the default configuration of the probe (Figure~\ref{fig:wearingTechProbe}A). Participants were then encouraged to experiment with customizing the setup to their own preferences, such as attaching straps in alternative positions (e.g., waist-to-shoulder). Both trainer and trainee wore the probe and explored three core feedback configurations:

\begin{enumerate}
    \item Inflatable air wedges (extrinsic, active pressure feedback)
    \item Rigid straps (intrinsic, non-elastic passive resistance)
    \item Stretchy straps (intrinsic, elastic passive resistance)
\end{enumerate}

This phase allowed participants to experience the tactile qualities of each configuration and develop an on-body sense of how feedback might support training. Since our focus was on personalizing feedback for trainees (\textbf{DG3}), trainers were not required to wear the probe beyond this stage.  

\subsubsection{Typical Training without Feedback}
To establish a baseline and observe natural trainer–trainee dynamics, the trainee removed or loosened the probe, and the trainer guided them through a typical coaching session using minimal weight. Heavier weights were available to explore more challenging scenarios, but their use was optional, supervised by the trainer, and contingent on the trainee’s comfort and consent. Participants were encouraged to take breaks or drink water during the study. This phase helped surface real-world challenges, and implicit feedback strategies.

\subsubsection{Co-Design Activity and Reflection}

\begin{figure}
    \centering
    \includegraphics[width=\linewidth]{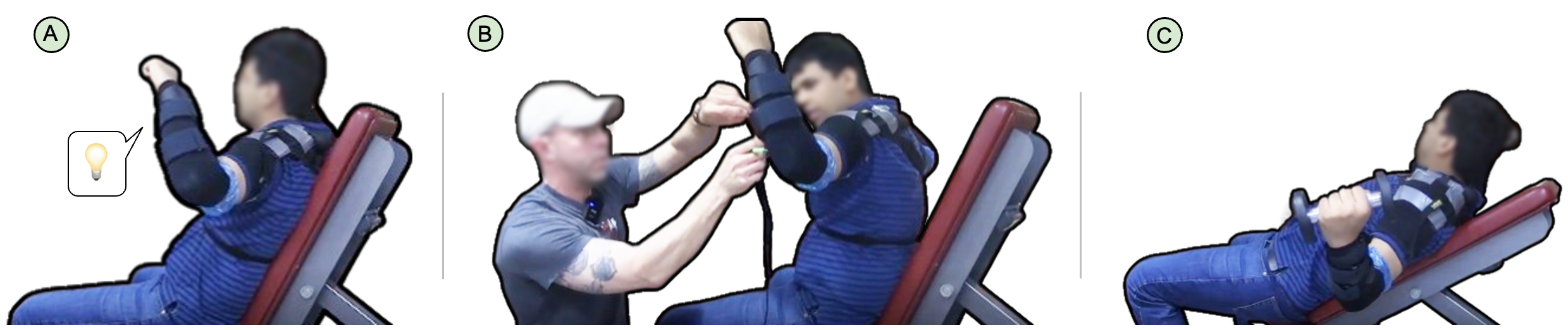}
    \caption{Example co-design prototyping flow from W1. (A) After performing a dumbbell incline fly, participant N1 identified the need to enforce a safe ROM. (B) Trainer E1 configured the probe by adding a rigid strap. (C) N1 repeated the exercise to experience the implemented feedback with the low-fi prototype.}
    \label{fig:codesflow}
\end{figure}

After the initial training, each trainer–trainee pair engaged in a collaborative co-design activity by sketching their ideas on the worksheet. They then configured the probe to reflect these design decisions, after which the trainee re-performed the exercise while wearing the configured probe. Participants reflected on how the feedback felt, discussed potential improvements, and recorded notes on the worksheet. If time permitted, they iterated on the design to better align with their intentions. The overall DIY process is demonstrated in Figure~\ref{fig:codesflow}

\subsection{Data Analysis}
\label{sec:codes-to-flexguard}

\subsubsection{Step 1: Code the Workshop Corpus}

We analyzed the audio and video recordings, transcripts, and annotated worksheets using reflexive thematic analysis \cite{reflexiveCoding}. Video was interpreted alongside transcripts and worksheets to examine movement demonstrations and embodied probe configurations \cite{videoanalysis}. The first author generated descriptive codes informed by prior sports-wearable frameworks \cite{sportsWearablesDesignSpace,mdpiReviewSportsWearable}. The first, fourth, and seventh authors jointly reviewed two workshops to establish a shared analytical orientation, after which the fourth and seventh authors reviewed the remaining workshops under the first author's supervision. Because each workshop paired a trainer and trainee, we retained participant role and distinguished trainer-provided coaching rationales, trainee accounts of embodied experience, and configurations negotiated jointly by each pair.

\subsubsection{Step 2: Consolidate Codes into Cross-Workshop Families}

We compared codes across exercises, participants, and probe configurations through affinity grouping and constant comparison \cite{affinityDiagramming,groundedTheory}. We consolidated wording variants only when they represented the same design-relevant property. For example, \textit{angle of motion}, \textit{unsafe ROM}, and \textit{elbow locking} formed a geometric family, whereas \textit{tempo}, \textit{swinging}, and \textit{excess momentum} formed a movement-dynamics family. We kept air-pressure, rigid-strap, and elastic-strap responses distinct because participants attributed different sources, sensations, and intervention roles to them.

\subsubsection{Step 3: Derive the Design Space Structure}

We treated each recurring cross-workshop family as a candidate design option and grouped options that answered the same design question into a shared dimension. We then grouped dimensions according to whether they described how a relevant movement condition is sensed and judged or how the resulting on-body feedback is delivered. This produced a three-level hierarchy following hierarchical affinity grouping and prior design-space methods \cite{affinityDiagramming,sportsWearablesDesignSpace}. The first level contains the two components \textit{Sensing \& Triggering} and \textit{Feedback Delivery}; the second contains eight design dimensions; and the third contains their corresponding options or values.

\subsubsection{Step 4: Refine Intersections and Boundaries}

We reapplied the candidate structure to the coded corpus, reviewed it with the broader author team, and compared it with recent design-space work \cite{sportsWearablesDesignSpace,nzWorkshopPaper}. We checked whether each option represented a reusable design decision across exercises and examined how values operated in combination. This process distinguished, for example, the functional location at which a condition is detected from the body location at which feedback is delivered. We retained exercise phase as a contextual index because it could alter several dimensions simultaneously, while safety, comfort, perceptibility, user agency, and feedback release were treated as constraints on complete configurations rather than additional dimensions.

\subsection{Synthesis of Co-Design Findings}
\label{sec:codesign-findings}

Across the workshops, trainers and trainees contributed complementary forms of knowledge to the design of on-body feedback. For example, during the dumbbell incline fly in W1, N1 identified the need to maintain a safe ROM, after which E1 translated that concern into a rigid-strap configuration that established a physical movement boundary. In W3, N3 reported feeling resistance specifically when moving the arm toward an undesired direction, demonstrating how the trainee's embodied response clarified whether the configured feedback communicated the intended correction. These episodes connected trainers' interpretations of movement with trainees' experiences of the resulting bodily feedback.

The trainer and trainee perspectives were aligned through iterative configuration and testing. In W6, E6 progressively increased air pressure as N6's arms drifted farther from the intended plane, translating the trainer's goal of correcting the deviation into feedback that initially preserved an opportunity for the trainee to self-correct before becoming stronger. In W8, after the trainee found the air wedge less restrictive than initially expected, they proposed inflating it during the raising phase and deflating it during the return, refining a general actuation concept into phase-specific feedback grounded in the felt movement. Such negotiations exposed recurring tensions between making feedback sufficiently noticeable or restrictive and preserving comfort, movement continuity, and opportunities for self-correction. They also showed that an appropriate intervention depended not only on the detected movement concern, but also on the trainee's body, the phase of the exercise, and how the feedback was experienced in practice.

Building from the initial codes generated through reflexive deductive and inductive coding of the workshop data, we consolidated related codes into recurring cross-workshop families and organized them according to two complementary parts of an on-body intervention. Families concerning what bodily or movement condition should be monitored, how that condition should be judged, and where it occurs formed the \textit{Sensing \& Triggering} component and its Trigger Metric, Trigger Policy, and Trigger Location dimensions. Families concerning how the bodily cue should be produced, modulated, timed, positioned, and functionally directed formed the \textit{Feedback Delivery} component and its Feedback Type, Feedback Intensity, Feedback Latency, Feedback Location, and Feedback Direction dimensions. Exercise phase provided context for combining these decisions, while safety, comfort, perceptibility, agency, and release constrained their application. Thus, the hierarchy was constructed by organizing code families according to the aspects of an on-body intervention that they described.

\section{FLEXGUARD}
\label{sec:flexguard-design-space}

\subsection{Structure and Terminology
\label{sec:flexguard-structure}}

We introduce \sys{}, a design space that models how on-body feedback can be sensed, triggered, and delivered to scaffold safety during strength training. \sys{} has a three-level structure. At the first level, two overarching components describe the major stages of an on-body intervention: \textit{Sensing \& Triggering}, which specifies how a safety-relevant condition is detected and judged, and \textit{Feedback Delivery}, which specifies how the resulting bodily cue is produced and applied. At the second level, these components contain eight design dimensions: Trigger Metric, Trigger Policy, Trigger Location, Feedback Type, Feedback Intensity, Feedback Latency, Feedback Location, and Feedback Direction. At the third level, each dimension contains a set of options or values, such as \ST{joint angle}, \ST{personalized}, \FD{elastic}, \FD{progressive}, and \FD{guiding}.

A configuration, or point in the design space, is formed by selecting and combining values across these dimensions. For example, one configuration may monitor \ST{joint angle} at a \ST{moving region} using an \ST{expert-based} policy and deliver \FD{intrinsic}, \FD{elastic}, \FD{progressive}, \FD{instant}, and \FD{guiding} feedback at the upper arm. Appendix~\ref{app:connecting-sensing-and-feedback} provides additional examples of how values across the two components can be combined.

We use five exercise phases---\textit{Initiation}, \textit{Execution}, \textit{Hold}, \textit{Return}, and \textit{Completion}---as a contextual index for applying these configurations. The \textit{Execution} and \textit{Return} phases may be either concentric or eccentric, depending on the exercise. Exercise phase is not an additional design dimension; rather, it provides context that may influence which values and combinations are appropriate. Similarly, safety, comfort, perceptibility, user agency, and the ability to release feedback are treated as constraints on configurations rather than as additional design dimensions.

Throughout the paper, we use \ST{blue styling} to denote dimensions and values within the \textit{Sensing \& Triggering} component and \FD{purple styling} to denote dimensions and values within the \textit{Feedback Delivery} component.

\subsection{Sensing \& Triggering}
\begin{figure}
    \centering
    \includegraphics[width=1\linewidth]{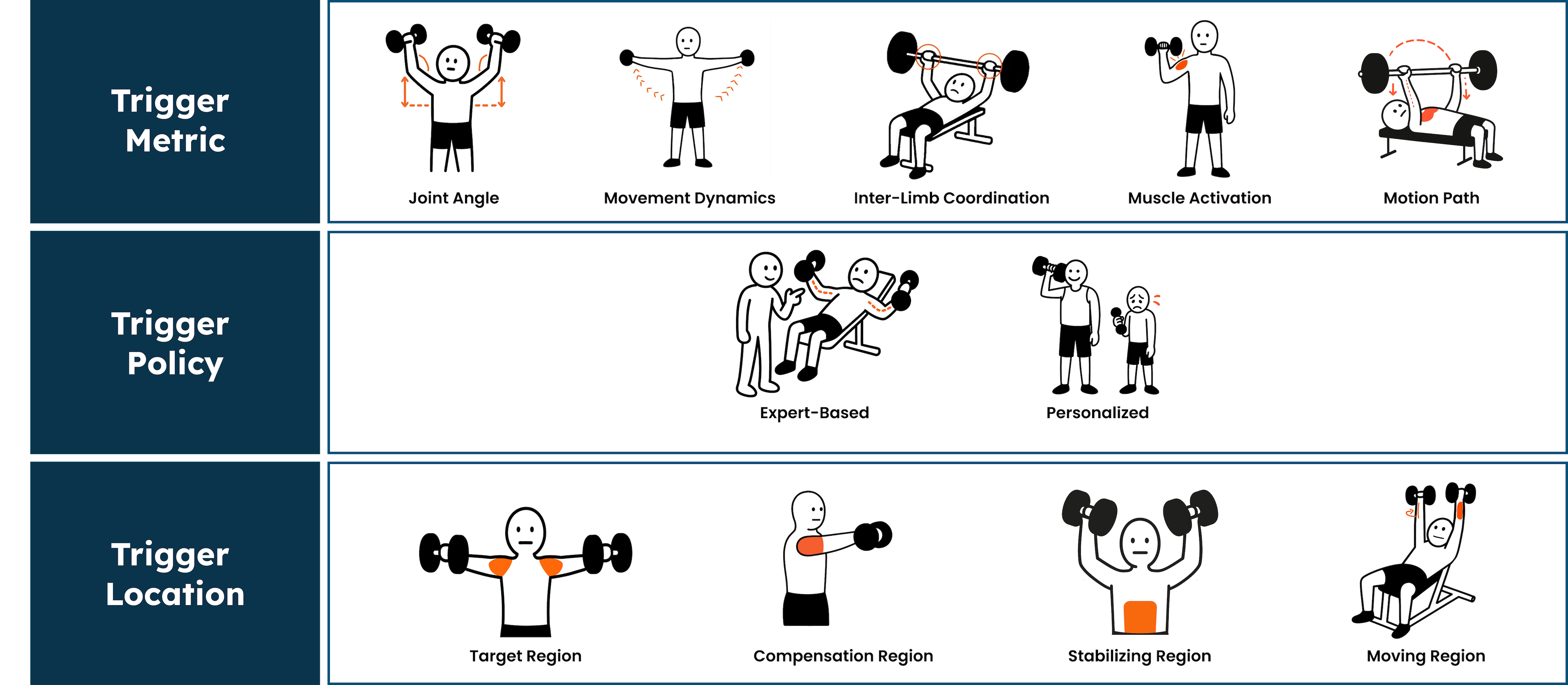}
    \caption{Design dimensions within the Sensing \& Triggering component of the \sys{} design space. 
    \textit{Trigger Metrics} capture what is being monitored, including \ST{joint angle}, \ST{movement dynamics}, \ST{inter-limb coordination}, \ST{muscle activation}, and \ST{motion path}. 
    \textit{Trigger Policies} determine when feedback is actuated or released, either through \ST{expert-based} knowledge (e.g., recommended ROM, standard joint angles) or through \ST{personalized} conditions (e.g., fatigue level, body type, or individual muscle engagement).
    \textit{Trigger Locations} describe where feedback is sensed, such as the \ST{target region} (primary muscles), \ST{compensation region} (unintended muscles taking over), \ST{stabilizing region} (muscles supporting posture), or \ST{moving region} (segments guiding trajectory).}
    \label{fig:sensingtriggering}
\end{figure}

Figure~\ref{fig:sensingtriggering} summarizes the design space for \textit{Sensing \& Triggering}, spanning \textit{Trigger Metric}, \textit{Trigger Policy}, and \textit{Trigger Location}. These dimensions define when and where the system should monitor to actuate feedback during strength training, as well as when to release it. Sensing must capture not only cues that warrant intervention but also signals indicating that feedback can be withdrawn once the user has corrected their form.

\subsubsection{Trigger Metric}
Trigger Metric answers what observable body state or movement property is evaluated. Its values are types of signals, so a system may monitor several metrics at once. Participants identified several physical characteristics that could serve as reliable triggers for feedback. Across multiple sessions, \ST{joint angle} was the most consistently mentioned marker (n=8), particularly when arms drifted out of alignment. As E3 noted, ``\textit{the only thing that has to be fixed is just the angle of motion}.'' Trainers also cautioned against end-range errors—``\textit{People often arch their back and then tend to raise their arms. I think that might be helpful but make sure that you don't go above range.}'' (W7). Others emphasized \ST{movement dynamics} (e.g., acceleration or tempo) as indicators of excessive momentum rather than controlled lifting; during a lateral raise, E9 explained to N9, ``\textit{Sometimes, I'm going too heavy and using too much momentum. For the last few reps, you can often try to get that extra set}.'' In W3, participants highlighted \ST{inter-limb coordination} (e.g., symmetric pacing/height in bench press) as a cue to synchronize both sides, noted with \textit{``It's a problem if the right hand is further than the left hand''}. In W8, feedback was tied to \ST{muscle activation}, with the trainer remarking, ``\textit{...almost feel like a little bit of pressure in the spot that you're trying to target... so you know that you're not using the wrong muscle}''. Three sessions (W2, W7, W9) discussed \ST{motion path}, noting that a shaky bar or wobbling dumbbell trajectory signals fatigue and breakdown of posture. For instance, participant noted \textit{``For this motion... If you do more up, it will hurt your joint.'' }

\begin{figure}
    \centering
    \includegraphics[height=0.20\textheight]{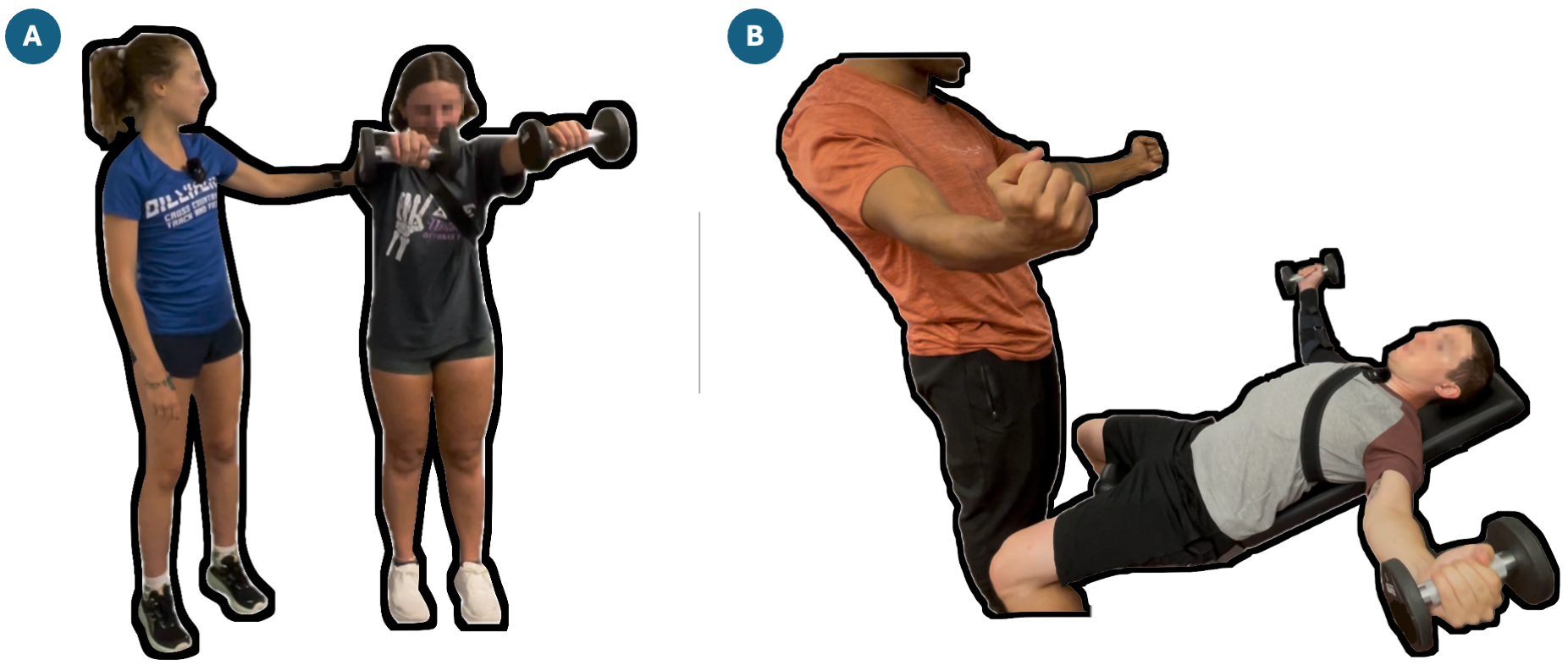}
    \caption{(A) Progressive support: E6 pumped additional air into the wedge as N6’s arms drifted out of plane during a dumbbell front hold, simulating stronger feedback over time to help complete the repetition. (B) Targeted activation: N9 demonstrated arching the back during a dumbbell incline fly to ensure chest muscles were activated.}
    \label{fig:s6s9codesfig}
\end{figure}

\begin{figure}
    \centering
    \includegraphics[height=0.22\textheight]{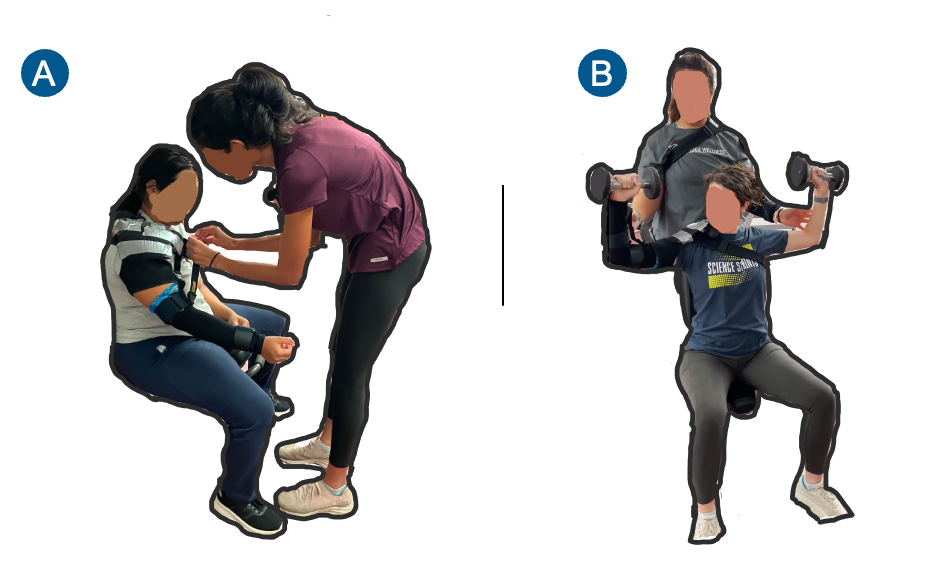}
    \caption{ (A) Trainer E7 adjusting and experimenting with different props alongside participant N7 during the co-design workshop. (B) Trainer E5 assisting trainee N5 by checking ROM and assessing prop comfort to determine appropriate feedback.}
    \label{fig:codesfig5}
\end{figure}

\subsubsection{Trigger Policy}
Trigger Policy specifies the decision rule that interprets one or more \textit{Trigger Metrics} and determines when feedback should begin, escalate, persist, or be released. Participants proposed rules for deciding when feedback should be actuated and when it should be withdrawn. Two values of Trigger Policy emerged. \ST{expert-based} policies draw on general training knowledge, such as recommended ROM, standard joint angles, and common coaching cues. Across 6 sessions, trainers tend to transfer their expertise verbally and give interventions when trainees drifted out of widely accepted guidance. For example, most trainers emphasized enforcing limits when elbows begin to lock during bench press, \textit{``You want to fully extend but you don't want to lock your elbows''}, or when it deviates from the shoulder-level path, \textit{``Sometimes people will lift them like this... we don't want that.''} In contrast, across 4 sessions, participants also highlight the need for \ST{personalized} policies that adapts the thresholds to individual trainee, accounting for factors such as fatigue, muscle engagement, or body type. For instance, feedback could be tuned to a user’s tolerance for fatigue or to whether the intended muscle is activating effectively, both of which can vary significantly across body types, as emphasized by E5 regarding N5's arms length, \textit{``Depending on how long this area is, this bubble needs to be further up as opposed to on my shoulder blade.''}. While all policies, including \ST{expert-based}, can ultimately be personalized, we make this distinction to separate those with only slight variations across most body types from those that are highly subjective.

\subsubsection{Trigger Location}
Trigger Location identifies the functional role of the region in which a condition is sensed. Several regions may be monitored for the same exercise. Rather than naming specific muscles, we describe locations in terms of their functional roles to allow for greater generalizability. Four values of Trigger Location emerged with a mean of 7 occurences by participants on each, corresponding to Figure~\ref{fig:sensingtriggering}. The \ST{target region} was a common focus, such as in the incline dumbbell fly: ``\textit{If you started overcompensating and using your wrong muscle, it would like identify that muscle and you’d know you don’t want [that]}'' (W6). Trainers also warned about overuse of \ST{compensation regions}, where non-target areas begin to take over the movement, as noted by E6 \textit{``If you feel an ache in another area, then you know that you're not doing it correctly''}. The importance of \ST{stabilizing regions}, particularly the upper back and core, was emphasized across sessions—``\textit{If it has a baseline measurement for  not engaging your core and then engaging your core, you'll able to find your center.}'' (W5). Finally, participants pointed to \ST{moving regions} (e.g., forearms and hands during an incline fly, or the wrists in a bench press) where feedback can guide trajectory directly. Although the muscle groups at these locations are not necessarily being worked, they act as reference points for maintaining correct movement.

\subsection{Feedback Delivery}

\begin{figure}
    \centering
    \includegraphics[width=1\linewidth]{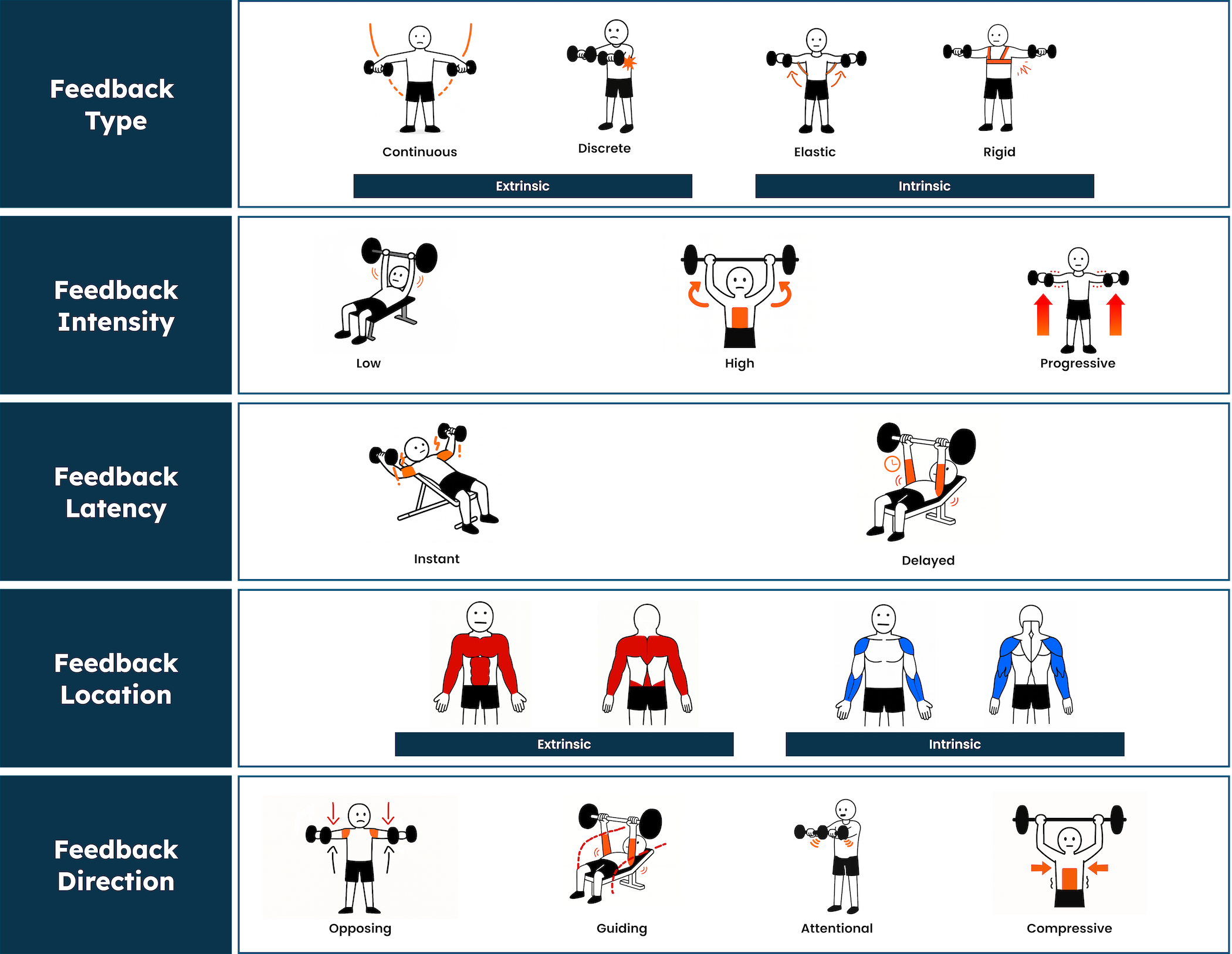}
    \caption{Design dimensions within the Feedback Delivery component of the \sys{} design space. 
    \textit{Feedback Type} distinguishes between \FD{extrinsic} (\FD{continuous} or \FD{discrete} actuation applied to the body) and \FD{intrinsic} feedback (\FD{elastic} or \FD{rigid} material-based feedback). 
    \textit{Feedback Intensity} ranges from \FD{low} to \FD{high}, with \FD{progressive} feedback that escalates as deviations persist. 
    \textit{Feedback Latency} describes whether cues are delivered \FD{instant}ly at the moment of deviation or \FD{delayed} to provide feedback at another time. 
    \textit{Feedback Location} indicates where feedback is applied on the body and varies by feedback type (see Appendix~\ref{app:body-zone-maps}). 
    \textit{Feedback Direction} specifies how corrective forces are oriented, including \FD{opposing}, \FD{guiding}, \FD{attentional}, and \FD{compressive} cues.}
    \label{fig:feedbackDeliveryDS}
\end{figure}

Feedback Delivery (Figure~\ref{fig:feedbackDeliveryDS}) captures how corrective cues are conveyed to the user once a triggering condition is met as defined in \textit{Sensing \& Triggering}. We define this not only as the onset of feedback, such as vibration, resistance, or pressure, but also its offset. Participants emphasized that the withdrawal of feedback was equally meaningful, since the absence of cues could serve as confirmation that posture had been corrected or movement had returned to the desired path. In this way, \textit{Feedback Delivery} encompasses both the activation and the release of feedback, as well as its modulation in intensity and timing.

\subsubsection{Feedback Type}
Feedback Type identifies the immediate source of the bodily sensation. \FD{extrinsic} feedback is generated by an actively controlled actuator whose output can be produced independently of movement, whereas \FD{intrinsic} feedback is experienced through the configured mechanical response of a worn structure as the body moves against or loads it. A system may combine several mechanisms, but each can be specified separately. 

\FD{extrinsic} feedback can be tuned to produce different actuation patterns. Participants described preferences for \FD{continuous} cues that guide movement throughout a repetition—``\textit{Imagine something helping draw a line for your hands to follow}'' (N2) which is a preference raised in 3 out of 9 sessions—as well as \FD{discrete} cues that act as one-time warnings when posture falters—``\textit{You shouldn’t nag them, just nudge once if they’re falling}'' (E1), emerged in 2 workshops (W1, W3).

\FD{intrinsic} feedback depends on how materials constrain or guide motion. We distinguish between \FD{rigid} and \FD{elastic} responses. \FD{rigid} feedback provides firm constraints, such as straps that enforce strict ROM limits: ``\textit{I could feel when I pulled wrong because the strap caught me}'' (W5). In contrast, \FD{elastic} feedback allows more flexibility, offering resistance that increases with deviation and guiding users back on track without completely halting movement. As E7 explained, ``\textit{So a strap that is going to help them maintain the stability. When I’m pressing it, I don’t want to go up and down… if the strap helps us to do that, then it’s pretty good.}'' Another participant added \textit{``Elastic is more flexible...it wouldn't stop me from doing the wrong movement. It would just notify me and I could readjust to my own.''} Comments favoring rigid constraints appeared at 5 sessions, while preferences for elastic guidance were raised in 3 sessions. 

\subsubsection{Feedback Intensity}
Feedback Intensity describes cue magnitude. \FD{low} and \FD{high} are fixed levels, while \FD{progressive} denotes a magnitude profile that changes as a deviation persists or worsens. \FD{low} intensity was favored for subtle prompts or early-stage correction (mentioned in 3 sessions). As E3 explains to N3, ``\textit{When someone exercises, they will not have concentration. When it comes to shaking (haptic feedback), it's not practical.}'' In contrast, \FD{high} intensity was valued for safety-critical moments, such as E6’s proposal for strong restricting feedback in the return phase of a lateral raise: ``\textit{If you started overcompensating and using wrong muscles, it would like identify that muscle and shift your arm back to make sure you're not using the wrong muscle. An ache in another area would let you know that you're not doing it correctly.}'' Participants also emphasized \FD{progressive} intensity from at least 3 sessions (feedback that begins subtly and escalates if the error persists or worsens) as in W6 where participant highlighted \textit{``Like slowly over time you can do less and less with the pressure.''}. This approach mirrors how a trainer might tolerate small deviations but intervene more forcefully over time as errors become more critical.

\subsubsection{Feedback Latency}
Feedback Latency is the interval between a \textit{Trigger Policy} becoming satisfied and the onset of feedback. \FD{instant} and \FD{delayed} describe onset timing. Immediacy is critical in situations involving high safety risk, where feedback must intervene before damage occurs, \textit{``If the strap helps make sure I don’t go too much in front or too much back… that prevents shoulder injury.''} Four sessions emphasized the need for immediate feedback especially for shoulder and elbow. At the same time, participants recognized the value of \FD{delayed} feedback that tolerates minor deviations and only escalates if errors persist. In W8, trainer states the need for guidance based on tempo not not urgent correction, \textit{``Maybe like click click click… you’d know you’re in the correct position… help with timing and slowing movement down.''} Such delayed feedback was seen as less disruptive and more supportive, as it felt less intimidating to users. Several also expressed preferences for \textit{progressive onset}, where feedback begins subtly and intensifies if the deviation continues. A more tolerable or delayed feedback were discussed in three sessions. Taken together, these insights suggest that \textit{Trigger Policies} must balance the safety benefits of \FD{instant} correction with the patience of \FD{delayed} feedback that reduces cognitive overload.

\subsubsection{Feedback Location}
Feedback Location specifies the body site where a cue is applied. This dimension describes where on the body feedback is delivered. From the co-design workshops, we derived body zone maps that highlight regions associated with feedback. \FD{extrinsic} feedback can be applied directly to the same muscle group identified by \textit{Sensing \& Triggering}, or redirected to a different site to signal correction. In W5, a participant remarks \textit{``Bench press on wrist is the only one I think uses resistance.''} A participant also annotated \textit{``A strap in the elbow bend would help''} from the co-design workshop in the dumbbell incline fly worksheet as shown in Figure ~\ref{fig:fullworksheet}. In contrast, \FD{intrinsic} feedback is delivered at the site of motion itself, where resistance provides direct guidance. For example, constraining elbow flare or stabilizing the shoulders. Several participants highlighted how resistance at the limb could guide technique, such as W4's observation that \textit{``If you tighten it from this point, there’s a chance that I won’t be able to drop my arm down,''} underscoring how intrinsic feedback constrains movement at the source.

\subsubsection{Feedback Direction}
Feedback Direction specifies the functional action of a cue relative to the intended movement or stabilization goal. This dimension describes how corrective forces or cues are oriented relative to the body. We identified four strategies. \FD{opposing} feedback was discussed in 6 of 9 sessions, to resist unwanted momentum, such as adding friction in a lateral raise to counter swinging as N3 marks \textit{``Right now I feel the restriction in the different direction. When I try to lean my arm toward my body, I feel the restriction this way.''}. 4 sessions mentioned \FD{guiding} feedback nudges the user toward the correct path, for example assisting with rep completion or steering the arms back into the proper motion plane, as N6 described when correcting form during a dumbbell incline fly, ``\textit{you know, to kind of like shift your arm back to make sure you're not using the wrong muscle.}''. \FD{attentional} feedback provides localized cues without displacement, such as vibration in a dumbbell front hold to warn when the arms drop below a safe angle as suggested by N1. Finally, \FD{compressive} feedback applies inward pressure, for instance stabilizing the core to maintain posture, which participants in three sessions (e.g., E3, E4, E5) emphasized with their trainees, ``\textit{always engage the core}.''

\subsection{Positioning and Applying \sys{}}
\label{sec:flexguard-positioning}

\subsubsection{Relation to Broader Sports-Wearable Design Spaces}

Table~\ref{tab:flexguard-positioning} positions \sys{} alongside the two closest sports-wearable design spaces. Turmo Vidal et al. provide a broad account of sports and fitness wearables across practice, technology, wearability, and wearable use in practice \cite{sportsWearablesDesignSpace}. Stawarz et al. focus on personalizing flexible wearable sensing through sensor-level, user-specific, contextual, and temporal adaptations \cite{nzWorkshopPaper}. \sys{} complements these spaces by specializing the point at which sensed movement becomes an on-body safety intervention.

The comparison positions \sys{} as complementary to these broader design spaces by operating at a finer level of resolution. Turmo Vidal et al. characterize design choices around input, output, wearability, and use in practice, while Stawarz et al. foreground sensor-level, user-specific, contextual, and temporal adaptation. \sys{} specializes these concerns for safety scaffolding in strength training by making explicit which movement metric is monitored, how and where an intervention is triggered, and how the resulting on-body feedback is configured through its type, intensity, latency, location, and direction.

\begin{table}[H]
\centering
\caption{Comparison of the design-space elements articulated by prior work and \sys{}.}
\label{tab:flexguard-positioning}
\begingroup
\scriptsize
\setlength{\tabcolsep}{2.5pt}
\renewcommand{\arraystretch}{1.08}
\arrayrulecolor{black!55}

\begin{tabularx}{\linewidth}{
|>{\columncolor{FGTableCluster}\Centering\arraybackslash}p{0.14\linewidth}
|>{\RaggedRight\arraybackslash}p{0.245\linewidth}
|>{\RaggedRight\arraybackslash}p{0.245\linewidth}
|>{\columncolor{FGTableGap}\RaggedRight\arraybackslash}X|
}
\hline

\rowcolor{FGTableHeader}
\textbf{Design component} &
\textbf{Turmo Vidal et al.~\cite{sportsWearablesDesignSpace}} &
\textbf{Stawarz et al.~\cite{nzWorkshopPaper}} &
\textbf{\sys{}} \\
\hline


\multirow[c]{3}{=}{%
    \makecell[c]{%
        \textbf{\textit{Sensing \&}}\\
        \textbf{\textit{Triggering}}
    }
}
&
\begin{itemize}[
    nosep,
    leftmargin=1.05em,
    labelsep=0.3em
]
    \item Implicit; Explicit
    \item Movement; Proxemics; Heart Rate; Pressure; Breath
    \item Gaze; Gestures; Muscle Engagement; Temperature; Touch
    \item Interaction with Screen; Voice; Scores; Video
\end{itemize}
&
\begin{itemize}[
    nosep,
    leftmargin=1.05em,
    labelsep=0.3em
]
    \item Sensor Selection
    \item Output Data Granularity
\end{itemize}
&
\textbf{Trigger Metric}
\begin{itemize}[
    nosep,
    leftmargin=1.05em,
    labelsep=0.3em,
    topsep=1pt
]
    \item \ST{joint angle}
    \item \ST{movement dynamics}
    \item \ST{inter-limb coordination}
    \item \ST{muscle activation}
    \item \ST{motion path}
\end{itemize}
\\
\cline{2-4}

&
\mbox{}
&
\begin{itemize}[
    nosep,
    leftmargin=1.05em,
    labelsep=0.3em
]
    \item User-Specific Adaptation
    \item Body Variability
    \item Skill Level
    \item Role
    \item Temporal Adaptation
\end{itemize}
&
\textbf{Trigger Policy}
\begin{itemize}[
    nosep,
    leftmargin=1.05em,
    labelsep=0.3em,
    topsep=1pt
]
    \item \ST{expert-based}
    \item \ST{personalized}
\end{itemize}
\\
\cline{2-4}

&
\begin{itemize}[
    nosep,
    leftmargin=1.05em,
    labelsep=0.3em
]
    \item Anchor Points: One; Many
    \item Location: Body; Equipment
    \item Strict; Lenient
\end{itemize}
&
\begin{itemize}[
    nosep,
    leftmargin=1.05em,
    labelsep=0.3em
]
    \item Sensor Placement
    \item Sport \& Body Motion Type
\end{itemize}
&
\textbf{Trigger Location}
\begin{itemize}[
    nosep,
    leftmargin=1.05em,
    labelsep=0.3em,
    topsep=1pt
]
    \item \ST{target region}
    \item \ST{compensation region}
    \item \ST{stabilizing region}
    \item \ST{moving region}
\end{itemize}
\\
\hline


\multirow[c]{5}{=}{%
    \makecell[c]{%
        \textbf{\textit{Feedback}}\\
        \textbf{\textit{Delivery}}
    }
}
&
\begin{itemize}[
    nosep,
    leftmargin=1.05em,
    labelsep=0.3em
]
    \item Visual; Auditory; Haptic
\end{itemize}
&
\mbox{}
&
\textbf{(On-Body) Feedback Type}
\begin{itemize}[
    nosep,
    leftmargin=1.05em,
    labelsep=0.3em,
    topsep=1pt
]
    \item \FD{extrinsic}: \FD{continuous}; \FD{discrete}
    \item \FD{intrinsic}: \FD{rigid}; \FD{elastic}
\end{itemize}
\\
\cline{2-4}

&
\mbox{}
&
\mbox{}
&
\textbf{(On-Body) Feedback Intensity}
\begin{itemize}[
    nosep,
    leftmargin=1.05em,
    labelsep=0.3em,
    topsep=1pt
]
    \item \FD{low}
    \item \FD{high}
    \item \FD{progressive}
\end{itemize}
\\
\cline{2-4}

&
\begin{itemize}[
    nosep,
    leftmargin=1.05em,
    labelsep=0.3em
]
    \item Interaction Time
    \item Synchronous; Asynchronous
\end{itemize}
&
\mbox{}
&
\textbf{(On-Body) Feedback Latency}
\begin{itemize}[
    nosep,
    leftmargin=1.05em,
    labelsep=0.3em,
    topsep=1pt
]
    \item \FD{instant}
    \item \FD{delayed}
\end{itemize}
\\
\cline{2-4}

&
\begin{itemize}[
    nosep,
    leftmargin=1.05em,
    labelsep=0.3em
]
    \item Location: Body; Equipment
    \item Anchor Points: One; Many
\end{itemize}
&
\mbox{}
&
\textbf{(On-Body) Feedback Location}
\begin{itemize}[
    nosep,
    leftmargin=1.05em,
    labelsep=0.3em,
    topsep=1pt
]
    \item Body Zones (Appendix~\ref{app:body-zone-maps})
\end{itemize}
\\
\cline{2-4}

&
\mbox{}
&
\mbox{}
&
\textbf{(On-Body) Feedback Direction}
\begin{itemize}[
    nosep,
    leftmargin=1.05em,
    labelsep=0.3em,
    topsep=1pt
]
    \item \FD{opposing}
    \item \FD{guiding}
    \item \FD{attentional}
    \item \FD{compressive}
\end{itemize}
\\
\hline

\end{tabularx}

\arrayrulecolor{black}
\endgroup
\end{table}

\subsubsection{Applying \sys{}}

In this section, we demonstrate how \sys{} can be applied for describing systems through its design dimensions. We show how relevant papers' sensing, triggering, and feedback configurations can be articulated within \sys{}.

FootStriker~\cite{FootStrikerIMWUT} detects heel strikes during running and applies electrical muscle stimulation to the calf to help users adjust their foot-strike pattern. In \sys{}, its sensing can be considered as \ST{movement dynamics} at a \ST{moving region} under an \ST{expert-based} policy, while its stimulation can be considered as \FD{extrinsic}, \FD{discrete}, \FD{instant}, and \FD{guiding} feedback.

ThermoFit~\cite{ThermoFitIMWUT} describes a fabrication pipeline for body-fitting smart orthoses with embedded sensing and configurable metamaterial structures. In its finger-extension splint application, an accelerometer at the fingertip monitors finger bending, which can be considered as \ST{joint angle} at a \ST{moving region} under an \ST{expert-based} rehabilitation policy. As the user bends the finger, the thermoplastic spring provides an immediate reaction force that assists its return toward extension. In \sys{}, this response can be described as \FD{intrinsic}, \FD{elastic}, \FD{instant}, \FD{progressive}, and \FD{opposing} feedback applied at the moving finger joint.

\section{SPEED DATING}
\label{sec:speed-dating}

\begin{figure}
    \centering
    \includegraphics[width=1\linewidth]{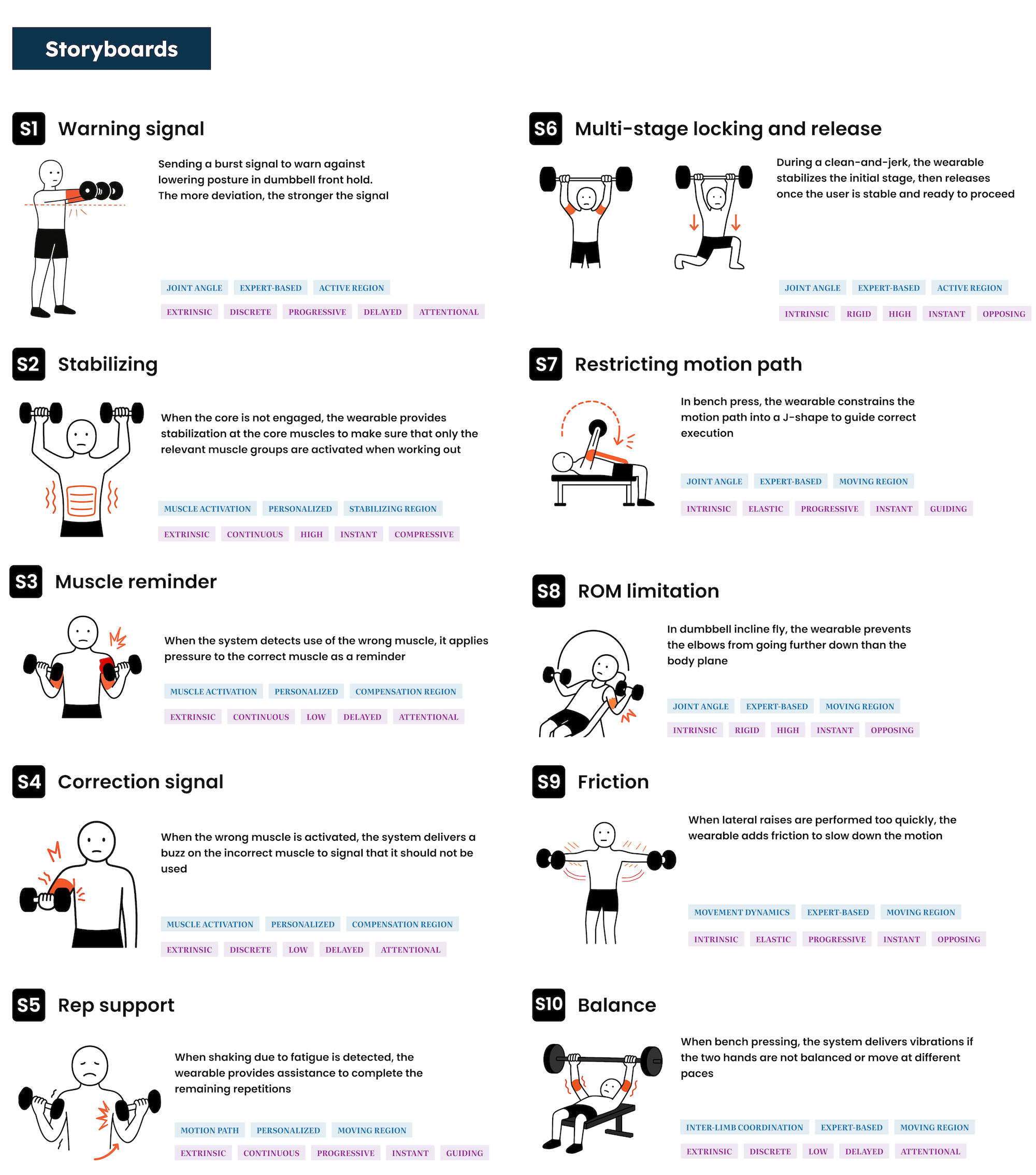}
    \caption{Compilation of ten storyboard images illustrating concepts of on-body feedback wearables scaffolding safety in strength training across different scenarios. Blue and purple labels indicate the instantiated design dimensions.}
    \label{fig:speeddatingstoryboards}
\end{figure}

To evaluate the \sys{} design space , we conducted a multi-stakeholder speed dating workshop \cite{PSD_AIED, PSD_CSCW}, in which participants from multiple stakeholder groups evaluated a set of storyboards from their respective perspectives.  Each storyboard depicted speculative scenarios combining different feedback modalities, body locations, exercise types, and sensing conditions. The goal was to validate and refine the design space with users and expert trainers, while surfacing the strengths and limitations of different design ideas across scenarios.

We developed ten storyboard scenarios (S1–S10) that instantiated different configurations across design dimensions within the \textit{Sensing \& Triggering} and \textit{Feedback Delivery} components (see Fig.~\ref{fig:speeddatingstoryboards}).

\subsection{Study Protocol}
\label{sec:speed-dating-overall-protocol}

To capture perspectives from both stakeholder groups, we first conducted workshops with beginner trainees to identify usability concerns and user needs. We then conducted workshops with expert trainers, focusing on safety considerations and scaffolding strategies while presenting the synthesized themes from the prior workshops. Our study was reviewed and approved by our institution’s IRB.

\subsubsection{Workshops with Beginner Trainees}
\label{sec:speed-dating-beginner-protocol}
We recruited seven beginners (B1–B7) with fewer than three months of consistent strength training experience to participate in three workshops (ages 24–34; 5 male, 2 female). Each workshop was conducted in our lab space, which was equipped with two pairs of dumbbells (2 and 12 pounds) and an adjustable workout bench. To establish context, participants were first asked to perform two exercises, dumbbell incline fly and lateral raise, by following short tutorial videos\footnote{https://www.youtube.com/@nutritioneering}, as these movements cover two primary degrees of freedom of the shoulder. Participants were also free to try additional exercises from the tutorial set if desired.

Following this exercise phase, participants were presented with a series of storyboards illustrating scenarios from the \sys{} design space. They were asked to reflect on whether the scenarios aligned with their needs, expectations, and perceived challenges in strength training. Participants could continue to access the equipment and tutorial videos throughout the session. Each workshop lasted approximately one hour, with the first 10 minutes dedicated to exercise practice and the remaining 50 minutes to storyboard discussion. Participants received \$20 in compensation for their time.

\subsubsection{Workshops with Expert Trainers}
\label{sec:speed-dating-expert-protocol}

We recruited six personal trainers for one-on-one online speed dating sessions (C1-C6; three male and three female). Each expert trainer session began with a short orientation to the project and the same storyboard set used in the beginner phase. For each storyboard, the trainer first provided an assessment of safety and coaching validity. We then showed a short summary of beginner trainee reactions from the first phase and asked the trainer to respond to trainee needs, identify points of agreement or tension, and propose revisions to the storyboard when needed. Each workshop lasted approximately one hour. Participants received \$25 in compensation for their time.

\subsection{Data Analysis}
\label{sec:speed-dating-data-analysis}

Two researchers analyzed the workshop transcripts in two passes. First, they organized evidence by stakeholder role: beginner trainees' accounts of interpretability, usefulness, comfort, and retained control; expert trainers' assessments of coaching validity, safety, and practical applicability; and points of alignment or tension between the two groups. Second, they mapped excerpts to the five open questions from F1--F5, asking whether a storyboard supported the relevance of a design dimension, bounded that dimension to a particular training situation, or suggested a refinement to \sys{}. The researchers compared the role- and finding-based summaries and resolved differences through discussion.

\subsection{Results}
\label{sec:speed-dating-results}

\subsubsection{Confirming Coverage and Refining the Design Space}

We first examined whether participants could identify a plausible use for the configurations instantiated across the ten storyboards and whether any part of \sys{} should be removed or refined. Participants identified potential value across the storyboard set. However, they repeatedly bounded that value by the context of training purpose, load, and perceived risk. For example, \textsc{S3} was considered useful for tutorial, but for the use case of post-set analysis, \textsc{S4} might be more appropriate to detect errors.

An example of design space refinement concerned \ST{inter-limb coordination}. We had questioned whether a wearable was needed because users can often feel an imbalance, yet B1, B2, B5, and B7 valued a cue that confirmed and localized the problem. As B5 explained, ``\textit{Although I can feel something is off, the vibration tells me exactly what's happening.}'' We therefore retained inter-limb coordination while clarifying that its role is to disambiguate a sensed problem rather than replace proprioception.

\subsubsection{F1: Complementary Roles for Feedback Mechanisms}

Participants assigned different roles to the two feedback types rather than treating them as interchangeable. Intrinsic resistance was interpreted as an embodied path or boundary: when discussing \textsc{S8}, B2 explained, ``\textit{If the material itself keeps me from going further, I immediately know what I'm doing wrong.}'' Extrinsic cues were interpreted more symbolically, such as vibration that draws attention to an asymmetry in \textsc{S10} or pressure that resembles a trainer's tap in \textsc{S3}. These responses supported retaining both types: intrinsic feedback can express \emph{how} movement should change, whereas extrinsic feedback can signal \emph{what} requires attention or \emph{when} to respond.
C3 made this distinction concrete when evaluating motion-path guidance: a ``\textit{flexible force or a slight nudge}'' could guide the movement itself, whereas the corresponding attentional cue should be ``\textit{something very light, like a soft buzz}'' that would not scare the trainee or cause them to drop the weight.

\subsubsection{F2: Coaching Priorities Shift Across Feedback Windows}

We first observed a common preference for minimizing feedback during active movement, particularly under load. This observation motivated us to define the notion of a \emph{feedback window}, which distinguishes the purpose and loading conditions of an interaction from the biomechanical \emph{movement phase}. Four feedback windows emerged from the study. During \emph{calibration}, C2 reframed progressive repetition support as an assessment procedure: gradually increasing ROM or load to estimate the trainee's current maximal capability and personalize later thresholds. During a \emph{tutorial or low-load practice}, repeated cues support both movement learning and interpretation of the feedback itself. For example, participants valued tempo feedback in \textsc{S9}, while noting that repeated reminders become less necessary once the movement is learned. During a \emph{loaded working set}, participants preferred a single subtle, actionable cue during safe execution---``\textit{It should be subtle so it doesn't make me jump mid-movement}'' (\textsc{S3})---but accepted stronger interventions when injury was perceived as imminent (\textsc{S8}). Finally, \emph{between sets}, participants preferred reviewing lower-priority explanations, muscle-engagement details, and summaries of repeated errors without competing with movement (C2, C3, and C5).

The stakeholder comparison also exposed a priority order. Trainers placed immediate safety and mechanically acceptable execution before tempo refinement, muscle-specific reminders, or motivation. C3 explained that ``\textit{safety is always going to be first}'' and ``\textit{the movement pattern is always the [next] priority},'' and C5 described stopping at ``\textit{form failure for beginners, not muscular failure},'' while beginners emphasized whether the selected cue was understandable and usable in the moment. Thus, broader sensing does not justify more simultaneous feedback; it requires selecting the issue that is most urgent, reliable, and actionable within the current feedback window.

\subsubsection{F3: Correctness Requires Expert Anchors and Personal Calibration}

Participants supported expert-based boundaries but resisted treating one movement template as universally correct. B1, B2, B5, and B6 considered expert ROM limits useful for escalating feedback near a boundary, while personalization was discussed in relation to fatigue, body proportions, injury status, exercise variation, and training goal. One beginner noted that ``\textit{Once I've learned and memorized the correct movement, a reminder might be less necessary.}'' Trainers likewise described acceptable variation as exercise- and goal-dependent rather than a single right-or-wrong posture. C1, C3, and C6 likewise emphasized person-specific ROM and path variation. Together, these responses support using expert knowledge as an initial anchor and calibration to adapt that anchor to the trainee's current body and objective.

\subsubsection{F4: A Small and Consistent Coaching Language}

Participants wanted cue meanings to remain stable and limited. B7 cautioned, ``\textit{The feedback has to be consistent. Otherwise I won't know what to do.}'' Trainers similarly warned that correcting several body regions at once could overwhelm novices and favored one encouraging, actionable cue over simultaneous messages. C3 warned that cueing the quads, glutes, and core could become ``\textit{very overwhelming very quickly}.'' In comparing muscle reminders with correction signals in \textsc{S3} and \textsc{S4}, trainers treated the reminder as more suitable during movement and the richer correction as more suitable for post-set explanation . These results bound F4: a wearable should use a compact vocabulary whose meaning is rehearsed and whose lower-priority detail can be deferred.

\subsubsection{F5: Preserve Movement Continuity, Agency, and Safe Release}

\begin{figure}[H]
    \centering
    \includegraphics[height=0.30\textheight]{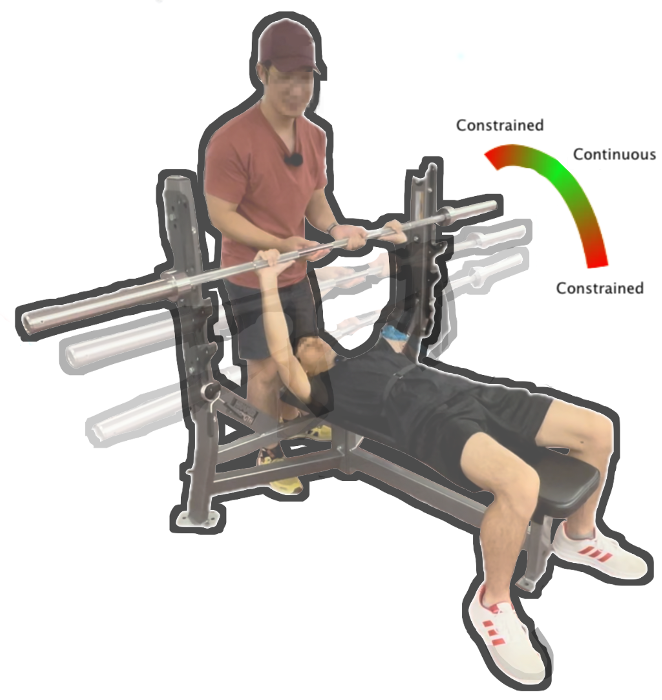}
    \caption{Illustration of how participants conceptualized movement continuity in on-body feedback. Feedback that allows movement to continue supports comfort and autonomy during execution, while feedback that interrupts or constrains movement becomes preferable at safety-critical moments, such as approaching ROM boundaries.}
    \label{fig:rig-elastic}
\end{figure}

Participants accepted intervention authority only when it remained proportionate and recoverable (Figure~\ref{fig:rig-elastic}). They preferred feedback that permitted self-correction during safe movement, but accepted stronger interruption near a perceived ROM or injury boundary. Locking in \textsc{S6} exposed the opposite case: B5 worried that a constrained user might be unable to respond safely if a weight fell, and B3 proposed an emergency release. Beginners also worried that excessive support could reduce opportunities to internalize correct movement, whereas tempo feedback in \textsc{S9} was seen as supporting learning without taking over the repetition. Trainers similarly preferred modifying the exercise or prompting self-correction before allowing a wearable to complete or override movement. C4 emphasized that ``\textit{we are always working toward autonomy},'' while C2 and C6 argued that assistance should wait until the trainee had a genuine chance to finish the repetition.

\section{DESIGN IMPLICATIONS}
\label{sec:design-implications}

We present five guidelines for applying the dimensions of \sys{}. Each guideline connects an unresolved question from the literature to empirical evidence and specifies how designers can combine values across design dimensions within the \textit{Sensing \& Triggering} and \textit{Feedback Delivery} components.

\subsection{Guideline 1: Use Extrinsic Cues to Communicate and Intrinsic Mechanics to Shape Movement}
\label{sec:design-implication-1}

F1 shows that feedback mechanisms are not interchangeable. Vibrotactile systems can encode spatial and temporal messages without continuously constraining movement \cite{VibroMap,HapticPilotIMWUT,VibraForge}, while electrical, pneumatic, and cable-driven systems can apply increasingly direct forces \cite{FootStrikerIMWUT,ForceJacket,Kinethreads}. Intrinsic mechanisms instead communicate through a worn structure's configured response, including orthoses, selectively stiffening garments, and passive material interfaces \cite{BistableOrthoses,selectivelyStiffening,ThermoFitIMWUT,wearableMaterialProps}. Table~\ref{tab:discussion-comparison} summarizes the resulting division.

Our studies suggest using extrinsic feedback when the primary need is explicit communication---for example, directing attention, marking timing, or signaling a state change---and intrinsic feedback when the movement itself should reveal a path, support, or boundary. A system can use both roles together, but the mechanisms should remain interpretable as separate choices: an extrinsic cue can identify \emph{what} needs attention, while intrinsic mechanics embody \emph{how} the trainee might respond.

\textbf{Guideline.} Select a feedback mechanism according to the role it must perform, not technical availability. Use extrinsic feedback primarily for explicit communication and intrinsic feedback primarily for predictable movement-coupled guidance, stabilization, or boundaries.

\begin{table}[H]
\centering
\caption{Complementary roles of extrinsic and intrinsic feedback.}
\label{tab:discussion-comparison}
\begingroup
\scriptsize
\setlength{\tabcolsep}{3pt}
\renewcommand{\arraystretch}{1.04}
\arrayrulecolor{black!55}
\newcommand{\FGTableList}[1]{\begin{itemize}[leftmargin=1.1em,nosep,topsep=0pt,partopsep=0pt]#1\end{itemize}}
\begin{tabularx}{\linewidth}{|>{\columncolor{FGTableCluster}\RaggedRight\arraybackslash}p{0.17\linewidth}|>{\RaggedRight\arraybackslash}X|>{\RaggedRight\arraybackslash}X|}
\hline
\rowcolor{FGTableHeader}
\textbf{Design concern} & \textbf{Extrinsic feedback} & \textbf{Intrinsic feedback} \\
\hline
\textbf{Primary role} & \FGTableList{\item Direct attention to a region or event.\item Communicate timing, state, urgency, or a symbolic instruction, as in trainer-like tapping \cite{TingleTouch}.} & \FGTableList{\item Embody a path, boundary, stabilization strategy, or graded resistance.\item Make the correction perceptible through the movement itself \cite{selectivelyStiffening,wearableMaterialProps}.} \\
\hline
\textbf{Temporal behavior} & \FGTableList{\item Can be triggered, patterned, and withdrawn independently of movement.\item Location, timing, and intensity can be authored dynamically \cite{VibraForge,HapticPilotIMWUT}.} & \FGTableList{\item Response is coupled to deformation, preload, and movement.\item Behavior often persists while the structure remains configured or loaded.} \\
\hline
\textbf{Drawbacks} & \FGTableList{\item The cue may be missed, mislocalized, or mapped inconsistently.\item Concurrent channels may interfere perceptually \cite{ieeeSkinStretchTesting}.} & \FGTableList{\item Fit, bulk, preload, or stiffness may alter comfort and effective loading.\item Constraint may reduce movement continuity or a safe recovery path \cite{FrozenSuit,funcNowWearableLater}.} \\
\hline
\end{tabularx}
\arrayrulecolor{black}
\endgroup
\end{table}

\subsection{Guideline 2: Arbitrate Coaching Priorities Before Actuating Feedback}
\label{sec:design-implication-2}

F2 shows that broader sensing creates a selection problem rather than automatically improving coaching. Wearables can recognize exercises and repetitions \cite{RecoFit,LiftSmartUbiComp,ProxiFitIMWUT}, estimate loading or muscle engagement \cite{ACLGuard,MuscleRehab}, and translate movement into descriptive feedback \cite{UbiPhysio}. When several signals deviate, the system must still rank them by severity, actionability, confidence, and relevance to the current coaching objective. We use \emph{feedback window} to distinguish this context from the biomechanical movement phases in \sys{}.

\subsubsection{Calibration and Assessment}

Calibration establishes what the system should treat as normal, coachable, or unsafe for a particular trainee. From speed dating, C2 reframed progressive repetition support as an assessment procedure: increase ROM or load incrementally to estimate current capability, rather than use assistance to complete a working set. The process should begin from expert-defined exercise and safety boundaries, use no or light load, obtain explicit consent from the user for physical guidance, and stop when control deteriorates.

\textbf{Guideline.} Prioritize feedback that establishes the trainee's baseline capability over feedback that improves immediate exercise performance. Begin from expert defaults, vary ROM or load only in controlled increments, distinguish assisted from independent capability, and store a personalized capability envelope with visible confidence and safety bounds.

\subsubsection{Tutorial and Low-Load Practice}

Tutorials and warm-up or low-load practice are appropriate for teaching movement paths and the wearable's cue vocabulary. More continuous or physically directive guidance can be acceptable here because the trainee is learning, the load is limited, and the feedback can be paused for explanation. Demonstration-based coaching systems and actuating-wearable research similarly treat early practice as a period for explicit guidance that should later fade \cite{MotionMA,basketballCoach,actuatingWearablesSkillLearning}.

\textbf{Guideline.} Prioritize feedback that teaches movement and cue meaning. Rehearse cue meanings, demonstrate movement paths, support deliberate self-correction, and require opt-in for forceful physical guidance.

\subsubsection{Loaded Working Sets}

During loaded repetitions, the highest priority is a recoverable movement path. Trainers in our studies placed immediate safety and substantial mechanical breakdown before tempo refinement, muscle-specific feedback, or motivation. Feedback should therefore communicate at most one familiar, immediately actionable issue; a surprising force or unfamiliar cue may startle the trainee or alter loading. Human supervision likewise involves deciding what is useful to address now rather than reporting every observable deviation \cite{fisher2023supervision,fisher2025supervision}. 

On-body assistance that actively supports movement, as illustrated in \textsc{S2}, can conceal pain or fatigue, create a false impression of independent capability, and foster reliance on the wearable. Systems should therefore avoid physically assisting the movement during loaded working sets under ordinary conditions. Forceful intervention should be reserved for narrowly defined severe hazards, such as an imminent barbell drop onto the trainee.

\textbf{Guideline.} Prioritize immediate safety feedback during loaded working sets. Avoid communicating issues that are not safety-related. Defer tempo, muscle-specific, and motivational feedback. Reserve strong forceful interruption for severe hazards.

\subsubsection{Between-Set Reflection}

Rest intervals provide an opportunity to review multiple issues that differ in priority, urgency, and topic. This period supports reflection, explanation, and planning for subsequent sets. The system can summarize persistent deviations, explain why cues were triggered, present muscle-engagement or trajectory evidence, and adjust coaching priorities for the next set. Prior work similarly supports using post-exercise reflection to help users review and understand their performance \cite{UbiPhysio,longTermPhyData}.

\textbf{Guideline.} Prioritize explanatory and performance-refinement feedback after the set. Present multi-metric summaries, muscle-engagement details, and repeated-error analyses between sets, and use the reflection to negotiate the next set's coaching objective.

\subsection{Guideline 3: Personalize Expert Guidance to the Individual}
\label{sec:design-implication-3}

F3 shows that no single reference fully defines correct movement for every trainee. Demonstrations and expert procedures provide valuable starting points \cite{MotionMA,basketballCoach}, but appropriate movement and feedback also depend on the trainee's body proportions, mobility, experience, injury history, fatigue, and many other factors. Intelligent wearables should therefore treat expert knowledge as an initial anchor rather than as a universal ground truth. Prior work on personalized coaching similarly emphasizes adapting support to changing capabilities, goals, and stakeholder needs instead of applying a fixed template to every user \cite{physiotherapyML,nzWorkshopPaper,phyActivityGoalsModeration}.

Personalization should influence both what the wearable considers relevant and how it responds. Expert knowledge can initialize exercise goals and recognized safety boundaries, while individual information should refine acceptable movement ranges, intervention thresholds, feedback priorities, and the type and intensity of feedback delivered. The same observed deviation may therefore warrant no intervention for one trainee, subtle guidance for another, or a stronger warning when it approaches that individual's safety boundary. These personalized settings should also evolve as the trainee's capability, fatigue, load, goals, or physical condition changes.

Because personalization can materially change when and how a wearable intervenes, these adaptations should remain visible and explainable to the trainee \cite{personalSensingNoti}. Systems should communicate which recommendations originate from expert knowledge, which have been personalized, and why a personalized threshold or feedback strategy has changed.

\textbf{Guideline.} Always personalize expert guidance to the individual trainee. Use expert knowledge as a starting point, not as a universal definition of correct movement, and adapt movement goals, intervention thresholds, feedback priorities, and feedback delivery to the trainee's body, capability, goals, preferences, and current condition. Update these adaptations as the trainee changes, and keep consequential personalization visible and explainable.

\subsection{Guideline 4: Use a Small, Stable Coaching Language}
\label{sec:design-implication-4}

F4 shows that successful actuation does not guarantee successful communication. Vibrotactile perception varies with body location, spacing, and posture \cite{VibroMap,HapticHead,HapticPilotIMWUT}; simultaneous stretch and squeeze can interfere \cite{ieeeSkinStretchTesting}, and temporally adjacent vibration can mask later stimuli \cite{ieeeVibration13}. A coaching language should therefore be designed around what trainees can reliably notice and distinguish.

The vocabulary should encode a small set of stable relationships among \emph{location}, \emph{action}, \emph{urgency}, and \emph{resolution}. A brief localized cue may mean \emph{attend here}; a rhythmic cue may mean \emph{adjust tempo}; elastic directional resistance may mean \emph{return toward this path}; increasing or rigid resistance may mean \emph{approaching a boundary}; and cue withdrawal may mean \emph{correction acknowledged}. The same cue should not silently change meaning across exercises, and one cue should not attempt to report every active muscle or detected deviation.

Meaning should be established through embodied rehearsal and stakeholder participation. Collaborative haptic authoring shows that designers rely on multisensory communication to refine intent \cite{CollabJam}, while sensory co-design and soma-based methods foreground bodily interpretation and personal boundaries \cite{codesIMWUT,SomaTheory,somaDesign}. Before a cue is used in a high-demand situation, the trainee should experience it, explain its intended meaning, and demonstrate the corresponding response.

\textbf{Guideline.} Keep cues spatially and directionally congruent, maintain stable meanings, acknowledge successful correction, and avoid simultaneous messages whose perceptual interaction has not been evaluated. If possible, introduce the cues for users to learn during practice sessions. 

\subsection{Guideline 5: Preserve Agency While Supporting Intended Action}
\label{sec:design-implication-5}

F5 shows that the degree of control exercised by an on-body system is itself a safety and learning decision. EMS and force-producing wearables can directly alter movement \cite{FootStrikerIMWUT,smartwatchEms,genEMG,Kinethreads}, raising the broader HCI question of how computers should exercise control over the human body \cite{grandChallengesCompControlBody}. During ordinary training, the wearable should support the trainee's intended action and ability to self-correct rather than perform the movement on the trainee's behalf. Higher-authority intervention should be limited to a predefined severe hazard or to situations in which the trainee has explicitly opted into physical guidance, such as rehabilitation or no- or low-load learning.

Physical assistance does not necessarily remove agency. A cue or resistance that helps the trainee maintain a voluntarily chosen movement path can preserve agency when the trainee still initiates the action, generates the effort, understands the system's influence, and can stop or override it. For example, a wearable may guide the arms toward an intended bench-press path without lifting the weight for the trainee. In contrast, a system reduces agency when it unexpectedly completes a repetition, forces the body into a posture, locks a joint, or prevents the trainee from using a safe recovery movement. The distinction is whether the system supports the trainee's intended action or replaces it.

Preserving agency requires the system's physical influence to be understandable, controllable, and easy to stop. The trainee should know when and why assistance is applied, retain the ability to refuse or override it, and have an immediate way to release any physical constraint. Assistance should increase gradually rather than shift unexpectedly into full control. When pain, injury, equipment failure, or an uncertain movement cannot be interpreted safely, the system should stop providing guidance and defer to the trainee, trainer, or clinician rather than autonomously completing the action.

\textbf{Guideline.} Preserve the trainee's agency by supporting their intended action without performing it for them. Keep physical assistance transparent, proportionate, and immediately releasable. Ensure that the trainee continues to initiate the movement and generate the required effort. 

\section{DISCUSSION}
\label{sec:discussion}

\subsection{Intelligence in Materials}
\label{sec:discussion-intelligence-in-materials}

Our findings show that there is a need for greater intelligence in materials themselves. Many participant needs pointed toward adaptive intrinsic feedback, where feedback is conveyed through material properties such as rigidity or elasticity. Unlike extrinsic actuation, which can be modulated digitally or mechanically, intrinsic feedback relies on the physical response of the material. To make such feedback adaptive, materials themselves must become ``smart'' and dynamically tunable. This suggests a path forward through mechano- or physical intelligence \cite{mechIntRoadmap}, where bistable, snap-through, and pressure-tunable mechanisms are engineered to reconfigure stiffness or shape on demand, supporting highly adaptive, wearable use cases~\cite{VabricBeads, FrozenSuit, wearableMaterialProps, SnapInflatables}.

\subsection{Dynamic Coaching Priorities}
\label{sec:discussion-dynamic-coaching-priorities}

Coaching needs change continuously during strength training. Different exercises, movement phases, repetitions, and fatigue levels require different forms of feedback \cite{sportsWearablesDesignSpace,nzWorkshopPaper}. From prior section, our findings suggest that future intelligent wearables should similarly move beyond static sensing-to-feedback mappings and adapt feedback according to the current training context \cite{genEMG,personalSensingNoti}. As an initial exploration of this direction, we developed a proof-of-concept wearable that instantiated selected dimensions of \sys{}. Survey responses and verbal feedback from six participants indicated that context-dependent feedback was generally perceived as appropriate and aligned with the changing demands of different movement phases and exercises. These preliminary findings provide early evidence that the design principles captured in \sys{} can be translated into future intelligent wearable systems with dynamic feedback modalities. We briefly summarize the study in the Appendix.

\subsection{Toward Embodied AI Training Partners}
\label{sec:discussion-toward-interactive-training-partner}

By situating safety scaffolding within a design space framework, \sys{} advances the conversation on how future sports and fitness wearables can evolve from passive tracking systems to embodied AI training partners that sense movement, interpret safety-relevant conditions, and respond directly through the body~\cite{MotionMA,FootStrikerIMWUT,UbiPhysio}. However, on-body feedback lacks the expressive interface available to screen-based systems, where text and visualizations can explicitly explain what went wrong and how to correct it. A wearable has to instead communicate the type, urgency, and direction of a correction through a limited vocabulary of bodily cues, including their location, intensity, timing, and direction. Our findings illustrate how wearables can move beyond performance monitoring to actively guide movement and support users in recognizing safety-relevant deviations during exercise, while still supporting motor learning. The idea has broader implications for adjacent domains such as rehabilitation, physical education, and recreational training, where embodied safety and skill acquisition are equally central~\cite{MuscleRehab}. In this way, \sys{} demonstrates how design spaces can serve as actionable frameworks for envisioning embodied AI systems that provide interpretable physical guidance while preserving user agency.

\section{LIMITATIONS AND FUTURE WORK}
\label{sec:limitations-and-future-work}

Our work carries several limitations that inform directions for future research. First, we investigated only upper-body movements. While these capture many common safety-sensitive patterns, they represent only a subset of the body. Some insights may generalize to other regions with similar roles, such as joints that govern ROM, supporting muscle groups, and related structures, but future work should extend our approach to lower-body and full-body exercises, where different biomechanical demands may surface new requirements for sensing and feedback design.  

Second, although this paper focuses on the design space and its multi-stakeholder evaluation, it does not include an embodied evaluation. As an initial exploration, we developed a proof-of-concept wearable prototype that instantiates selected dimensions of \sys{}. We briefly summarize this prototype in the Appendix~\ref{app:initial-study-with-poc}. Future work should develop higher-fidelity wearable implementations that realize a broader range of the proposed design dimensions and evaluate them in real-world training settings.

Future implementations should build on this foundation by automating configuration through adaptive calibration, onboard sensing, or machine learning to reduce setup burden and enable seamless, on-the-fly personalization. Such systems should remain lightweight and body-conforming to support strength training. This direction would move beyond design-space validation toward deployable wearable systems capable of delivering continuous, real-time guidance in everyday training settings.

\section{CONCLUSION}
\label{sec:conclusion}

This paper presented \sys{}, a design space for on-body wearable feedback to scaffold safety in strength training, derived from nine co-design workshops with trainer–trainee pairs using a sensory probe that explored both extrinsic and intrinsic modalities. \sys{} is articulated through two components---\textit{Sensing \& Triggering} and \textit{Feedback Delivery}---and their eight design dimensions, providing designers with a framework to relate safety-relevant movement risks to appropriate forms of on-body feedback. We evaluated and refined this design space through a multi-stakeholder speed dating study involving beginner trainees and experienced trainers. The study validated the design dimensions while informing design implications for future wearable systems. Together, these findings demonstrate that \sys{} offers a valid and useful framework for designing on-body feedback systems that scaffold trainees’ awareness of and support for safety-relevant movement during strength training.

\begin{acks}
This work was supported by the Center for Human-Computer Interaction at Virginia Tech (CHCI@VT). We thank our participants for their valuable insights. We also thank Utkarsha Mohan, Xiaohang Tang, and Tong Wu for generously lending study equipment.
\end{acks}

\bibliographystyle{ACM-Reference-Format}
\bibliography{sample-base}

@String{BIT = "{BIT}" }

@String{Computing = "Computing" }

@String{Computer = "{IEEE} Computer" }

@String{Springer = "Springer-Verlag" }

@inproceedings{VabricBeads,
author = {Pardomuan, Jefferson and Miyafuji, Shio and Takahashi, Nobuhiro and Koike, Hideki},
title = {VabricBeads: Variable Stiffness Structured Fabric using Artificial Muscle in Woven Beads},
year = {2024},
isbn = {9798400703300},
publisher = {Association for Computing Machinery},
address = {New York, NY, USA},
url = {https://doi.org/10.1145/3613904.3642401},
doi = {10.1145/3613904.3642401},
booktitle = {Proceedings of the 2024 CHI Conference on Human Factors in Computing Systems},
articleno = {335},
numpages = {17},
location = {Honolulu, HI, USA},
series = {CHI '24}
}

@inproceedings{PneumAct,
author = {G\"{u}nther, Sebastian and Makhija, Mohit and M\"{u}ller, Florian and Sch\"{o}n, Dominik and M\"{u}hlh\"{a}user, Max and Funk, Markus},
title = {PneumAct: Pneumatic Kinesthetic Actuation of Body Joints in Virtual Reality Environments},
year = {2019},
isbn = {9781450358507},
publisher = {Association for Computing Machinery},
address = {New York, NY, USA},
url = {https://doi.org/10.1145/3322276.3322302},
doi = {10.1145/3322276.3322302},
booktitle = {Proceedings of the 2019 on Designing Interactive Systems Conference},
pages = {227–240},
numpages = {14},
location = {San Diego, CA, USA},
series = {DIS '19}
}

@inproceedings{wearableMaterialProps,
author = {Lin, Yuyu and Guner, Hatice Gokcen and Gu, Jianzhe and Prashant, Sonia and Ion, Alexandra},
title = {Wearable Material Properties: Passive Wearable Microstructures as Adaptable Interfaces for the Physical Environment},
year = {2025},
isbn = {9798400713941},
publisher = {Association for Computing Machinery},
address = {New York, NY, USA},
url = {https://doi.org/10.1145/3706598.3714215},
doi = {10.1145/3706598.3714215},
booktitle = {Proceedings of the 2025 CHI Conference on Human Factors in Computing Systems},
articleno = {305},
numpages = {16},
location = {
},
series = {CHI '25}
}

@inproceedings{FrozenSuit,
author = {Al Maimani, Ahmed and Roudaut, Anne},
title = {Frozen Suit: Designing a Changeable Stiffness Suit and its Application to Haptic Games},
year = {2017},
isbn = {9781450346559},
publisher = {Association for Computing Machinery},
address = {New York, NY, USA},
url = {https://doi.org/10.1145/3025453.3025655},
doi = {10.1145/3025453.3025655},
booktitle = {Proceedings of the 2017 CHI Conference on Human Factors in Computing Systems},
pages = {2440–2448},
numpages = {9},
location = {Denver, Colorado, USA},
series = {CHI '17}
}

@inproceedings{ForceJacket,
author = {Delazio, Alexandra and Nakagaki, Ken and Klatzky, Roberta L. and Hudson, Scott E. and Lehman, Jill Fain and Sample, Alanson P.},
title = {Force Jacket: Pneumatically-Actuated Jacket for Embodied Haptic Experiences},
year = {2018},
isbn = {9781450356206},
publisher = {Association for Computing Machinery},
address = {New York, NY, USA},
url = {https://doi.org/10.1145/3173574.3173894},
doi = {10.1145/3173574.3173894},
booktitle = {Proceedings of the 2018 CHI Conference on Human Factors in Computing Systems},
pages = {1–12},
numpages = {12},
location = {Montreal QC, Canada},
series = {CHI '18}
}

@inproceedings{KiriHapticSwatches,
author = {Chang, Zekun and Ta, Tung D. and Narumi, Koya and Kim, Heeju and Okuya, Fuminori and Li, Dongchi and Kato, Kunihiro and Qi, Jie and Miyamoto, Yoshinobu and Saito, Kazuya and Kawahara, Yoshihiro},
title = {Kirigami Haptic Swatches: Design Methods for Cut-and-Fold Haptic Feedback Mechanisms},
year = {2020},
isbn = {9781450367080},
publisher = {Association for Computing Machinery},
address = {New York, NY, USA},
url = {https://doi.org/10.1145/3313831.3376655},
doi = {10.1145/3313831.3376655},
booktitle = {Proceedings of the 2020 CHI Conference on Human Factors in Computing Systems},
pages = {1–12},
numpages = {12},
location = {Honolulu, HI, USA},
series = {CHI '20}
}

@inproceedings{milliMorph,
author = {Lu, Qiuyu and Ou, Jifei and Wilbert, Jo\~{a}o and Haben, Andr\'{e} and Mi, Haipeng and Ishii, Hiroshi},
title = {milliMorph -- Fluid-Driven Thin Film Shape-Change Materials for Interaction Design},
year = {2019},
isbn = {9781450368162},
publisher = {Association for Computing Machinery},
address = {New York, NY, USA},
url = {https://doi.org/10.1145/3332165.3347956},
doi = {10.1145/3332165.3347956},
booktitle = {Proceedings of the 32nd Annual ACM Symposium on User Interface Software and Technology},
pages = {663–672},
numpages = {10},
location = {New Orleans, LA, USA},
series = {UIST '19}
}

@inproceedings{ParametricHaptics,
author = {Han, Violet Yinuo and Boadi-Agyemang, Abena and Lin, Yuyu and Lindlbauer, David and Ion, Alexandra},
title = {Parametric Haptics: Versatile Geometry-based Tactile Feedback Devices},
year = {2023},
isbn = {9798400701320},
publisher = {Association for Computing Machinery},
address = {New York, NY, USA},
url = {https://doi.org/10.1145/3586183.3606766},
doi = {10.1145/3586183.3606766},
booktitle = {Proceedings of the 36th Annual ACM Symposium on User Interface Software and Technology},
articleno = {65},
numpages = {13},
location = {San Francisco, CA, USA},
series = {UIST '23}
}

@article{laminarJamming,
  title={Mechanically versatile soft machines through laminar jamming},
  author={Narang, Yashraj S and Vlassak, Joost J and Howe, Robert D},
  journal={Advanced Functional Materials},
  volume={28},
  number={17},
  pages={1707136},
  year={2018},
  publisher={Wiley Online Library}
}

@article{
oriArtificialMuscle,
author = {Shuguang Li  and Daniel M. Vogt  and Daniela Rus  and Robert J. Wood },
title = {Fluid-driven origami-inspired artificial muscles},
journal = {Proceedings of the National Academy of Sciences},
volume = {114},
number = {50},
pages = {13132-13137},
year = {2017},
doi = {10.1073/pnas.1713450114},
URL = {https://www.pnas.org/doi/abs/10.1073/pnas.1713450114},
eprint = {https://www.pnas.org/doi/pdf/10.1073/pnas.1713450114}}

@article{fishKiri,
  title={Bioinspired kirigami fish-based highly stretched wearable biosensor for human biochemical--physiological hybrid monitoring},
  author={Gao, Bingbing and Elbaz, Abdelrahman and He, Zhenzhu and Xie, Zhuoying and Xu, Hua and Liu, Songqin and Su, Enben and Liu, Hong and Gu, Zhongze},
  journal={Advanced Materials Technologies},
  volume={3},
  number={4},
  pages={1700308},
  year={2018},
  publisher={Wiley Online Library}
}

@article{umichShoulder,
  title={Developable rotationally symmetric Kirigami-based structures as sensor platforms},
  author={Evke, Erin E and Meli, Dilara and Shtein, Max},
  journal={Advanced Materials Technologies},
  volume={4},
  number={12},
  pages={1900563},
  year={2019},
  publisher={Wiley Online Library}
}

@inproceedings{FlexTure,
author = {Johnson, Tate and Patel, Dinesh K and Yang, Humphrey and Civici, Umut Serdar and Fernandes Minori, Adriane and Yao, Lining},
title = {FlexTure: Designing Configurable and Dynamic Surface Features},
year = {2023},
isbn = {9781450398930},
publisher = {Association for Computing Machinery},
address = {New York, NY, USA},
url = {https://doi.org/10.1145/3563657.3595995},
doi = {10.1145/3563657.3595995},
booktitle = {Proceedings of the 2023 ACM Designing Interactive Systems Conference},
pages = {580–593},
numpages = {14},
location = {Pittsburgh, PA, USA},
series = {DIS '23}
}

@inproceedings{WaxpaperSeq,
author = {Wu, Di and Guan, Emily and Zhang, Yunjia and Lai, Hsuanju and Lu, Qiuyu and Yao, Lining},
title = {Waxpaper Actuator: Sequentially and Conditionally Programmable Wax Paper for Morphing Interfaces},
year = {2024},
isbn = {9798400703300},
publisher = {Association for Computing Machinery},
address = {New York, NY, USA},
url = {https://doi.org/10.1145/3613904.3642373},
doi = {10.1145/3613904.3642373},
booktitle = {Proceedings of the 2024 CHI Conference on Human Factors in Computing Systems},
articleno = {865},
numpages = {16},
location = {Honolulu, HI, USA},
series = {CHI '24}
}

@inproceedings{ConeAct,
author = {Lin, Yuyu and Gonzalez, Jesse T and Cui, Zhitong and Banka, Yash Rajeev and Ion, Alexandra},
title = {ConeAct: A Multistable Actuator for Dynamic Materials},
year = {2024},
isbn = {9798400703300},
publisher = {Association for Computing Machinery},
address = {New York, NY, USA},
url = {https://doi.org/10.1145/3613904.3642949},
doi = {10.1145/3613904.3642949},
booktitle = {Proceedings of the 2024 CHI Conference on Human Factors in Computing Systems},
articleno = {324},
numpages = {16},
location = {Honolulu, HI, USA},
series = {CHI '24}
}

@inproceedings{humanScaleRobotic,
author = {Gonzalez, Jesse T and Prashant, Sonia and Tayal, Sapna and Kedia, Juhi and Ion, Alexandra and Hudson, Scott E},
title = {Constraint-Driven Robotic Surfaces, At Human-Scale},
year = {2023},
isbn = {9798400701320},
publisher = {Association for Computing Machinery},
address = {New York, NY, USA},
url = {https://doi.org/10.1145/3586183.3606740},
doi = {10.1145/3586183.3606740},
booktitle = {Proceedings of the 36th Annual ACM Symposium on User Interface Software and Technology},
articleno = {53},
numpages = {12},
location = {San Francisco, CA, USA},
series = {UIST '23}
}

@article{
intelligentTextiles,
author = {Yunzhu Li  and Yiyue Luo },
title = {Intelligent textiles are looking bright},
journal = {Science},
volume = {384},
number = {6691},
pages = {29-30},
year = {2024},
doi = {10.1126/science.ado5922},
URL = {https://www.science.org/doi/abs/10.1126/science.ado5922},
eprint = {https://www.science.org/doi/pdf/10.1126/science.ado5922}}

@article{glovePiano,
  title={Adaptive tactile interaction transfer via digitally embroidered smart gloves},
  author={Luo, Yiyue and Liu, Chao and Lee, Young Joong and DelPreto, Joseph and Wu, Kui and Foshey, Michael and Rus, Daniela and Palacios, Tom{\'a}s and Li, Yunzhu and Torralba, Antonio and others},
  journal={Nature communications},
  volume={15},
  number={1},
  pages={868},
  year={2024},
  publisher={Nature Publishing Group UK London}
}

@article{fiberFabric,
  title={Highly oriented electrospun P (VDF-TrFE) fibers via mechanical stretching for wearable motion sensing},
  author={Ma, Shaoyang and Ye, Tao and Zhang, Ting and Wang, Zhe and Li, Kaiwei and Chen, Ming and Zhang, Jing and Wang, Zhixun and Ramakrishna, Seeram and Wei, Lei},
  journal={Advanced Materials Technologies},
  volume={3},
  number={7},
  pages={1800033},
  year={2018},
  publisher={Wiley Online Library}
}

@misc{mimickMuscle,
      title={Patterned Structure Muscle : Arbitrary Shaped Wire-driven Artificial Muscle Utilizing Anisotropic Flexible Structure for Musculoskeletal Robots}, 
      author={Shunnosuke Yoshimura and Akihiro Miki and Kazuhiro Miyama and Yuta Sahara and Kento Kawaharazuka and Kei Okada and Masayuki Inaba},
      year={2024},
      eprint={2410.07682},
      archivePrefix={arXiv},
      primaryClass={cs.RO},
      url={https://arxiv.org/abs/2410.07682}, 
}

@ARTICLE{kneeStraigtener,
  author={Park, Junghoon and Choi, Junhwan and Kim, Sangjoon J. and Seo, Kap-Ho and Kim, Jung},
  journal={IEEE Robotics and Automation Letters}, 
  title={Design of an Inflatable Wrinkle Actuator With Fast Inflation/Deflation Responses for Wearable Suits}, 
  year={2020},
  volume={5},
  number={3},
  pages={3799-3805},
  doi={10.1109/LRA.2020.2976299}}

@ARTICLE{oriHapticsReview,
  author={Iiyoshi, Ken and Lee, Jong Hyun and Dalaq, Ahmed S. and Khazaaleh, Shadi and Daqaq, Mohammed F. and Korres, Georgios and Eid, Mohamad},
  journal={IEEE Access}, 
  title={Origami-Inspired Haptics: A Literature Review}, 
  year={2024},
  volume={12},
  number={},
  pages={33309-33327},
  doi={10.1109/ACCESS.2024.3371480}}

@inproceedings{HapticHead,
author = {Kaul, Oliver Beren and Rohs, Michael},
title = {HapticHead: A Spherical Vibrotactile Grid around the Head for 3D Guidance in Virtual and Augmented Reality},
year = {2017},
isbn = {9781450346559},
publisher = {Association for Computing Machinery},
address = {New York, NY, USA},
url = {https://doi.org/10.1145/3025453.3025684},
doi = {10.1145/3025453.3025684},
booktitle = {Proceedings of the 2017 CHI Conference on Human Factors in Computing Systems},
pages = {3729–3740},
numpages = {12},
location = {Denver, Colorado, USA},
series = {CHI '17}
}

@ARTICLE{AeroHaptix,
  author={Huang, Bingjian and Wang, Zhecheng and Cheng, Qilong and Ren, Siyi and Cai, Hanfeng and Valdivia, Antonio Alvarez and Mahadevan, Karthik and Wigdor, Daniel},
  journal={IEEE Robotics and Automation Letters}, 
  title={AeroHaptix: A Wearable Vibrotactile Feedback System for Enhancing Collision Avoidance in UAV Teleoperation}, 
  year={2025},
  volume={10},
  number={5},
  pages={4260-4267},
  doi={10.1109/LRA.2025.3548866}}

@inproceedings{VibraForge,
author = {Huang, Bingjian and Ren, Siyi and Luo, Yuewen and Cheng, Qilong and Cai, Hanfeng and Sang, Yeqi and Sousa, Mauricio and Dietz, Paul H and Wigdor, Daniel},
title = {VibraForge: A Scalable Prototyping Toolkit For Creating Spatialized Vibrotactile Feedback Systems},
year = {2025},
isbn = {9798400713941},
publisher = {Association for Computing Machinery},
address = {New York, NY, USA},
url = {https://doi.org/10.1145/3706598.3714273},
doi = {10.1145/3706598.3714273},
booktitle = {Proceedings of the 2025 CHI Conference on Human Factors in Computing Systems},
articleno = {1137},
numpages = {18},
location = {
},
series = {CHI '25}
}

@inproceedings{purnendu,
author = {Purnendu and Acome, Eric and Keplinger, Christoph and Gross, Mark D and Bruns, Carson and Leithinger, Daniel},
title = {Soft Electrohydraulic Actuators for Origami Inspired Shape-Changing Interfaces},
year = {2021},
isbn = {9781450380959},
publisher = {Association for Computing Machinery},
address = {New York, NY, USA},
url = {https://doi.org/10.1145/3411763.3451590},
doi = {10.1145/3411763.3451590},
booktitle = {Extended Abstracts of the 2021 CHI Conference on Human Factors in Computing Systems},
articleno = {377},
numpages = {6},
location = {Yokohama, Japan},
series = {CHI EA '21}
}

@inproceedings{MagKnitic,
author = {Luo, Yiyue and Zhu, Junyi and Wu, Kui and Honnet, Cedric and Mueller, Stefanie and Matusik, Wojciech},
title = {MagKnitic: Machine-knitted Passive and Interactive Haptic Textiles with Integrated Binary Sensing},
year = {2023},
isbn = {9798400701320},
publisher = {Association for Computing Machinery},
address = {New York, NY, USA},
url = {https://doi.org/10.1145/3586183.3606765},
doi = {10.1145/3586183.3606765},
booktitle = {Proceedings of the 36th Annual ACM Symposium on User Interface Software and Technology},
articleno = {66},
numpages = {13},
location = {San Francisco, CA, USA},
series = {UIST '23}
}

@inproceedings{Zishi,
author = {Wang, Qi and Toeters, Marina and Chen, Wei and Timmermans, Annick and Markopoulos, Panos},
title = {Zishi: A Smart Garment for Posture Monitoring},
year = {2016},
isbn = {9781450340823},
publisher = {Association for Computing Machinery},
address = {New York, NY, USA},
url = {https://doi.org/10.1145/2851581.2890262},
doi = {10.1145/2851581.2890262},
booktitle = {Proceedings of the 2016 CHI Conference Extended Abstracts on Human Factors in Computing Systems},
pages = {3792–3795},
numpages = {4},
location = {San Jose, California, USA},
series = {CHI EA '16}
}

@INPROCEEDINGS{Bellowband,
  author={Young, Eric M. and Memar, Amirhossein H. and Agarwal, Priyanshu and Colonnese, Nick},
  booktitle={2019 IEEE World Haptics Conference (WHC)}, 
  title={Bellowband: A Pneumatic Wristband for Delivering Local Pressure and Vibration}, 
  year={2019},
  volume={},
  number={},
  pages={55-60},
  doi={10.1109/WHC.2019.8816075}}

@inproceedings{ASTEROIDS,
author = {Li, Jiannan and Sousa, Maur\'{\i}cio and Li, Chu and Liu, Jessie and Chen, Yan and Balakrishnan, Ravin and Grossman, Tovi},
title = {ASTEROIDS: Exploring Swarms of Mini-Telepresence Robots for Physical Skill Demonstration},
year = {2022},
isbn = {9781450391573},
publisher = {Association for Computing Machinery},
address = {New York, NY, USA},
url = {https://doi.org/10.1145/3491102.3501927},
doi = {10.1145/3491102.3501927},
booktitle = {Proceedings of the 2022 CHI Conference on Human Factors in Computing Systems},
articleno = {111},
numpages = {14},
location = {New Orleans, LA, USA},
series = {CHI '22}
}

@inproceedings{InstructAR,
author = {Keelawat, Panayu and Suzuki, Ryo},
title = {Transforming Procedural Instructions into In-Situ Augmented Reality Guides with InstructAR},
year = {2024},
isbn = {9798400707186},
publisher = {Association for Computing Machinery},
address = {New York, NY, USA},
url = {https://doi.org/10.1145/3672539.3686321},
doi = {10.1145/3672539.3686321},
booktitle = {Adjunct Proceedings of the 37th Annual ACM Symposium on User Interface Software and Technology},
articleno = {70},
numpages = {3},
location = {Pittsburgh, PA, USA},
series = {UIST Adjunct '24}
}

@inproceedings{TheYogaVR,
author = {Jo, Hye-Young and Seidel, Laurenz and Pahud, Michel and Sinclair, Mike and Bianchi, Andrea},
title = {FlowAR: How Different Augmented Reality Visualizations of Online Fitness Videos Support Flow for At-Home Yoga Exercises},
year = {2023},
isbn = {9781450394215},
publisher = {Association for Computing Machinery},
address = {New York, NY, USA},
url = {https://doi.org/10.1145/3544548.3580897},
doi = {10.1145/3544548.3580897},
booktitle = {Proceedings of the 2023 CHI Conference on Human Factors in Computing Systems},
articleno = {469},
numpages = {17},
location = {Hamburg, Germany},
series = {CHI '23}
}

@article{dias2017influence,
  title={Influence of a personal trainer on self-selected loading during resistance exercise},
  author={Dias, Marcelo RC and Sim{\~a}o, Roberto F and Saavedra, Francisco JF and Ratamess, Nicholas A},
  journal={The Journal of Strength \& Conditioning Research},
  volume={31},
  number={7},
  pages={1925--1930},
  year={2017},
  publisher={LWW}
}

@incollection{rao2024formfit,
  title={FormFit: A physical exercise form correction system using BlazePose},
  author={Rao, SK and Savalgi, R and Suresh, VK and Pyati, N and Krishnamurthy, S},
  booktitle={Computer Science Engineering},
  pages={490--497},
  year={2024},
  publisher={CRC Press}
}

@inproceedings{sittingReview,
author = {Krauter, Christian and Angerbauer, Katrin and Sousa Calepso, Aim\'{e}e and Achberger, Alexander and Mayer, Sven and Sedlmair, Michael},
title = {Sitting Posture Recognition and Feedback: A Literature Review},
year = {2024},
isbn = {9798400703300},
publisher = {Association for Computing Machinery},
address = {New York, NY, USA},
url = {https://doi.org/10.1145/3613904.3642657},
doi = {10.1145/3613904.3642657},
booktitle = {Proceedings of the 2024 CHI Conference on Human Factors in Computing Systems},
articleno = {943},
numpages = {20},
location = {Honolulu, HI, USA},
series = {CHI '24}
}

@inproceedings{tactilePose,
author = {Seong, Minwoo and Kim, Gwangbin and Lee, Jaehee and DelPreto, Joseph and Matusik, Wojciech and Rus, Daniela and Kim, SeungJun},
title = {Intelligent Seat: Tactile Signal-Based 3D Sitting Pose Inference},
year = {2024},
isbn = {9798400710582},
publisher = {Association for Computing Machinery},
address = {New York, NY, USA},
url = {https://doi.org/10.1145/3675094.3678374},
doi = {10.1145/3675094.3678374},
booktitle = {Companion of the 2024 on ACM International Joint Conference on Pervasive and Ubiquitous Computing},
pages = {791–796},
numpages = {6},
location = {Melbourne VIC, Australia},
series = {UbiComp '24}
}

@article{mechIntRoadmap,
	author={Alu, Andrea and Arrieta, Andres F and Del Dottore, Emanuela and Dickey, Michael D and Ferracin, Samuele and Harne, Ryan L and Hauser, Helmut and He, Qiguang and Hopkins, Jonathan Brigham and Hyatt, Lance and Li, Suyi and Mariani, Stefano and Mazzolai, Barbara and Mondini, Alessio and Pal, Aniket and Preston, Daniel J and Rajappan, Anoop and Raney, Jordan R. and Reis, Pedro and Sarles, Stephen Andrew and Sitti, Metin and Ubamanyu, Uba K. and Hecke, Martin van and Wang, Kon-Well},
	title={Roadmap on embodying mechano-intelligence and computing in functional materials and structures},
	journal={Smart Materials and Structures},
	url={http://iopscience.iop.org/article/10.1088/1361-665X/adb7aa},
	year={2025}
}

@inproceedings{wearableWisdom,
author = {Pataranutaporn, Pat and Vega G\'{a}lvez, Tom\'{a}s and Yoo, Lisa and Chhetri, Abishkar and Maes, Pattie},
title = {Wearable Wisdom: An Intelligent Audio-Based System for Mediating Wisdom and Advice},
year = {2020},
isbn = {9781450368193},
publisher = {Association for Computing Machinery},
address = {New York, NY, USA},
url = {https://doi.org/10.1145/3334480.3383092},
doi = {10.1145/3334480.3383092},
booktitle = {Extended Abstracts of the 2020 CHI Conference on Human Factors in Computing Systems},
pages = {1–8},
numpages = {8},
location = {Honolulu, HI, USA},
series = {CHI EA '20}
}

@inbook{OpenDanceLab,
author = {Archiwaranguprok, Chayapatr and Pataranutaporn, Pat and Mano, Phoomparin and Bhongse-Tong, Piyaporn and Maes, Pattie and Klunchun, Pichet},
title = {Open Dance Lab: Digital Platform for Examining, Experimenting, and Evolving Intangible Cultural Heritage},
year = {2025},
isbn = {9798400710933},
publisher = {Association for Computing Machinery},
address = {New York, NY, USA},
url = {https://doi.org/10.1145/3677389.3702602},
booktitle = {Proceedings of the 24th ACM/IEEE Joint Conference on Digital Libraries},
articleno = {45},
numpages = {12}
}

@inproceedings{nzWorkshopPaper,
author = {Stawarz, Katarzyna and Everitt, Aluna and Bowen, Judy},
title = {Exploring Opportunities for Flexible Wearables to Support Physical Training},
year = {2025},
isbn = {9798400714856},
publisher = {Association for Computing Machinery},
address = {New York, NY, USA},
url = {https://doi.org/10.1145/3715336.3735703},
doi = {10.1145/3715336.3735703},
booktitle = {Proceedings of the 2025 ACM Designing Interactive Systems Conference},
pages = {2156–2170},
numpages = {15},
location = {
},
series = {DIS '25}
}

@article{fisher2025supervision,
  title={Supervision During Strength Training—the Interplay with Facilitation, Feedback and Attentional Focus: A Narrative Review},
  author={Fisher, James P},
  journal={Sports Medicine},
  pages={1--9},
  year={2025},
  publisher={Springer}
}

@article{teslasuitSensors,
  title={Recognizing full-body exercise execution errors using the teslasuit},
  author={Caserman, Polona and Krug, Clemens and G{\"o}bel, Stefan},
  journal={Sensors},
  volume={21},
  number={24},
  pages={8389},
  year={2021},
  publisher={MDPI}
}

@inproceedings{sportsWearablesDesignSpace,
author = {Turmo Vidal, Laia and Zhu, Hui and Waern, Annika and M\'{a}rquez Segura, Elena},
title = {The Design Space of Wearables for Sports and Fitness Practices},
year = {2021},
isbn = {9781450380966},
publisher = {Association for Computing Machinery},
address = {New York, NY, USA},
url = {https://doi.org/10.1145/3411764.3445700},
doi = {10.1145/3411764.3445700},
booktitle = {Proceedings of the 2021 CHI Conference on Human Factors in Computing Systems},
articleno = {267},
numpages = {14},
location = {Yokohama, Japan},
series = {CHI '21}
}

@inproceedings{grandChallengesSportsHCI,
author = {Elvitigala, Don Samitha and Karahano\u{g}lu, Arma\u{g}an and Matviienko, Andrii and Turmo Vidal, Laia and Postma, Dees and Jones, Michael D and Montoya, Maria F. and Harrison, Daniel and Elb\ae{}k, Lars and Daiber, Florian and Burr, Lisa Anneke and Patibanda, Rakesh and Buono, Paolo and H\"{a}m\"{a}l\"{a}inen, Perttu and Van Delden, Robby and Bernhaupt, Regina and Ren, Xipei and Van Rheden, Vincent and Zambetta, Fabio and Van Den Hoven, Elise and Lallemand, Carine and Reidsma, Dennis and Mueller, Florian ‘Floyd’},
title = {Grand Challenges in SportsHCI},
year = {2024},
isbn = {9798400703300},
publisher = {Association for Computing Machinery},
address = {New York, NY, USA},
url = {https://doi.org/10.1145/3613904.3642050},
doi = {10.1145/3613904.3642050},
booktitle = {Proceedings of the 2024 CHI Conference on Human Factors in Computing Systems},
articleno = {312},
numpages = {20},
location = {Honolulu, HI, USA},
series = {CHI '24}
}

@article{selectivelyStiffening,
  title={Selectively stiffening garments enabled by cellular composites},
  author={Kwon, Junghan and Choi, Inrak and Park, Myungsun and Moon, Jeongin and Jeong, Bomin and Pathak, Prabhat and Ahn, Jooeun and Park, Yong-Lae},
  journal={Advanced Materials Technologies},
  volume={7},
  number={9},
  pages={2101543},
  year={2022},
  publisher={Wiley Online Library}
}

@inproceedings{hapticScores,
author = {Escudero, Julio Andres},
title = {Haptic Scores: Non-Visual Ways of Experiencing Choreography},
year = {2024},
isbn = {9798400710582},
publisher = {Association for Computing Machinery},
address = {New York, NY, USA},
url = {https://doi.org/10.1145/3675094.3678368},
doi = {10.1145/3675094.3678368},
booktitle = {Companion of the 2024 on ACM International Joint Conference on Pervasive and Ubiquitous Computing},
pages = {281–285},
numpages = {5},
location = {Melbourne VIC, Australia},
series = {UbiComp '24}
}

@inproceedings{mocoWorkshop,
author = {Tajadura-Jimenez, Ana and Ley-Flores, Judith and Valdiviezo, Omar and Singh, Aneesha and Sanchez-Martin, Milagrosa and Diaz Duran, Joaquin and M\'{a}rquez Segura, Elena},
title = {Exploring the Design Space for Body Transformation Wearables to Support Physical Activity through Sensitizing and Bodystorming},
year = {2022},
isbn = {9781450387163},
publisher = {Association for Computing Machinery},
address = {New York, NY, USA},
url = {https://doi.org/10.1145/3537972.3538001},
doi = {10.1145/3537972.3538001},
booktitle = {Proceedings of the 8th International Conference on Movement and Computing},
articleno = {23},
numpages = {9},
location = {Chicago, IL, USA},
series = {MOCO '22}
}

@online{catapult,
  author =    {Catapult},
  title =  {Capture, Analyze, Plan, Share every aspect of performance},
  year = 2025,
  url =    {https://www.catapult.com},
  lastaccessed = {Jul 15, 2025}
  }

@inproceedings{bodymapsPaper,
author = {Turmo Vidal, Laia and Li, Yinchu and Stojanov, Martin and Johansson, Karin B and Tylstedt, Beatrice and Eklund, Lina},
title = {Towards Advancing Body Maps as Research Tool for Interaction Design},
year = {2023},
isbn = {9781450399777},
publisher = {Association for Computing Machinery},
address = {New York, NY, USA},
url = {https://doi.org/10.1145/3569009.3573838},
doi = {10.1145/3569009.3573838},
booktitle = {Proceedings of the Seventeenth International Conference on Tangible, Embedded, and Embodied Interaction},
articleno = {20},
numpages = {14},
location = {Warsaw, Poland},
series = {TEI '23}
}

@ARTICLE{torqueFeedback,
  author={Kim, Hubert and Asbeck, Alan T.},
  journal={IEEE Access}, 
  title={The Effects of Torque Magnitude and Stiffness in Arm Guidance Through Joint Torque Feedback}, 
  year={2022},
  volume={10},
  number={},
  pages={5842-5854},
  doi={10.1109/ACCESS.2022.3141981}}

@inproceedings{basketballCoach,
author = {Weng, Jian-Jia and Ku, Calvin and Wang, Jo Chien and Cheng, Chih-Jen and Lin, Tica and Su, Yu-An and Tsai, Tsung-Hsun and Lin, You-Yi and Ku, Lun-Wei and Chu, Hung-Kuo and Hu, Min-Chun},
title = {Bridging Coaching Knowledge and AI Feedback to Enhance Motor Learning in Basketball Shooting Mechanics Through a Knowledge-Based SOP Framework},
year = {2025},
isbn = {9798400713941},
publisher = {Association for Computing Machinery},
address = {New York, NY, USA},
url = {https://doi.org/10.1145/3706598.3713324},
doi = {10.1145/3706598.3713324},
booktitle = {Proceedings of the 2025 CHI Conference on Human Factors in Computing Systems},
articleno = {989},
numpages = {20},
location = {
},
series = {CHI '25}
}

@inproceedings{fingerTipVRForce,
author = {Hashimoto, Takeru and Hirao, Yutaro},
title = {Selfrionette: A Fingertip Force-Input Controller for Continuous Full-Body Avatar Manipulation and Diverse Haptic Interactions},
year = {2024},
isbn = {9798400706288},
publisher = {Association for Computing Machinery},
address = {New York, NY, USA},
url = {https://doi.org/10.1145/3654777.3676409},
doi = {10.1145/3654777.3676409},
booktitle = {Proceedings of the 37th Annual ACM Symposium on User Interface Software and Technology},
articleno = {137},
numpages = {14},
location = {Pittsburgh, PA, USA},
series = {UIST '24}
}

@inproceedings{pilatesCorrection,
author = {Turmo Vidal, Laia and M\'{a}rquez Segura, Elena and Waern, Annika},
title = {Movement Correction in Instructed Fitness Training: Design Recommendations and Opportunities},
year = {2018},
isbn = {9781450351980},
publisher = {Association for Computing Machinery},
address = {New York, NY, USA},
url = {https://doi.org/10.1145/3196709.3196789},
doi = {10.1145/3196709.3196789},
booktitle = {Proceedings of the 2018 Designing Interactive Systems Conference},
pages = {1041–1054},
numpages = {14},
location = {Hong Kong, China},
series = {DIS '18}
}

@inproceedings{vizCgDeadLifts,
author = {Elvitigala, Don Samitha and Matthies, Denys J.C. and David, L\"{o}ic and Weerasinghe, Chamod and Nanayakkara, Suranga},
title = {GymSoles: Improving Squats and Dead-Lifts by Visualizing the User's Center of Pressure},
year = {2019},
isbn = {9781450359702},
publisher = {Association for Computing Machinery},
address = {New York, NY, USA},
url = {https://doi.org/10.1145/3290605.3300404},
doi = {10.1145/3290605.3300404},
booktitle = {Proceedings of the 2019 CHI Conference on Human Factors in Computing Systems},
pages = {1–12},
numpages = {12},
location = {Glasgow, Scotland Uk},
series = {CHI '19}
}

@inproceedings{LegolasGolf,
author = {Ko, Kangbeen and Oh, Minwoo and Seong, Minwoo and Kim, SeungJun},
title = {LEGOLAS: Learning \& Enhancing Golf Skills through LLM-Augmented System},
year = {2025},
isbn = {9798400713958},
publisher = {Association for Computing Machinery},
address = {New York, NY, USA},
url = {https://doi.org/10.1145/3706599.3720141},
doi = {10.1145/3706599.3720141},
booktitle = {Proceedings of the Extended Abstracts of the CHI Conference on Human Factors in Computing Systems},
articleno = {368},
numpages = {10},
location = {
},
series = {CHI EA '25}
}

@inproceedings{MotionMA,
author = {Velloso, Eduardo and Bulling, Andreas and Gellersen, Hans},
title = {MotionMA: motion modelling and analysis by demonstration},
year = {2013},
isbn = {9781450318990},
publisher = {Association for Computing Machinery},
address = {New York, NY, USA},
url = {https://doi.org/10.1145/2470654.2466171},
doi = {10.1145/2470654.2466171},
booktitle = {Proceedings of the SIGCHI Conference on Human Factors in Computing Systems},
pages = {1309–1318},
numpages = {10},
location = {Paris, France},
series = {CHI '13}
}

@inproceedings{JetUnit,
author = {Zhang, Zining and Li, Jiasheng and Yan, Zeyu and Nishida, Jun and Peng, Huaishu},
title = {JetUnit: Rendering Diverse Force Feedback in Virtual Reality Using Water Jets},
year = {2024},
isbn = {9798400706288},
publisher = {Association for Computing Machinery},
address = {New York, NY, USA},
url = {https://doi.org/10.1145/3654777.3676440},
doi = {10.1145/3654777.3676440},
booktitle = {Proceedings of the 37th Annual ACM Symposium on User Interface Software and Technology},
articleno = {136},
numpages = {15},
location = {Pittsburgh, PA, USA},
series = {UIST '24}
}

@inproceedings{skinStretching,
author = {Lin, Yilong and Zhang, Peng and Ofek, Eyal and Je, Seungwoo},
title = {ArmDeformation: Inducing the Sensation of Arm Deformation in Virtual Reality Using Skin-Stretching},
year = {2024},
isbn = {9798400703300},
publisher = {Association for Computing Machinery},
address = {New York, NY, USA},
url = {https://doi.org/10.1145/3613904.3642518},
doi = {10.1145/3613904.3642518},
booktitle = {Proceedings of the 2024 CHI Conference on Human Factors in Computing Systems},
articleno = {407},
numpages = {18},
location = {Honolulu, HI, USA},
series = {CHI '24}
}

@inproceedings{ImpactVest,
author = {Tsai, Hsin-Ruey and Liao, Yu-So and Tsai, Chieh},
title = {ImpactVest: Rendering Spatio-Temporal Multilevel Impact Force Feedback on Body in VR},
year = {2022},
isbn = {9781450391573},
publisher = {Association for Computing Machinery},
address = {New York, NY, USA},
url = {https://doi.org/10.1145/3491102.3501971},
doi = {10.1145/3491102.3501971},
booktitle = {Proceedings of the 2022 CHI Conference on Human Factors in Computing Systems},
articleno = {356},
numpages = {11},
location = {New Orleans, LA, USA},
series = {CHI '22}
}

@inproceedings{somaDesign,
author = {Tennent, Paul and Marshall, Joe and Tsaknaki, Vasiliki and Windlin, Charles and H\"{o}\"{o}k, Kristina and Alfaras, Miquel},
title = {Soma Design and Sensory Misalignment},
year = {2020},
isbn = {9781450367080},
publisher = {Association for Computing Machinery},
address = {New York, NY, USA},
url = {https://doi.org/10.1145/3313831.3376812},
doi = {10.1145/3313831.3376812},
booktitle = {Proceedings of the 2020 CHI Conference on Human Factors in Computing Systems},
pages = {1–12},
numpages = {12},
location = {Honolulu, HI, USA},
series = {CHI '20}
}

@inproceedings{motionlessMovement,
author = {Ding, Yuran and Sabnis, Nihar and Strohmeier, Paul},
title = {Motionless Movement: Towards Vibrotactile Kinesthetic Displays},
year = {2024},
isbn = {9798400703300},
publisher = {Association for Computing Machinery},
address = {New York, NY, USA},
url = {https://doi.org/10.1145/3613904.3642499},
doi = {10.1145/3613904.3642499},
booktitle = {Proceedings of the 2024 CHI Conference on Human Factors in Computing Systems},
articleno = {418},
numpages = {16},
location = {Honolulu, HI, USA},
series = {CHI '24}
}

@article{MLMovementDes,
author = {Gillies, Marco},
title = {Understanding the Role of Interactive Machine Learning in Movement Interaction Design},
year = {2019},
issue_date = {February 2019},
publisher = {Association for Computing Machinery},
address = {New York, NY, USA},
volume = {26},
number = {1},
issn = {1073-0516},
url = {https://doi.org/10.1145/3287307},
doi = {10.1145/3287307},
journal = {ACM Trans. Comput.-Hum. Interact.},
month = feb,
articleno = {5},
numpages = {34}
}

@inproceedings{bjornDemo,
author = {Hartmann, Bj\"{o}rn and Abdulla, Leith and Mittal, Manas and Klemmer, Scott R.},
title = {Authoring sensor-based interactions by demonstration with direct manipulation and pattern recognition},
year = {2007},
isbn = {9781595935939},
publisher = {Association for Computing Machinery},
address = {New York, NY, USA},
url = {https://doi.org/10.1145/1240624.1240646},
doi = {10.1145/1240624.1240646},
booktitle = {Proceedings of the SIGCHI Conference on Human Factors in Computing Systems},
pages = {145–154},
numpages = {10},
location = {San Jose, California, USA},
series = {CHI '07}
}

@inproceedings{BodyLights,
author = {Turmo Vidal, Laia and Zhu, Hui and Riego-Delgado, Abraham},
title = {BodyLights: Open-Ended Augmented Feedback to Support Training Towards a Correct Exercise Execution},
year = {2020},
isbn = {9781450367080},
publisher = {Association for Computing Machinery},
address = {New York, NY, USA},
url = {https://doi.org/10.1145/3313831.3376268},
doi = {10.1145/3313831.3376268},
booktitle = {Proceedings of the 2020 CHI Conference on Human Factors in Computing Systems},
pages = {1–14},
numpages = {14},
location = {Honolulu, HI, USA},
series = {CHI '20}
}

@inproceedings{physiotherapyML,
author = {Turmo Vidal, Laia and Waern, Annika and Cabanas-Vald\'{e}s, Rosa and van Loo, Lauren and Li, Yinchu and Meenaakshisundaram, Karthik Venkataraman},
title = {Towards Personalized Physiotherapy through Interactive Machine Learning: A Conceptual Infrastructure Design for In-Clinic and Out-of-Clinic Support},
year = {2025},
isbn = {9798400713941},
publisher = {Association for Computing Machinery},
address = {New York, NY, USA},
url = {https://doi.org/10.1145/3706598.3713823},
doi = {10.1145/3706598.3713823},
booktitle = {Proceedings of the 2025 CHI Conference on Human Factors in Computing Systems},
articleno = {670},
numpages = {19},
location = {
},
series = {CHI '25}
}

@inproceedings{phyTrainingVisual,
author = {Semeraro, Alessandra and Turmo Vidal, Laia},
title = {Visualizing Instructions for Physical Training: Exploring Visual Cues to Support Movement Learning from Instructional Videos},
year = {2022},
isbn = {9781450391573},
publisher = {Association for Computing Machinery},
address = {New York, NY, USA},
url = {https://doi.org/10.1145/3491102.3517735},
doi = {10.1145/3491102.3517735},
booktitle = {Proceedings of the 2022 CHI Conference on Human Factors in Computing Systems},
articleno = {90},
numpages = {16},
location = {New Orleans, LA, USA},
series = {CHI '22}
}

@inproceedings{RecoFit,
author = {Morris, Dan and Saponas, T. Scott and Guillory, Andrew and Kelner, Ilya},
title = {RecoFit: using a wearable sensor to find, recognize, and count repetitive exercises},
year = {2014},
isbn = {9781450324731},
publisher = {Association for Computing Machinery},
address = {New York, NY, USA},
url = {https://doi.org/10.1145/2556288.2557116},
doi = {10.1145/2556288.2557116},
booktitle = {Proceedings of the SIGCHI Conference on Human Factors in Computing Systems},
pages = {3225–3234},
numpages = {10},
location = {Toronto, Ontario, Canada},
series = {CHI '14}
}

@article{intelligentWearableSurvey,
author = {Yang, Luyao and Amin, Osama and Shihada, Basem},
title = {Intelligent Wearable Systems: Opportunities and Challenges in Health and Sports},
year = {2024},
issue_date = {July 2024},
publisher = {Association for Computing Machinery},
address = {New York, NY, USA},
volume = {56},
number = {7},
issn = {0360-0300},
url = {https://doi.org/10.1145/3648469},
doi = {10.1145/3648469},
journal = {ACM Comput. Surv.},
month = apr,
articleno = {190},
numpages = {42}
}

@inproceedings{LiftSmartUbiComp,
author = {Kowsar, Yousef and Velloso, Eduardo and Kulik, Lars and Leckie, Christopher},
title = {LiftSmart: a monitoring and warning wearable for weight trainers},
year = {2019},
isbn = {9781450368698},
publisher = {Association for Computing Machinery},
address = {New York, NY, USA},
url = {https://doi.org/10.1145/3341162.3343795},
doi = {10.1145/3341162.3343795},
booktitle = {Adjunct Proceedings of the 2019 ACM International Joint Conference on Pervasive and Ubiquitous Computing and Proceedings of the 2019 ACM International Symposium on Wearable Computers},
pages = {298–301},
numpages = {4},
location = {London, United Kingdom},
series = {UbiComp/ISWC '19 Adjunct}
}

@inproceedings{danceON,
author = {Payne, William Christopher and Bergner, Yoav and West, Mary Etta and Charp, Carlie and Shapiro, R. Benjamin and Szafir, Danielle Albers and Taylor, Edd V. and DesPortes, Kayla},
title = {danceON: Culturally Responsive Creative Computing},
year = {2021},
isbn = {9781450380966},
publisher = {Association for Computing Machinery},
address = {New York, NY, USA},
url = {https://doi.org/10.1145/3411764.3445149},
doi = {10.1145/3411764.3445149},
booktitle = {Proceedings of the 2021 CHI Conference on Human Factors in Computing Systems},
articleno = {96},
numpages = {16},
location = {Yokohama, Japan},
series = {CHI '21}
}

@ARTICLE{exoRobotDemoLearning,
AUTHOR={Lauretti, Clemente  and Cordella, Francesca  and Ciancio, Anna Lisa  and Trigili, Emilio  and Catalan, Jose Maria  and Badesa, Francisco Javier  and Crea, Simona  and Pagliara, Silvio Marcello  and Sterzi, Silvia  and Vitiello, Nicola  and Garcia Aracil, Nicolas  and Zollo, Loredana },
         
TITLE={Learning by Demonstration for Motion Planning of Upper-Limb Exoskeletons},
        
JOURNAL={Frontiers in Neurorobotics},
        
VOLUME={Volume 12 - 2018},

YEAR={2018},

URL={https://www.frontiersin.org/journals/neurorobotics/articles/10.3389/fnbot.2018.00005},

DOI={10.3389/fnbot.2018.00005},

ISSN={1662-5218}}

@inproceedings{Rapido,
author = {Leiva, Germ\'{a}n and Gr\o{}nb\ae{}k, Jens Emil and Klokmose, Clemens Nylandsted and Nguyen, Cuong and Kazi, Rubaiat Habib and Asente, Paul},
title = {Rapido: Prototyping Interactive AR Experiences through Programming by Demonstration},
year = {2021},
isbn = {9781450386357},
publisher = {Association for Computing Machinery},
address = {New York, NY, USA},
url = {https://doi.org/10.1145/3472749.3474774},
doi = {10.1145/3472749.3474774},
booktitle = {The 34th Annual ACM Symposium on User Interface Software and Technology},
pages = {626–637},
numpages = {12},
location = {Virtual Event, USA},
series = {UIST '21}
}

@inproceedings{Mimic,
author = {Mahadevan, Karthik and Chen, Yan and Cakmak, Maya and Tang, Anthony and Grossman, Tovi},
title = {Mimic: In-Situ Recording and Re-Use of Demonstrations to Support Robot Teleoperation},
year = {2022},
isbn = {9781450393201},
publisher = {Association for Computing Machinery},
address = {New York, NY, USA},
url = {https://doi.org/10.1145/3526113.3545639},
doi = {10.1145/3526113.3545639},
booktitle = {Proceedings of the 35th Annual ACM Symposium on User Interface Software and Technology},
articleno = {40},
numpages = {13},
location = {Bend, OR, USA},
series = {UIST '22}
}

@inproceedings{assistanceOrDisruption,
author = {Pu, Kevin and Lazaro, Daniel and Arawjo, Ian and Xia, Haijun and Xiao, Ziang and Grossman, Tovi and Chen, Yan},
title = {Assistance or Disruption? Exploring and Evaluating the Design and Trade-offs of Proactive AI Programming Support},
year = {2025},
isbn = {9798400713941},
publisher = {Association for Computing Machinery},
address = {New York, NY, USA},
url = {https://doi.org/10.1145/3706598.3713357},
doi = {10.1145/3706598.3713357},
booktitle = {Proceedings of the 2025 CHI Conference on Human Factors in Computing Systems},
articleno = {152},
numpages = {21},
location = {
},
series = {CHI '25}
}

@inproceedings{headVRCorrection,
author = {Dang, Minh Duc and Luong, Duy Phuoc and Napier, Christopher and Kim, Lawrence H},
title = {Co-Design \& Evaluation of Visual Interventions for Head Posture Correction in Virtual Reality Games},
year = {2025},
isbn = {9798400713941},
publisher = {Association for Computing Machinery},
address = {New York, NY, USA},
url = {https://doi.org/10.1145/3706598.3713177},
doi = {10.1145/3706598.3713177},
booktitle = {Proceedings of the 2025 CHI Conference on Human Factors in Computing Systems},
articleno = {460},
numpages = {15},
location = {
},
series = {CHI '25}
}

@inproceedings{SnapInflatables,
author = {Yang, Yue and Ren, Lei and Chen, Chuang and Hu, Bin and Zhang, Zhuoyi and Li, Xinyan and Shen, Yanchen and Zhu, Kuangqi and Ji, Junzhe and Zhang, Yuyang and Ni, Yongbo and Wu, Jiayi and Wang, Qi and Wu, Jiang and Sun, Lingyun and Tao, Ye and Wang, Guanyun},
title = {SnapInflatables: Designing Inflatables with Snap-through Instability for Responsive Interaction},
year = {2024},
isbn = {9798400703300},
publisher = {Association for Computing Machinery},
address = {New York, NY, USA},
url = {https://doi.org/10.1145/3613904.3642933},
doi = {10.1145/3613904.3642933},
booktitle = {Proceedings of the 2024 CHI Conference on Human Factors in Computing Systems},
articleno = {342},
numpages = {15},
location = {Honolulu, HI, USA},
series = {CHI '24}
}

@inproceedings{EOrthosis,
author = {Yang, Yue and Ren, Lei and Chen, Chuang and Wang, Xinyue and Fan, Yitao and Shao, Yilin and Zhu, Kuangqi and Li, Jiaji and Wang, Qi and Sun, Lingyun and Tao, Ye and Wang, Guanyun},
title = {E-Orthosis: Augmenting Off-the-Shelf Orthoses with Electronics},
year = {2023},
isbn = {9781450394215},
publisher = {Association for Computing Machinery},
address = {New York, NY, USA},
url = {https://doi.org/10.1145/3544548.3581471},
doi = {10.1145/3544548.3581471},
booktitle = {Proceedings of the 2023 CHI Conference on Human Factors in Computing Systems},
articleno = {625},
numpages = {15},
location = {Hamburg, Germany},
series = {CHI '23}
}

@inproceedings{Embrogami,
author = {Jiang, Yu and Haynes, Alice C and Pourjafarian, Narjes and Borchers, Jan and Steimle, J\"{u}rgen},
title = {Embrogami: Shape-Changing Textiles with Machine Embroidery},
year = {2024},
isbn = {9798400706288},
publisher = {Association for Computing Machinery},
address = {New York, NY, USA},
url = {https://doi.org/10.1145/3654777.3676431},
doi = {10.1145/3654777.3676431},
booktitle = {Proceedings of the 37th Annual ACM Symposium on User Interface Software and Technology},
articleno = {63},
numpages = {15},
location = {Pittsburgh, PA, USA},
series = {UIST '24}
}

@inproceedings{ReKnitCare,
author = {Hu, Hongci and Jiang, Mengqi and Lin, Kai and Jiang, Kinor Shou-xiang and Bai, Ziqian},
title = {ReKnit-Care: A Seamless-Knitted Sensing Glove for Sensory Rehabilitation and Adaptive Haptic Feedback},
year = {2025},
isbn = {9798400711978},
publisher = {Association for Computing Machinery},
address = {New York, NY, USA},
url = {https://doi.org/10.1145/3689050.3705979},
doi = {10.1145/3689050.3705979},
booktitle = {Proceedings of the Nineteenth International Conference on Tangible, Embedded, and Embodied Interaction},
articleno = {58},
numpages = {7},
location = {
},
series = {TEI '25}
}

@inproceedings{PneuMa,
author = {Saini, Aryan and Patibanda, Rakesh and Overdevest, Nathalie and Van Den Hoven, Elise and Mueller, Florian ‘Floyd’},
title = {PneuMa: Designing Pneumatic Bodily Extensions for Supporting Movement in Everyday Life},
year = {2024},
isbn = {9798400704024},
publisher = {Association for Computing Machinery},
address = {New York, NY, USA},
url = {https://doi.org/10.1145/3623509.3633349},
doi = {10.1145/3623509.3633349},
booktitle = {Proceedings of the Eighteenth International Conference on Tangible, Embedded, and Embodied Interaction},
articleno = {1},
numpages = {16},
location = {Cork, Ireland},
series = {TEI '24}
}

@article{actuatingWearablesSkillLearning,
  title={Actuating wearables for motor skill learning: A constructive design research perspective},
  author={Markopoulos, Panos and Wang, Qi and Tomico, Oscar and Goveia da Rocha, Bruna and ten Bh{\"o}mer, Martijn and Giacolini, Luca and Palaima, Mantas and Virtala, Nita},
  journal={Design for Health},
  volume={4},
  number={2},
  pages={231--251},
  year={2020},
  publisher={Taylor \& Francis}
}

@online{allonic,
  author =    {allonic},
  title =  {allonic: Robotics by the meter},
  year = 2025,
  url =    {https://www.allonic.co},
  lastaccessed = {Jul 15, 2025}
  }

@INPROCEEDINGS{sourcePaperAllonic,
  author={Zhe Xu and Todorov, Emanuel},
  booktitle={2016 IEEE International Conference on Robotics and Automation (ICRA)}, 
  title={Design of a highly biomimetic anthropomorphic robotic hand towards artificial limb regeneration}, 
  year={2016},
  volume={},
  number={},
  pages={3485-3492},
  doi={10.1109/ICRA.2016.7487528}}

@inproceedings{ShapeTex,
author = {Du, Jiachun and Markopoulos, Panos and Wang, Qi and Toeters, Marina and Gong, Ting},
title = {ShapeTex: Implementing Shape-Changing Structures in Fabric for Wearable Actuation},
year = {2018},
isbn = {9781450355681},
publisher = {Association for Computing Machinery},
address = {New York, NY, USA},
url = {https://doi.org/10.1145/3173225.3173245},
doi = {10.1145/3173225.3173245},
booktitle = {Proceedings of the Twelfth International Conference on Tangible, Embedded, and Embodied Interaction},
pages = {166–176},
numpages = {11},
location = {Stockholm, Sweden},
series = {TEI '18}
}

@article{ardalanPaper,
  title={Dynamically adaptive soft metamaterial for wearable human--machine interfaces},
  author={Tanriverdi, Ugur and Senesi, Guglielmo and Asfour, Tarek and Kurt, Hasan and Smith, Sabrina L and Toderita, Diana and Shalhoub, Joseph and Burgess, Laura and Bull, Anthony MJ and G{\"u}der, Firat},
  journal={Nature Communications},
  volume={16},
  number={1},
  pages={2621},
  year={2025},
  publisher={Nature Publishing Group UK London}
}

@article{powerToThePeople,
author = {Amershi, Saleema and Cakmak, Maya and Knox, W. Bradley and Kulesza, Todd},
title = {Power to the People: The Role of Humans in Interactive Machine Learning},
journal = {AI Magazine},
volume = {35},
number = {4},
pages = {105-120},
doi = {https://doi.org/10.1609/aimag.v35i4.2513},
url = {https://onlinelibrary.wiley.com/doi/abs/10.1609/aimag.v35i4.2513},
eprint = {https://onlinelibrary.wiley.com/doi/pdf/10.1609/aimag.v35i4.2513},
year = {2014}
}

@article{commonInjuries,
  title={Weightlifting movements: do the benefits outweigh the risks?},
  author={Hedrick, Allen and Wada, Hiroaki},
  journal={Strength \& Conditioning Journal},
  volume={30},
  number={6},
  pages={26--35},
  year={2008},
  publisher={LWW}
}

@article{barnes2021peak,
  title={Peak power output and onset of muscle activation during high pull exercise},
  author={Barnes, Matthew J and Petterson, Ashley and Cochrane, Darryl J},
  journal={The Journal of Strength \& Conditioning Research},
  volume={35},
  number={3},
  pages={675--679},
  year={2021},
  publisher={LWW}
}

@article{alqarni2019common,
  title={Common injuries in resistance training},
  author={Alqarni, Ahmed Mohammed},
  journal={Saudi Journal of Sports Medicine},
  volume={19},
  number={2},
  pages={38--42},
  year={2019},
  publisher={Medknow}
}

@article{2010youngatheletes,
  title={Resistance training among young athletes: safety, efficacy and injury prevention effects},
  author={Faigenbaum, Avery D and Myer, Gregory D},
  journal={British journal of sports medicine},
  volume={44},
  number={1},
  pages={56--63},
  year={2010},
  publisher={British Association of Sport and Excercise Medicine}
}

@article{fisher2023supervision,
  title={Supervision during resistance training: A comparison of trainer and trainee perceptions},
  author={Fisher, James and Androulakis-Korakakis, Patroklos and Giessing, J{\"u}rgen and Helms, Eric and Schoenfeld, Brad and Smith, Dave and Winett, Richard},
  journal={International Journal of Strength and Conditioning},
  volume={3},
  number={1},
  year={2023},
  publisher={International Universities Strength and Conditioning Association}
}

@article{aslam2025neuromuscular,
  title={Neuromuscular adaptations to resistance training in elite versus recreational athletes},
  author={Aslam, Sumaira and Habyarimana, Jean De Dieu and Bin, Shi Yong},
  journal={Frontiers in Physiology},
  volume={16},
  pages={1598149},
  year={2025},
  publisher={Frontiers Media SA}
}

@article{latella2019differences,
  title={Differences in strength performance between novice and elite athletes: Evidence from powerlifters},
  author={Latella, Christopher and Van Den Hoek, Daniel and Teo, Wei-Peng},
  journal={The Journal of Strength \& Conditioning Research},
  volume={33},
  pages={S103--S112},
  year={2019},
  publisher={LWW}
}

@article{chernozub2024correlation,
  title={Correlation between load volume and indicators of adaptive body changes in untrained young men participating in fitness},
  author={Chernozub, Andrei and Hlukhov, Ivan and Drobot, Kateryna and Synytsia, Andrii and Rymyk, Roman and Pyatnychuk, Halyna and Leshchak, Oleksandr and Malanyuk, Lyubomyr and Potop, Vladimir},
  journal={Journal of Physical Education and Sport},
  pages={321--328},
  year={2024}
}

@article{2020progOverload,
  title={How much? How fast? How soon? Three simple concepts for progressing training loads to minimize injury risk and enhance performance},
  author={Gabbett, Tim J},
  journal={Journal of orthopaedic \& sports physical therapy},
  volume={50},
  number={10},
  pages={570--573},
  year={2020},
  publisher={JOSPT, Inc. JOSPT, 1033 North Fairfax Street, Suite 304, Alexandria, VA~…}
}

@article{beltsOnLowerBack,
  title={Effect of weightlifting belts on lumbar biomechanics and muscle activity in deadlift and squat},
  author={Herbaut, Alexis and Tuloup, Edouard},
  journal={Sports Engineering},
  volume={28},
  number={1},
  pages={1--9},
  year={2025},
  publisher={Springer}
}

@article{kneeSleeves,
  title={Neoprene knee sleeves of varying tightness augment barbell squat one repetition maximum performance without improving other indices of muscular strength, power, or endurance},
  author={Machek, Steven B and Cardaci, Thomas D and Wilburn, Dylan T and Cholewinski, Mitchell C and Latt, Scarlett Lin and Harris, Dillon R and Willoughby, Darryn S},
  journal={The Journal of Strength \& Conditioning Research},
  volume={35},
  pages={S6--S15},
  year={2021},
  publisher={LWW}
}

@article{coldPack,
  title={The use of a cold pack during resistance exercises is effective for reducing intramuscular oxygenation and increasing myoelectric activity},
  author={Goto, Masahiro and Yamashina, Yoshihiro and Takada, Akihiro and Kikuchi, Yui and Hamaoka, Takafumi and Terada, Shigeru},
  journal={Journal of Physical Therapy Science},
  volume={34},
  number={4},
  pages={335--340},
  year={2022},
  publisher={The Society of Physical Therapy Science}
}

@article{squatSuit,
  title={How technological impacts on performance have been managed in elite sport: a powerlifting example},
  author={Latella, Christopher and Garrett, Joel and van den Hoek, Daniel},
  journal={Journal of Applied Physiology},
  volume={137},
  number={4},
  pages={821--822},
  year={2024},
  publisher={American Physiological Society Rockville, MD}
}

@article{stretchyKneeElsevier,
title = {A garment-integrated textile stitch-based strain sensor device, IoT-Enabled for enhanced wearable sportswear applications},
journal = {Results in Engineering},
volume = {23},
pages = {102794},
year = {2024},
issn = {2590-1230},
doi = {https://doi.org/10.1016/j.rineng.2024.102794},
url = {https://www.sciencedirect.com/science/article/pii/S2590123024010491},
author = {Akshaya Kumar Aliyana and Danying Yang and Orathai Tangsirinaruenart and George K. Stylios}
}

@inproceedings{Weight-Mate,
author = {Paay, Jeni and Kjeldskov, Jesper and Sorensen, Frederick and Jensen, Thomas and Tirosh, Oren},
title = {Weight-Mate: Adaptive Training Support for Weight Lifting},
year = {2020},
isbn = {9781450376969},
publisher = {Association for Computing Machinery},
address = {New York, NY, USA},
url = {https://doi.org/10.1145/3369457.3369466},
doi = {10.1145/3369457.3369466},
booktitle = {Proceedings of the 31st Australian Conference on Human-Computer-Interaction},
pages = {95–105},
numpages = {11},
location = {Fremantle, WA, Australia},
series = {OzCHI '19}
}

@article{xu2025smart,
  title={Smart textiles for personalized sports and healthcare},
  author={Xu, Ziao and Zhang, Chentian and Wang, Faqiang and Yu, Jianyong and Yang, Gang and Surmenev, Roman A and Li, Zhaoling and Ding, Bin},
  journal={Nano-Micro Letters},
  volume={17},
  number={1},
  pages={1--39},
  year={2025},
  publisher={Springer}
}

@inproceedings{TrainerTap,
author = {Jo, Hye-Young and Wie, Chan Hu and Jang, Yejin and Kim, Dong-Uk and Son, Yurim and Kim, Yoonji},
title = {TrainerTap: Weightlifting Support System Prototype Simulating Personal Trainer's Tactile and Auditory Guidance},
year = {2023},
isbn = {9798400700965},
publisher = {Association for Computing Machinery},
address = {New York, NY, USA},
url = {https://doi.org/10.1145/3586182.3616644},
doi = {10.1145/3586182.3616644},
booktitle = {Adjunct Proceedings of the 36th Annual ACM Symposium on User Interface Software and Technology},
articleno = {14},
numpages = {3},
location = {San Francisco, CA, USA},
series = {UIST '23 Adjunct}
}

@ARTICLE{YogaVI,
  
AUTHOR={Islam, Md Shafiqul  and Lee, Sang Won  and Harden, Samantha M.  and Lim, Sol },
         
TITLE={Effects of vibrotactile feedback on yoga practice},
        
JOURNAL={Frontiers in Sports and Active Living},
        
VOLUME={Volume 4 - 2022},

YEAR={2022},

URL={https://www.frontiersin.org/journals/sports-and-active-living/articles/10.3389/fspor.2022.1005003},

DOI={10.3389/fspor.2022.1005003},

ISSN={2624-9367}}

@inproceedings{Springlets,
author = {Hamdan, Nur Al-huda and Wagner, Adrian and Voelker, Simon and Steimle, J\"{u}rgen and Borchers, Jan},
title = {Springlets: Expressive, Flexible and Silent On-Skin Tactile Interfaces},
year = {2019},
isbn = {9781450359702},
publisher = {Association for Computing Machinery},
address = {New York, NY, USA},
url = {https://doi.org/10.1145/3290605.3300718},
doi = {10.1145/3290605.3300718},
booktitle = {Proceedings of the 2019 CHI Conference on Human Factors in Computing Systems},
pages = {1–14},
numpages = {14},
location = {Glasgow, Scotland Uk},
series = {CHI '19}
}

@inproceedings{expressiveRobotLLM,
author = {Mahadevan, Karthik and Chien, Jonathan and Brown, Noah and Xu, Zhuo and Parada, Carolina and Xia, Fei and Zeng, Andy and Takayama, Leila and Sadigh, Dorsa},
title = {Generative Expressive Robot Behaviors using Large Language Models},
year = {2024},
isbn = {9798400703225},
publisher = {Association for Computing Machinery},
address = {New York, NY, USA},
url = {https://doi.org/10.1145/3610977.3634999},
doi = {10.1145/3610977.3634999},
booktitle = {Proceedings of the 2024 ACM/IEEE International Conference on Human-Robot Interaction},
pages = {482–491},
numpages = {10},
location = {Boulder, CO, USA},
series = {HRI '24}
}

@inproceedings{MuscleRehab,
author = {Zhu, Junyi and Lei, Yuxuan and Shah, Aashini and Schein, Gila and Ghaednia, Hamid and Schwab, Joseph and Harteveld, Casper and Mueller, Stefanie},
title = {MuscleRehab: Improving Unsupervised Physical Rehabilitation by Monitoring and Visualizing Muscle Engagement},
year = {2022},
isbn = {9781450393201},
publisher = {Association for Computing Machinery},
address = {New York, NY, USA},
url = {https://doi.org/10.1145/3526113.3545705},
doi = {10.1145/3526113.3545705},
booktitle = {Proceedings of the 35th Annual ACM Symposium on User Interface Software and Technology},
articleno = {33},
numpages = {14},
location = {Bend, OR, USA},
series = {UIST '22}
}

@Article{FSRandMMG,
AUTHOR = {Esposito, Daniele and Andreozzi, Emilio and Fratini, Antonio and Gargiulo, Gaetano D and Savino, Sergio and Niola, Vincenzo and Bifulco, Paolo},
TITLE = {A Piezoresistive Sensor to Measure Muscle Contraction and Mechanomyography},
JOURNAL = {Sensors},
VOLUME = {18},
YEAR = {2018},
NUMBER = {8},
ARTICLE-NUMBER = {2553},
URL = {https://www.mdpi.com/1424-8220/18/8/2553},
PubMedID = {30081541},
ISSN = {1424-8220},
DOI = {10.3390/s18082553}
}

@inproceedings{mmgTei,
author = {Rudolph, Julius Cosmo Romeo and Holman, David and De Araujo, Bruno and Jota, Ricardo and Wigdor, Daniel and Savage, Valkyrie},
title = {Sensing Hand Interactions with Everyday Objects by Profiling Wrist Topography},
year = {2022},
isbn = {9781450391474},
publisher = {Association for Computing Machinery},
address = {New York, NY, USA},
url = {https://doi.org/10.1145/3490149.3501320},
doi = {10.1145/3490149.3501320},
booktitle = {Proceedings of the Sixteenth International Conference on Tangible, Embedded, and Embodied Interaction},
articleno = {14},
numpages = {14},
location = {Daejeon, Republic of Korea},
series = {TEI '22}
}

@inproceedings{XRVisualDesSpace,
author = {Yu, Xingyao and Lee, Benjamin and Sedlmair, Michael},
title = {Design Space of Visual Feedforward And Corrective Feedback in XR-Based Motion Guidance Systems},
year = {2024},
isbn = {9798400703300},
publisher = {Association for Computing Machinery},
address = {New York, NY, USA},
url = {https://doi.org/10.1145/3613904.3642143},
doi = {10.1145/3613904.3642143},
booktitle = {Proceedings of the 2024 CHI Conference on Human Factors in Computing Systems},
articleno = {723},
numpages = {15},
location = {Honolulu, HI, USA},
series = {CHI '24}
}

@inproceedings{CollabJam,
author = {Wittchen, Dennis and Ramian, Alexander and Sabnis, Nihar and B\"{o}hme, Richard and Chlebowski, Christopher and Freitag, Georg and Fruchard, Bruno and Degraen, Donald},
title = {CollabJam: Studying Collaborative Haptic Experience Design for On-Body Vibrotactile Patterns},
year = {2025},
isbn = {9798400713941},
publisher = {Association for Computing Machinery},
address = {New York, NY, USA},
url = {https://doi.org/10.1145/3706598.3713469},
doi = {10.1145/3706598.3713469},
booktitle = {Proceedings of the 2025 CHI Conference on Human Factors in Computing Systems},
articleno = {1132},
numpages = {20},
location = {
},
series = {CHI '25}
}

@inproceedings{snowboardVibro,
author = {Spelmezan, Daniel and Jacobs, Mareike and Hilgers, Anke and Borchers, Jan},
title = {Tactile motion instructions for physical activities},
year = {2009},
isbn = {9781605582467},
publisher = {Association for Computing Machinery},
address = {New York, NY, USA},
url = {https://doi.org/10.1145/1518701.1519044},
doi = {10.1145/1518701.1519044},
booktitle = {Proceedings of the SIGCHI Conference on Human Factors in Computing Systems},
pages = {2243–2252},
numpages = {10},
location = {Boston, MA, USA},
series = {CHI '09}
}

@inproceedings{mitTunableSpaceSuit,
author = {Tessmer, Lavender and Goldstein, Ganit and Herrera-Arcos, Guillermo and Korolovych, Volodymyr and Bellisle, Rachel and Paige, Cody and Shallal, Christopher and Sahasrabudhe, Atharva and Herr, Hugh and Boriskina, Svetlana and Newman, Dava and Tibbits, Skylar},
year = {2022},
month = {10},
pages = {},
title = {3D Knit Spacesuit Sleeve with Multifunctional Fibers and Tunable Compression}
}

@inproceedings{avaTTAR,
author = {Ma, Dizhi and Hu, Xiyun and Shi, Jingyu and Patel, Mayank and Jain, Rahul and Liu, Ziyi and Zhu, Zhengzhe and Ramani, Karthik},
title = {avaTTAR: Table Tennis Stroke Training with Embodied and Detached Visualization in Augmented Reality},
year = {2024},
isbn = {9798400706288},
publisher = {Association for Computing Machinery},
address = {New York, NY, USA},
url = {https://doi.org/10.1145/3654777.3676400},
doi = {10.1145/3654777.3676400},
booktitle = {Proceedings of the 37th Annual ACM Symposium on User Interface Software and Technology},
articleno = {35},
numpages = {16},
location = {Pittsburgh, PA, USA},
series = {UIST '24}
}

@inproceedings{SkinMorph,
author = {Kao, Hsin-Liu (Cindy) and Bamforth, Miren and Kim, David and Schmandt, Chris},
title = {Skinmorph: texture-tunable on-skin interface through thin, programmable gel},
year = {2018},
isbn = {9781450359672},
publisher = {Association for Computing Machinery},
address = {New York, NY, USA},
url = {https://doi.org/10.1145/3267242.3267262},
doi = {10.1145/3267242.3267262},
booktitle = {Proceedings of the 2018 ACM International Symposium on Wearable Computers},
pages = {196–203},
numpages = {8},
location = {Singapore, Singapore},
series = {ISWC '18}
}

@inproceedings{OmniFiber,
author = {Kilic Afsar, Ozgun and Shtarbanov, Ali and Mor, Hila and Nakagaki, Ken and Forman, Jack and Modrei, Karen and Jeong, Seung Hee and Hjort, Klas and H\"{o}\"{o}k, Kristina and Ishii, Hiroshi},
title = {OmniFiber: Integrated Fluidic Fiber Actuators for Weaving Movement based Interactions into the ‘Fabric of Everyday Life’},
year = {2021},
isbn = {9781450386357},
publisher = {Association for Computing Machinery},
address = {New York, NY, USA},
url = {https://doi.org/10.1145/3472749.3474802},
doi = {10.1145/3472749.3474802},
booktitle = {The 34th Annual ACM Symposium on User Interface Software and Technology},
pages = {1010–1026},
numpages = {17},
location = {Virtual Event, USA},
series = {UIST '21}
}

@inproceedings{FlexTruss,
author = {Sun, Lingyun and Li, Jiaji and Chen, Yu and Yang, Yue and Yu, Zhi and Luo, Danli and Gu, Jianzhe and Yao, Lining and Tao, Ye and Wang, Guanyun},
title = {FlexTruss: A Computational Threading Method for Multi-material, Multi-form and Multi-use Prototyping},
year = {2021},
isbn = {9781450380966},
publisher = {Association for Computing Machinery},
address = {New York, NY, USA},
url = {https://doi.org/10.1145/3411764.3445311},
doi = {10.1145/3411764.3445311},
booktitle = {Proceedings of the 2021 CHI Conference on Human Factors in Computing Systems},
articleno = {432},
numpages = {12},
location = {Yokohama, Japan},
series = {CHI '21}
}

@inproceedings{BistableOrthoses,
  author    = {Lin, Yuyu and Zhu, Dian and Naidu, Anoushka and Yu, Kenneth and Harper, Deon and Halilaj, Eni and Weber, Douglas and Kenney, Deborah Ellen and Popchak, Adam J. and Baratz, Mark and Ion, Alexandra},
  title     = {Personalized Bistable Orthoses for Rehabilitation of Finger Joints},
  booktitle = {Proceedings of the 38th Annual ACM Symposium on User Interface Software and Technology},
  series    = {UIST '25},
  year      = {2025},
  location  = {Busan, Republic of Korea},
  publisher = {Association for Computing Machinery},
  address   = {New York, NY, USA},
  doi       = {10.1145/3746059.3747643},
  url       = {https://doi.org/10.1145/3746059.3747643},
  month     = sep,
  isbn      = {979-8-4007-2033-0}
}

@inproceedings{thermalBreathing,
author = {Brooks, Jas and Mazursky, Alex and Hixon, Janice and Lopes, Pedro},
title = {Augmented Breathing via Thermal Feedback in the Nose},
year = {2024},
isbn = {9798400706288},
publisher = {Association for Computing Machinery},
address = {New York, NY, USA},
url = {https://doi.org/10.1145/3654777.3676438},
doi = {10.1145/3654777.3676438},
booktitle = {Proceedings of the 37th Annual ACM Symposium on User Interface Software and Technology},
articleno = {26},
numpages = {11},
location = {Pittsburgh, PA, USA},
series = {UIST '24}
}

@inproceedings{smartwatchEms,
author = {Takahashi, Akifumi and Tanaka, Yudai and Tamhane, Archit and Shen, Alan and Teng, Shan-Yuan and Lopes, Pedro},
title = {Can a Smartwatch Move Your Fingers? Compact and Practical Electrical Muscle Stimulation in a Smartwatch},
year = {2024},
isbn = {9798400706288},
publisher = {Association for Computing Machinery},
address = {New York, NY, USA},
url = {https://doi.org/10.1145/3654777.3676373},
doi = {10.1145/3654777.3676373},
booktitle = {Proceedings of the 37th Annual ACM Symposium on User Interface Software and Technology},
articleno = {2},
numpages = {15},
location = {Pittsburgh, PA, USA},
series = {UIST '24}
}

@inproceedings{ThermoVR,
author = {Peiris, Roshan Lalintha and Peng, Wei and Chen, Zikun and Chan, Liwei and Minamizawa, Kouta},
title = {ThermoVR: Exploring Integrated Thermal Haptic Feedback with Head Mounted Displays},
year = {2017},
isbn = {9781450346559},
publisher = {Association for Computing Machinery},
address = {New York, NY, USA},
url = {https://doi.org/10.1145/3025453.3025824},
doi = {10.1145/3025453.3025824},
booktitle = {Proceedings of the 2017 CHI Conference on Human Factors in Computing Systems},
pages = {5452–5456},
numpages = {5},
location = {Denver, Colorado, USA},
series = {CHI '17}
}

@inproceedings{danceVisionDesignSpace,
author = {Lee, Soohwan and Hwang, Seoyeong and Oakley, Ian and Lee, Kyungho},
title = {Expanding the Design Space of Vision-based Interactive Systems for Group Dance Practice},
year = {2024},
isbn = {9798400705830},
publisher = {Association for Computing Machinery},
address = {New York, NY, USA},
url = {https://doi.org/10.1145/3643834.3661568},
doi = {10.1145/3643834.3661568},
booktitle = {Proceedings of the 2024 ACM Designing Interactive Systems Conference},
pages = {2768–2787},
numpages = {20},
location = {Copenhagen, Denmark},
series = {DIS '24}
}

@inproceedings{hereAndNow,
author = {Zhou, Qiushi and Grebel, Louise and Irlitti, Andrew and Minaai, Julie Ann and Goncalves, Jorge and Velloso, Eduardo},
title = {Here and Now: Creating Improvisational Dance Movements with a Mixed Reality Mirror},
year = {2023},
isbn = {9781450394215},
publisher = {Association for Computing Machinery},
address = {New York, NY, USA},
url = {https://doi.org/10.1145/3544548.3580666},
doi = {10.1145/3544548.3580666},
booktitle = {Proceedings of the 2023 CHI Conference on Human Factors in Computing Systems},
articleno = {183},
numpages = {16},
location = {Hamburg, Germany},
series = {CHI '23}
}

@article{reflexiveCoding,
author = {Virginia Braun and Victoria Clarke},
title = {Reflecting on reflexive thematic analysis},
journal = {Qualitative Research in Sport, Exercise and Health},
volume = {11},
number = {4},
pages = {589--597},
year = {2019},
publisher = {Routledge},
doi = {10.1080/2159676X.2019.1628806},
URL = { https://doi.org/10.1080/2159676X.2019.1628806 },
eprint = { https://doi.org/10.1080/2159676X.2019.1628806 }
}

@article{mdpiReviewSportsWearable,
  title={Review on Wearable Technology in sports: Concepts, Challenges and opportunities},
  author={Se{\c{c}}kin, Ahmet {\c{C}}a{\u{g}}da{\c{s}} and Ate{\c{s}}, Bahar and Se{\c{c}}kin, Mine},
  journal={Applied sciences},
  volume={13},
  number={18},
  pages={10399},
  year={2023},
  publisher={MDPI}
}

@inbook{videoanalysis,
author = {Knoblauch, Hubert and T., René and Schnettler, Bernt},
year = {2014},
month = {07},
pages = {},
title = {Video Analysis and Videography Qualitative Methods}
}

@misc{workoutcool2025,
  author       = {{Snouzy}},
  title        = {Modern open‐source fitness coaching platform. Create workout plans, track progress, and access a comprehensive exercise database},
  howpublished = {\url{https://www.workout.cool/}},
  year         = {2025},
  note         = {Accessed: 2025-09-04}
}

@inproceedings{rockClimbing,
author = {Kosmalla, Felix and Wiehr, Frederik and Daiber, Florian and Kr\"{u}ger, Antonio and L\"{o}chtefeld, Markus},
title = {ClimbAware: Investigating Perception and Acceptance of Wearables in Rock Climbing},
year = {2016},
isbn = {9781450333627},
publisher = {Association for Computing Machinery},
address = {New York, NY, USA},
url = {https://doi.org/10.1145/2858036.2858562},
doi = {10.1145/2858036.2858562},
booktitle = {Proceedings of the 2016 CHI Conference on Human Factors in Computing Systems},
pages = {1097–1108},
numpages = {12},
location = {San Jose, California, USA},
series = {CHI '16}
}

@article{Subtletee,
author = {Wo\'{z}niak, Miko\l{}aj P. and Dominiak, Julia and Pieprzowski, Micha\l{} and \L{}ado\'{n}ski, Piotr and Grudzie\'{n}, Krzysztof and Lischke, Lars and Romanowski, Andrzej and Wo\'{z}niak, Pawe\l{} W.},
title = {Subtletee: Augmenting Posture Awareness for Beginner Golfers},
year = {2020},
issue_date = {November 2020},
publisher = {Association for Computing Machinery},
address = {New York, NY, USA},
volume = {4},
number = {ISS},
url = {https://doi.org/10.1145/3427332},
doi = {10.1145/3427332},
journal = {Proc. ACM Hum.-Comput. Interact.},
month = nov,
articleno = {204},
numpages = {24}
}

@inproceedings{mediapipeaccuracy,
  title={Accuracy evaluation of 3D pose estimation with mediapipe pose for physical exercises},
  author={Dill, Sebastian and R{\"o}sch, Andreas and Rohr, Maurice and G{\"u}ney, G{\"o}khan and De Witte, Luisa and Schwartz, Elias and Hoog Antink, Christoph},
  booktitle={Current Directions in Biomedical Engineering},
  volume={9},
  number={1},
  pages={563--566},
  year={2023},
  organization={De Gruyter}
}

@inproceedings{SomaTheory,
author = {H\"{o}\"{o}k, Kristina and Hummels, Caroline and Isbister, Katherine and Marti, Patrizia and M\'{a}rquez Segura, Elena and Jonsson, Martin and Mueller, Florian 'Floyd' and Sanches, Pedro A.N. and Schiphorst, Thecla and St\r{a}hl, Anna and Svanaes, Dag and Trotto, Ambra and Petersen, Marianne Graves and Lim, Youn-kyung},
title = {Soma-Based Design Theory},
year = {2017},
isbn = {9781450346566},
publisher = {Association for Computing Machinery},
address = {New York, NY, USA},
url = {https://doi.org/10.1145/3027063.3027082},
doi = {10.1145/3027063.3027082},
booktitle = {Proceedings of the 2017 CHI Conference Extended Abstracts on Human Factors in Computing Systems},
pages = {550–557},
numpages = {8},
location = {Denver, Colorado, USA},
series = {CHI EA '17}
}

@article{HapticSnakes,
author = {Al-Sada, Mohammed and Jiang, Keren and Ranade, Shubhankar and Kalkattawi, Mohammed and Nakajima, Tatsuo},
title = {HapticSnakes: multi-haptic feedback wearable robots for immersive virtual reality},
year = {2020},
issue_date = {Jun 2020},
publisher = {Springer-Verlag},
address = {Berlin, Heidelberg},
volume = {24},
number = {2},
issn = {1359-4338},
url = {https://doi.org/10.1007/s10055-019-00404-x},
doi = {10.1007/s10055-019-00404-x},
journal = {Virtual Real.},
month = jun,
pages = {191–209},
numpages = {19}
}

@inproceedings{Gaiters,
author = {Wang, Chi and Huang, Da-Yuan and Hsu, Shuo-Wen and Lin, Cheng-Lung and Chiu, Yeu-Luen and Hou, Chu-En and Chen, Bing-Yu},
title = {Gaiters: Exploring Skin Stretch Feedback on Legs for Enhancing Virtual Reality Experiences},
year = {2020},
isbn = {9781450367080},
publisher = {Association for Computing Machinery},
address = {New York, NY, USA},
url = {https://doi.org/10.1145/3313831.3376865},
doi = {10.1145/3313831.3376865},
booktitle = {Proceedings of the 2020 CHI Conference on Human Factors in Computing Systems},
pages = {1–14},
numpages = {14},
location = {Honolulu, HI, USA},
series = {CHI '20}
}

@inproceedings{PneuMod,
author = {Zhang, Bowen and Sra, Misha},
title = {PneuMod: A Modular Haptic Device with Localized Pressure and Thermal Feedback},
year = {2021},
isbn = {9781450390927},
publisher = {Association for Computing Machinery},
address = {New York, NY, USA},
url = {https://doi.org/10.1145/3489849.3489857},
doi = {10.1145/3489849.3489857},
booktitle = {Proceedings of the 27th ACM Symposium on Virtual Reality Software and Technology},
articleno = {30},
numpages = {7},
location = {Osaka, Japan},
series = {VRST '21}
}

@inproceedings{SomaBits,
author = {Windlin, Charles and H\"{o}\"{o}k, Kristina and Laaksolahti, Jarmo},
title = {SKETCHING SOMA BITS},
year = {2022},
isbn = {9781450393584},
publisher = {Association for Computing Machinery},
address = {New York, NY, USA},
url = {https://doi.org/10.1145/3532106.3533510},
doi = {10.1145/3532106.3533510},
booktitle = {Proceedings of the 2022 ACM Designing Interactive Systems Conference},
pages = {1758–1772},
numpages = {15},
location = {Virtual Event, Australia},
series = {DIS '22}
}

@article{mediaPipeLatency,
  title={Human pose estimation using mediapipe pose and optimization method based on a humanoid model},
  author={Kim, Jong-Wook and Choi, Jin-Young and Ha, Eun-Ju and Choi, Jae-Ho},
  journal={Applied sciences},
  volume={13},
  number={4},
  pages={2700},
  year={2023},
  publisher={MDPI}
}

@article{codesIMWUT,
author = {Ley-Flores, Judith and Vidal, Laia Turmo and Segura, Elena M\'{a}rquez and Singh, Aneesha and Bevilacqua, Frederic and Cuadrado, Francisco and Dur\'{a}n, Joaqu\'{\i}n Roberto D\'{\i}az and Valdiviezo-Hern\'{a}ndez, Omar and S\'{a}nchez-Martin, Milagrosa and Tajadura-Jim\'{e}nez, Ana},
title = {Co-Designing Sensory Feedback for Wearables to Support Physical Activity through Body Sensations},
year = {2024},
issue_date = {March 2024},
publisher = {Association for Computing Machinery},
address = {New York, NY, USA},
volume = {8},
number = {1},
url = {https://doi.org/10.1145/3643499},
doi = {10.1145/3643499},
journal = {Proc. ACM Interact. Mob. Wearable Ubiquitous Technol.},
month = mar,
articleno = {40},
numpages = {31}
}

@article{HapticPilotIMWUT,
      author = {Sung, Youjin and Kim, Rachel and Song, Kun Woo and Shao, Yitian and Yoon, Sang Ho},
      title = {HapticPilot: Authoring In-situ Hand Posture-Adaptive Vibrotactile Feedback for Virtual Reality},
      year = {2024},
      issue_date = {December 2023},
      publisher = {Association for Computing Machinery},
      address = {New York, NY, USA},
      volume = {7},
      number = {4},
      url = {https://doi.org/10.1145/3631453},
      doi = {10.1145/3631453},
      journal = {Proc. ACM Interact. Mob. Wearable Ubiquitous Technol.},
      month = jan,
      articleno = {179},
      numpages = {28}
}

@article{FootStrikerIMWUT,
author = {Hassan, Mahmoud and Daiber, Florian and Wiehr, Frederik and Kosmalla, Felix and Kr\"{u}ger, Antonio},
title = {FootStriker: An EMS-based Foot Strike Assistant for Running},
year = {2017},
issue_date = {March 2017},
publisher = {Association for Computing Machinery},
address = {New York, NY, USA},
volume = {1},
number = {1},
url = {https://doi.org/10.1145/3053332},
doi = {10.1145/3053332},
journal = {Proc. ACM Interact. Mob. Wearable Ubiquitous Technol.},
month = mar,
articleno = {2},
numpages = {18}
}

@article{aSpireIMWUT,
author = {Choi, Kyung Yun and ElHaouij, Neska and Lee, Jinmo and Picard, Rosalind W. and Ishii, Hiroshi},
title = {Design and Evaluation of a Clippable and Personalizable Pneumatic-haptic Feedback Device for Breathing Guidance},
year = {2022},
issue_date = {March 2022},
publisher = {Association for Computing Machinery},
address = {New York, NY, USA},
volume = {6},
number = {1},
url = {https://doi.org/10.1145/3517234},
doi = {10.1145/3517234},
journal = {Proc. ACM Interact. Mob. Wearable Ubiquitous Technol.},
month = mar,
articleno = {6},
numpages = {36}
}

@article{ThermoFitIMWUT,
author = {Wang, Guanyun and Yang, Yue and Guo, Mengyan and Zhu, Kuangqi and Yan, Zihan and Cui, Qiang and Zhou, Zihong and Ji, Junzhe and Li, Jiaji and Luo, Danli and Pan, Deying and Fan, Yitao and Han, Teng and Tao, Ye and Sun, Lingyun},
title = {ThermoFit: Thermoforming Smart Orthoses via Metamaterial Structures for Body-Fitting and Component-Adjusting},
year = {2023},
issue_date = {March 2023},
publisher = {Association for Computing Machinery},
address = {New York, NY, USA},
volume = {7},
number = {1},
url = {https://doi.org/10.1145/3580806},
doi = {10.1145/3580806},
journal = {Proc. ACM Interact. Mob. Wearable Ubiquitous Technol.},
month = mar,
articleno = {31},
numpages = {27}
}

@article{ProxiFitIMWUT,
author = {Kim, Jiha and Nam, Younho and Lee, Jungeun and Suh, Young-Joo and Hwang, Inseok},
title = {ProxiFit: Proximity Magnetic Sensing Using a Single Commodity Mobile toward Holistic Weight Exercise Monitoring},
year = {2023},
issue_date = {September 2023},
publisher = {Association for Computing Machinery},
address = {New York, NY, USA},
volume = {7},
number = {3},
url = {https://doi.org/10.1145/3610920},
doi = {10.1145/3610920},
journal = {Proc. ACM Interact. Mob. Wearable Ubiquitous Technol.},
month = sep,
articleno = {105},
numpages = {32}
}

@article{SeamFitIMWUT,
author = {Yu, Tianhong Catherine and Zhang, Manru Mary and Malenab, Luis Miguel and Lee, Chi-Jung and Jiang, Jacky Hao and Zhang, Ruidong and Guimbreti\`{e}re, Fran\c{c}ois and Zhang, Cheng},
title = {SeamFit: Towards Practical Smart Clothing for Automatic Exercise Logging},
year = {2025},
issue_date = {March 2025},
publisher = {Association for Computing Machinery},
address = {New York, NY, USA},
volume = {9},
number = {1},
url = {https://doi.org/10.1145/3712287},
doi = {10.1145/3712287},
journal = {Proc. ACM Interact. Mob. Wearable Ubiquitous Technol.},
month = mar,
articleno = {24},
numpages = {22}
}

@inproceedings{ShapeKit,
author = {Zhou, Ran and Ding, Jianru and Gao, Chenfeng and Qian, Wanli and Erickson, Benjamin and Balaam, Madeline and Leithinger, Daniel and Nakagaki, Ken},
title = {Shape-Kit: A Design Toolkit for Crafting On-Body Expressive Haptics},
year = {2025},
isbn = {9798400713941},
publisher = {Association for Computing Machinery},
address = {New York, NY, USA},
url = {https://doi.org/10.1145/3706598.3713981},
doi = {10.1145/3706598.3713981},
booktitle = {Proceedings of the 2025 CHI Conference on Human Factors in Computing Systems},
articleno = {504},
numpages = {26},
location = {
},
series = {CHI '25}
}

@article{VibroMap,
author = {Elsayed, Hesham and Weigel, Martin and M\"{u}ller, Florian and Schmitz, Martin and Marky, Karola and G\"{u}nther, Sebastian and Riemann, Jan and M\"{u}hlh\"{a}user, Max},
title = {VibroMap: Understanding the Spacing of Vibrotactile Actuators across the Body},
year = {2020},
issue_date = {December 2020},
publisher = {Association for Computing Machinery},
address = {New York, NY, USA},
volume = {4},
number = {4},
url = {https://doi.org/10.1145/3432189},
doi = {10.1145/3432189},
journal = {Proc. ACM Interact. Mob. Wearable Ubiquitous Technol.},
month = dec,
articleno = {125},
numpages = {16}
}

@article{smartRingsIMWUT,
author = {Wang, Zeyu and Yu, Ruotong and Wang, Xiangyang and Ding, Jiexin and Tang, Jiankai and Fang, Jun and He, Zhe and Li, Zhuojun and R\"{o}ddiger, Tobias and Xu, Weiye and Zhang, Xiyuxing and Gao, Huan-ang and Gao, Nan and Yu, Chun and Shi, Yuanchun and Wang, Yuntao},
title = {Computing with Smart Rings: A Systematic Literature Review},
year = {2025},
issue_date = {September 2025},
publisher = {Association for Computing Machinery},
address = {New York, NY, USA},
volume = {9},
number = {3},
url = {https://doi.org/10.1145/3749480},
doi = {10.1145/3749480},
journal = {Proc. ACM Interact. Mob. Wearable Ubiquitous Technol.},
month = sep,
articleno = {137},
numpages = {54}
}

@article{ModElecIMWUT,
author = {He, Liang and Wittkopf, Jarrid A. and Jun, Ji Won and Erickson, Kris and Ballagas, Rafael Tico},
title = {ModElec: A Design Tool for Prototyping Physical Computing Devices Using Conductive 3D Printing},
year = {2022},
issue_date = {Dec 2021},
publisher = {Association for Computing Machinery},
address = {New York, NY, USA},
volume = {5},
number = {4},
url = {https://doi.org/10.1145/3495000},
doi = {10.1145/3495000},
journal = {Proc. ACM Interact. Mob. Wearable Ubiquitous Technol.},
month = dec,
articleno = {159},
numpages = {20}
}

@inproceedings{FluxLab,
  author = {Lee, Hsuanling and Yu, Jiakun and Zheng, Shurui and Wu, Te-Yan and He, Liang},
  title = {FluxLab: Creating 3D Printable Shape-Changing Devices with Integrated Deformation Sensing},
  booktitle = {Proceedings of the Twentieth International Conference on Tangible, Embedded, and Embodied Interaction},
  series = {TEI '26},
  year = {2026},
  publisher = {Association for Computing Machinery},
  address = {New York, NY, USA},
  location = {Chicago, IL, USA},
  pages = {1--12},
  numpages = {12},
  doi = {10.1145/3731459.3773331},
  url = {https://doi.org/10.1145/3731459.3773331},
}

@online{miro,
  author =    {Miro},
  title =  {Where teams and their agents think, plan and build together},
  year = 2026,
  url =    {https://miro.com},
  lastaccessed = {Jul 27, 2026}
  }

@InProceedings{PSD_AIED,
author="Holstein, Kenneth
and McLaren, Bruce M.
and Aleven, Vincent",
editor="Isotani, Seiji
and Mill{\'a}n, Eva
and Ogan, Amy
and Hastings, Peter
and McLaren, Bruce
and Luckin, Rose",
title="Designing for Complementarity: Teacher and Student Needs for Orchestration Support in AI-Enhanced Classrooms",
booktitle="Artificial Intelligence in Education",
year="2019",
publisher="Springer International Publishing",
address="Cham",
pages="157--171",
isbn="978-3-030-23204-7"
}

@article{PSD_CSCW,
  title={Balancing Teacher and Student Agency: Co-Orchestration Tool Design Supporting Real-Time Dynamic Pairing},
  author={Yang, Kexin Bella and Liu, Menghan and Xu, Liyi and Rummel, Nikol and Aleven, Vincent},
  journal={arXiv preprint arXiv:2605.18761},
  year={2026}
}

@book{affinityDiagramming,
  title={Contextual design: defining customer-centered systems},
  author={Holtzblatt, Karen and Beyer, Hugh},
  year={1997},
  publisher={Elsevier}
}

@book{groundedTheory,
  title={Discovery of grounded theory: Strategies for qualitative research},
  author={Glaser, Barney and Strauss, Anselm},
  year={2017},
  publisher={Routledge}
}

@article{abductiveSynthesis,
  title={Theory construction in qualitative research: From grounded theory to abductive analysis},
  author={Timmermans, Stefan and Tavory, Iddo},
  journal={Sociological theory},
  volume={30},
  number={3},
  pages={167--186},
  year={2012},
  publisher={Sage publications Sage CA: Los Angeles, CA}
}

@inproceedings{ImaginaryJoint,
author = {Takashita, Shuto and Steimle, J\"{u}rgen and Inami, Masahiko},
title = {Imaginary Joint: Proprioceptive Feedback for Virtual Body Extensions via Skin Stretch},
year = {2025},
isbn = {9798400720376},
publisher = {Association for Computing Machinery},
address = {New York, NY, USA},
url = {https://doi.org/10.1145/3746059.3747800},
doi = {10.1145/3746059.3747800},
booktitle = {Proceedings of the 38th Annual ACM Symposium on User Interface Software and Technology},
articleno = {63},
numpages = {15},
location = {
},
series = {UIST '25}
}

@article{UbiPhysio,
author = {Wang, Chongyang and Feng, Yuan and Zhong, Lingxiao and Zhu, Siyi and Zhang, Chi and Zheng, Siqi and Liang, Chen and Wang, Yuntao and He, Chengqi and Yu, Chun and Shi, Yuanchun},
title = {UbiPhysio: Support Daily Functioning, Fitness, and Rehabilitation with Action Understanding and Feedback in Natural Language},
year = {2024},
issue_date = {March 2024},
publisher = {Association for Computing Machinery},
address = {New York, NY, USA},
volume = {8},
number = {1},
url = {https://doi.org/10.1145/3643552},
doi = {10.1145/3643552},
journal = {Proc. ACM Interact. Mob. Wearable Ubiquitous Technol.},
month = mar,
articleno = {20},
numpages = {27}
}

@article{personalSensingNoti,
author = {Newn, Joshua and Kelly, Ryan M. and D'Alfonso, Simon and Lederman, Reeva},
title = {Examining and Promoting Explainable Recommendations for Personal Sensing Technology Acceptance},
year = {2022},
issue_date = {September 2022},
publisher = {Association for Computing Machinery},
address = {New York, NY, USA},
volume = {6},
number = {3},
url = {https://doi.org/10.1145/3550297},
doi = {10.1145/3550297},
journal = {Proc. ACM Interact. Mob. Wearable Ubiquitous Technol.},
month = sep,
articleno = {133},
numpages = {27}
}

@article{longTermPhyData,
author = {Tang, Lie Ming and Kay, Judy},
title = {Harnessing Long Term Physical Activity Data—How Long-term Trackers Use Data and How an Adherence-based Interface Supports New Insights},
year = {2017},
issue_date = {June 2017},
publisher = {Association for Computing Machinery},
address = {New York, NY, USA},
volume = {1},
number = {2},
url = {https://doi.org/10.1145/3090091},
doi = {10.1145/3090091},
journal = {Proc. ACM Interact. Mob. Wearable Ubiquitous Technol.},
month = jun,
articleno = {26},
numpages = {28}
}

@article{lowcostFitnessWearables,
author = {Kouaho, Whitney-Jocelyn and Epstein, Daniel A.},
title = {Investigating Perspectives of and Experiences with Low Cost Commercial Fitness Wearables},
year = {2024},
issue_date = {December 2024},
publisher = {Association for Computing Machinery},
address = {New York, NY, USA},
volume = {8},
number = {4},
url = {https://doi.org/10.1145/3699740},
doi = {10.1145/3699740},
journal = {Proc. ACM Interact. Mob. Wearable Ubiquitous Technol.},
month = nov,
articleno = {193},
numpages = {22}
}

@article{phyActivityGoalsModeration,
author = {Alqahtani, Deemah and Jay, Caroline and Vigo, Markel},
title = {The Effect of Goal Moderation on the Achievement and Satisfaction of Physical Activity Goals},
year = {2020},
issue_date = {December 2020},
publisher = {Association for Computing Machinery},
address = {New York, NY, USA},
volume = {4},
number = {4},
url = {https://doi.org/10.1145/3432209},
doi = {10.1145/3432209},
journal = {Proc. ACM Interact. Mob. Wearable Ubiquitous Technol.},
month = dec,
articleno = {116},
numpages = {18}
}

@inproceedings{funcNowWearableLater,
author = {Pettys-Baker, Robert and Clarke, Megan E. and Holschuh, Brad},
title = {Functional Now, Wearable Later: Examining the Design Practices of Wearable Technologists},
year = {2024},
isbn = {9798400710599},
publisher = {Association for Computing Machinery},
address = {New York, NY, USA},
url = {https://doi.org/10.1145/3675095.3676615},
doi = {10.1145/3675095.3676615},
booktitle = {Proceedings of the 2024 ACM International Symposium on Wearable Computers},
pages = {71–81},
numpages = {11},
location = {Melbourne VIC, Australia},
series = {ISWC '24}
}

@article{ACLGuard,
author = {Yang, Baichen and Zhang, Xinyi and He, Xin and Xu, Chi and Xie, Wentao and Liang, Zuru and Yung, Shu-hang, Patrick and Zhang, Qian},
title = {ACLGuard: Physics-Aware Knee Loading Monitoring System for Anterior Cruciate Ligament Injury Prevention Training},
year = {2025},
issue_date = {December 2025},
publisher = {Association for Computing Machinery},
address = {New York, NY, USA},
volume = {9},
number = {4},
url = {https://doi.org/10.1145/3770692},
doi = {10.1145/3770692},
journal = {Proc. ACM Interact. Mob. Wearable Ubiquitous Technol.},
month = dec,
articleno = {231},
numpages = {28}
}

@inproceedings{MagTex,
author = {Zhang, Sen and Miao, Yuxuan and Taylor, Jazlin and Luo, Yiyue},
title = {MagTex: Machine-Knitted Magnetoactive Textiles for Bidirectional Human-Machine Interface},
year = {2025},
isbn = {9798400720376},
publisher = {Association for Computing Machinery},
address = {New York, NY, USA},
url = {https://doi.org/10.1145/3746059.3747648},
doi = {10.1145/3746059.3747648},
booktitle = {Proceedings of the 38th Annual ACM Symposium on User Interface Software and Technology},
articleno = {134},
numpages = {15},
location = {
},
series = {UIST '25}
}

@inproceedings{BIT,
author = {Xu, Weiye and Li, Tony and Wang, Yuntao and Yang, Xing-Dong and Wu, Te-Yen},
title = {BIT: Battery-free, IC-less and Wireless Smart Textile Interface and Sensing System},
year = {2025},
isbn = {9798400713941},
publisher = {Association for Computing Machinery},
address = {New York, NY, USA},
url = {https://doi.org/10.1145/3706598.3713100},
doi = {10.1145/3706598.3713100},
booktitle = {Proceedings of the 2025 CHI Conference on Human Factors in Computing Systems},
articleno = {926},
numpages = {18},
location = {
},
series = {CHI '25}
}

@inproceedings{IntelliLining,
author = {Ghane Ezabadi, Mahdie and Nittala, Aditya Shekhar and Yang, Xing-Dong and Wu, Te-Yen},
title = {IntelliLining: Activity Sensing through Textile Interlining Sensors Using TENGs},
year = {2025},
isbn = {9798400713941},
publisher = {Association for Computing Machinery},
address = {New York, NY, USA},
url = {https://doi.org/10.1145/3706598.3713167},
doi = {10.1145/3706598.3713167},
booktitle = {Proceedings of the 2025 CHI Conference on Human Factors in Computing Systems},
articleno = {439},
numpages = {14},
location = {
},
series = {CHI '25}
}

@inproceedings{Kinethreads,
author = {Shen, Vivian and Harrison, Chris},
title = {Kinethreads: Soft Full-Body Haptic Exosuit using Low-Cost Motor-Pulley Mechanisms},
year = {2025},
isbn = {9798400720376},
publisher = {Association for Computing Machinery},
address = {New York, NY, USA},
url = {https://doi.org/10.1145/3746059.3747755},
doi = {10.1145/3746059.3747755},
booktitle = {Proceedings of the 38th Annual ACM Symposium on User Interface Software and Technology},
articleno = {1},
numpages = {16},
location = {
},
series = {UIST '25}
}

@inproceedings{reelFeel,
author = {DeVrio, Nathan and Harrison, Chris},
title = {Reel Feel: Rich Haptic XR Experiences Using an Active, Worn, Multi-String Device},
year = {2025},
isbn = {9798400713941},
publisher = {Association for Computing Machinery},
address = {New York, NY, USA},
url = {https://doi.org/10.1145/3706598.3713615},
doi = {10.1145/3706598.3713615},
booktitle = {Proceedings of the 2025 CHI Conference on Human Factors in Computing Systems},
articleno = {610},
numpages = {20},
location = {
},
series = {CHI '25}
}

@inproceedings{genEMG,
author = {Ho, Yun and Nith, Romain and Jiang, Peili and He, Steven and Felalaga, Bruno and Teng, Shan-Yuan and Seeralan, Rhea and Lopes, Pedro},
title = {Generative Muscle Stimulation: Providing Users with Physical Assistance by Constraining Multimodal-AI with Embodied Knowledge},
year = {2026},
isbn = {9798400722783},
publisher = {Association for Computing Machinery},
address = {New York, NY, USA},
url = {https://doi.org/10.1145/3772318.3790817},
doi = {10.1145/3772318.3790817},
booktitle = {Proceedings of the 2026 CHI Conference on Human Factors in Computing Systems},
articleno = {548},
numpages = {22},
location = {
},
series = {CHI '26}
}

@inproceedings{Mod2Hap,
author = {Heo, Yong Hae and Lee, Seok Hun and Yoon, Hongmin and Lee, Jaeyeon and Kim, Sang-Youn},
title = {Mod2Hap: Two Level Modular Haptic System for Body- and Hand- Interactions using Magnetorheological Fluid},
year = {2025},
isbn = {9798400720376},
publisher = {Association for Computing Machinery},
address = {New York, NY, USA},
url = {https://doi.org/10.1145/3746059.3747712},
doi = {10.1145/3746059.3747712},
booktitle = {Proceedings of the 38th Annual ACM Symposium on User Interface Software and Technology},
articleno = {58},
numpages = {11},
location = {
},
series = {UIST '25}
}

@inproceedings{TingleTouch,
author = {Kim, Dong-Uk and Jo, Hye-Young and Kim, Hankyung and Suzuki, Ryo and Je, Seungwoo and Kim, Yoonji},
title = {TingleTouch: Touch Guidance through Electrical Stimulation in Resistance Training},
year = {2026},
isbn = {9798400722783},
publisher = {Association for Computing Machinery},
address = {New York, NY, USA},
url = {https://doi.org/10.1145/3772318.3791646},
doi = {10.1145/3772318.3791646},
booktitle = {Proceedings of the 2026 CHI Conference on Human Factors in Computing Systems},
articleno = {230},
numpages = {20},
location = {
},
series = {CHI '26}
}

@inproceedings{danceInflatables,
author = {Chen, Yulu and Chen, Chuang and Yang, Yue and Tao, Ye and Wang, Guanyun},
title = {Dancing with Inflatables: Design Bistable Inflatable Wearables for Choreographic Tools},
year = {2025},
isbn = {9798400720369},
publisher = {Association for Computing Machinery},
address = {New York, NY, USA},
url = {https://doi.org/10.1145/3746058.3758387},
doi = {10.1145/3746058.3758387},
booktitle = {Adjunct Proceedings of the 38th Annual ACM Symposium on User Interface Software and Technology},
articleno = {134},
numpages = {3},
location = {
},
series = {UIST Adjunct '25}
}

@inproceedings{grandChallengesCompControlBody,
author = {‘Floyd’ Mueller, Florian and Berthouze, Nadia and Sra, Misha and Gonzalez-Franco, Mar and Pohl, Henning and Boll, Susanne and Byrne, Richard and Caetano, Arthur and Inami, Masahiko and Knibbe, Jarrod and Kristensson, Per Ola and Li, Xiang and Li, Zhuying and Marshall, Joe and Petersen Matjeka, Louise and Nygren, Minna and Patibanda, Rakesh and Price, Sara and Reiterer, Harald and Saini, Aryan and Schneider, Oliver and Shahu, Ambika and Toups Dugas, Phoebe O. and Elvitigala, Don Samitha},
title = {Grand Challenges around Designing Computers’ Control Over Our Bodies},
year = {2026},
isbn = {9798400722783},
publisher = {Association for Computing Machinery},
address = {New York, NY, USA},
url = {https://doi.org/10.1145/3772318.3790606},
doi = {10.1145/3772318.3790606},
booktitle = {Proceedings of the 2026 CHI Conference on Human Factors in Computing Systems},
articleno = {742},
numpages = {22},
location = {
},
series = {CHI '26}
}

@ARTICLE{ieeeVibration1,
  author={de Vlam, Valerie and Wiertlewski, Michaël and Vardar, Yasemin},
  journal={IEEE Transactions on Haptics}, 
  title={Focused Vibrotactile Stimuli From a Wearable Sparse Array of Actuators}, 
  year={2023},
  volume={16},
  number={4},
  pages={511-517},
  doi={10.1109/TOH.2023.3270362}}

@ARTICLE{ieeeVibration2,
  author={Casalino, Andrea and Messeri, Costanza and Pozzi, Maria and Zanchettin, Andrea Maria and Rocco, Paolo and Prattichizzo, Domenico},
  journal={IEEE Robotics and Automation Letters}, 
  title={Operator Awareness in Human–Robot Collaboration Through Wearable Vibrotactile Feedback}, 
  year={2018},
  volume={3},
  number={4},
  pages={4289-4296},
  doi={10.1109/LRA.2018.2865034}}

@ARTICLE{ieeeVibration3,
  author={Barontini, Federica and Catalano, Manuel G. and Pallottino, Lucia and Leporini, Barbara and Bianchi, Matteo},
  journal={IEEE Transactions on Haptics}, 
  title={Integrating Wearable Haptics and Obstacle Avoidance for the Visually Impaired in Indoor Navigation: A User-Centered Approach}, 
  year={2021},
  volume={14},
  number={1},
  pages={109-122},
  doi={10.1109/TOH.2020.2996748}}

@ARTICLE{ieeeVibration4,
  author={Pacchierotti, Claudio and Sinclair, Stephen and Solazzi, Massimiliano and Frisoli, Antonio and Hayward, Vincent and Prattichizzo, Domenico},
  journal={IEEE Transactions on Haptics}, 
  title={Wearable Haptic Systems for the Fingertip and the Hand: Taxonomy, Review, and Perspectives}, 
  year={2017},
  volume={10},
  number={4},
  pages={580-600},
  doi={10.1109/TOH.2017.2689006}}

@ARTICLE{ieeeVibration5,
  author={Coles, Timothy R. and Meglan, Dwight and John, Nigel W.},
  journal={IEEE Transactions on Haptics}, 
  title={The Role of Haptics in Medical Training Simulators: A Survey of the State of the Art}, 
  year={2011},
  volume={4},
  number={1},
  pages={51-66},
  doi={10.1109/TOH.2010.19}}

@INPROCEEDINGS{ieeeVibration6,
  author={Lee, Hojeong and Kim, Eunho and Yoon, Sang Ho},
  booktitle={2025 IEEE World Haptics Conference (WHC)}, 
  title={3D Shape Perception Through Spatiotemporal Vibrotactile Patterns with Kinesthetic Feedback}, 
  year={2025},
  volume={},
  number={},
  pages={439-445},
  doi={10.1109/WHC64065.2025.11123386}}

@ARTICLE{ieeeVibration7,
  author={Pra, Yuri De and Papetti, Stefano and Järveläinen, Hanna and Bianchi, Matteo and Fontana, Federico},
  journal={IEEE Transactions on Haptics}, 
  title={Effects of Vibration Direction and Pressing Force on Finger Vibrotactile Perception and Force Control}, 
  year={2023},
  volume={16},
  number={1},
  pages={23-32},
  doi={10.1109/TOH.2022.3225714}}

@ARTICLE{ieeeVibration8,
  author={Flores, German and Kurniawan, Sri and Manduchi, Roberto and Martinson, Eric and Morales, Lourdes M. and Sisbot, Emrah Akin},
  journal={IEEE Transactions on Haptics}, 
  title={Vibrotactile Guidance for Wayfinding of Blind Walkers}, 
  year={2015},
  volume={8},
  number={3},
  pages={306-317},
  doi={10.1109/TOH.2015.2409980}}

@ARTICLE{ieeeVibration9,
  author={Deo, Darrel R. and Rezaii, Paymon and Hochberg, Leigh R. and Okamura, Allison M. and Shenoy, Krishna V. and Henderson, Jaimie M.},
  journal={IEEE Transactions on Haptics}, 
  title={Effects of Peripheral Haptic Feedback on Intracortical Brain-Computer Interface Control and Associated Sensory Responses in Motor Cortex}, 
  year={2021},
  volume={14},
  number={4},
  pages={762-775},
  doi={10.1109/TOH.2021.3072615}}

@ARTICLE{ieeeVibration10,
  author={Moriyama, Taha and Kajimoto, Hiroyuki},
  journal={IEEE Transactions on Haptics}, 
  title={Wearable Haptic Device Presenting Sensations of Fingertips to the Forearm}, 
  year={2022},
  volume={15},
  number={1},
  pages={91-96},
  doi={10.1109/TOH.2022.3143663}}

@ARTICLE{ieeeVibration11,
  author={De-Pra, Yuri and Catrambone, Vincenzo and van-Wassenhove, Virginie and Moscatelli, Alessandro and Valenza, Gaetano and Bianchi, Matteo},
  journal={IEEE Transactions on Haptics}, 
  title={Altering Time Perception in Virtual Reality Through Multimodal Visual-Tactile Kappa Effect}, 
  year={2023},
  volume={16},
  number={4},
  pages={518-523},
  doi={10.1109/TOH.2023.3270639}}

@INPROCEEDINGS{ieeeVibration12,
  author={Erwin, Andrew and Sup, Frank},
  booktitle={2014 IEEE Haptics Symposium (HAPTICS)}, 
  title={Design and perceptibility of a wearable haptic device using low-frequency stimulations on the forearm}, 
  year={2014},
  volume={},
  number={},
  pages={505-508},
  doi={10.1109/HAPTICS.2014.6775507}}

@INPROCEEDINGS{ieeeVibration13,
  author={Schwendicke, Anna and Dou, Jing and Altinsoy, M. Ercan},
  booktitle={2017 IEEE World Haptics Conference (WHC)}, 
  title={Temporal masking characteristics of whole body vibration perception}, 
  year={2017},
  volume={},
  number={},
  pages={478-483},
  doi={10.1109/WHC.2017.7989948}}

@ARTICLE{ieeeVibration14,
  author={Vyas, Preeti and Al Taha, Feras and Blum, Jeffrey R. and Weill-Duflos, Antoine and Cooperstock, Jeremy R.},
  journal={IEEE Transactions on Haptics}, 
  title={Ten Little Fingers, Ten Little Toes: Can Toes Match Fingers for Haptic Discrimination?}, 
  year={2020},
  volume={13},
  number={1},
  pages={130-136},
  doi={10.1109/TOH.2020.2966969}}

@ARTICLE{ieeeVibration15,
  author={Luo, Hu and Fu, Yuxiang and Ding, Ning and Dong, Chenyu and Zhang, Yuru and Wang, Dangxiao},
  journal={IEEE Transactions on Haptics}, 
  title={Hap-pulse: A Wearable Vibrotactile Glove for Medical Pulse Wave Rendering}, 
  year={2022},
  volume={15},
  number={2},
  pages={280-291},
  doi={10.1109/TOH.2022.3157832}}

@ARTICLE{ieeeVibration16,
  author={Fino, Nathaniel and Jumet, Barclay and Zook, Zane A. and Preston, Daniel J. and O'Malley, Marcia K.},
  journal={IEEE Transactions on Haptics}, 
  title={Mechanofluidic Instability-Driven Wearable Textile Vibrotactor}, 
  year={2023},
  volume={16},
  number={4},
  pages={530-535},
  doi={10.1109/TOH.2023.3271128}}

@ARTICLE{ieeeEMSTENS1,
  author={Kourtesis, Panagiotis and Argelaguet, Ferran and Vizcay, Sebastian and Marchal, Maud and Pacchierotti, Claudio},
  journal={IEEE Transactions on Haptics}, 
  title={Electrotactile Feedback Applications for Hand and Arm Interactions: A Systematic Review, Meta-Analysis, and Future Directions}, 
  year={2022},
  volume={15},
  number={3},
  pages={479-496},
  doi={10.1109/TOH.2022.3189866}}

@ARTICLE{ieeeEMSTENS2,
  author={Vrontos, Apostolos and Nitsch, Verena and Mertens, Alexander and Brandl, Christopher},
  journal={IEEE Transactions on Haptics}, 
  title={Characterizing EMS-Induced Haptic Weight Perceptions: Effects of Posture, Augmentation Parameters, and User Variability}, 
  year={2026},
  volume={19},
  number={1},
  pages={205-216},
  doi={10.1109/TOH.2026.3663216}}

@ARTICLE{ieeeEMSTENS3,
  author={Lee, Jungpyo and McPherson, Andrew I. W. and Huang, Haoxiang and Yu, Licheng and Gloumakov, Yuri and Stuart, Hannah S.},
  journal={IEEE Transactions on Neural Systems and Rehabilitation Engineering}, 
  title={Expanding Functional Workspace for People With C5-C7 Spinal Cord Injury With Supernumerary Dorsal Grasping}, 
  year={2025},
  volume={33},
  number={},
  pages={22-33},
  doi={10.1109/TNSRE.2024.3514135}}

@ARTICLE{ieeeEMSTENS4,
  author={Ackermann, D. Michael and Foldes, Emily L. and Bhadra, Niloy and Kilgore, Kevin L.},
  journal={IEEE Transactions on Neural Systems and Rehabilitation Engineering}, 
  title={Effect of Bipolar Cuff Electrode Design on Block Thresholds in High-Frequency Electrical Neural Conduction Block}, 
  year={2009},
  volume={17},
  number={5},
  pages={469-477},
  doi={10.1109/TNSRE.2009.2034069}}

@ARTICLE{ieeePneum1,
  author={Talhan, Aishwari and Yoo, Yongjae and Cooperstock, Jeremy R.},
  journal={IEEE Transactions on Haptics}, 
  title={Soft Pneumatic Haptic Wearable to Create the Illusion of Human Touch}, 
  year={2024},
  volume={17},
  number={2},
  pages={177-190},
  doi={10.1109/TOH.2023.3305495}}

@ARTICLE{ieeeMech1,
  author={Bark, Karlin and Wheeler, Jason and Shull, Pete and Savall, Joan and Cutkosky, Mark},
  journal={IEEE Transactions on Haptics}, 
  title={Rotational Skin Stretch Feedback: A Wearable Haptic Display for Motion}, 
  year={2010},
  volume={3},
  number={3},
  pages={166-176},
  doi={10.1109/TOH.2010.21}}

@ARTICLE{ieeeMech2,
  author={Shull, Peter B. and Tan, Tian and Culbertson, Heather and Zhu, Xiangyang and Okamura, Allison M.},
  journal={IEEE Transactions on Haptics}, 
  title={Resonant Frequency Skin Stretch for Wearable Haptics}, 
  year={2019},
  volume={12},
  number={3},
  pages={247-256},
  doi={10.1109/TOH.2019.2917072}}

@ARTICLE{ieeeMech3,
  author={Chinello, Francesco and Pacchierotti, Claudio and Bimbo, Joao and Tsagarakis, Nikos G. and Prattichizzo, Domenico},
  journal={IEEE Robotics and Automation Letters}, 
  title={Design and Evaluation of a Wearable Skin Stretch Device for Haptic Guidance}, 
  year={2018},
  volume={3},
  number={1},
  pages={524-531},
  doi={10.1109/LRA.2017.2766244}}

@ARTICLE{ieeeMech4,
  author={Aggravi, Marco and Pausé, Florent and Giordano, Paolo Robuffo and Pacchierotti, Claudio},
  journal={IEEE Robotics and Automation Letters}, 
  title={Design and Evaluation of a Wearable Haptic Device for Skin Stretch, Pressure, and Vibrotactile Stimuli}, 
  year={2018},
  volume={3},
  number={3},
  pages={2166-2173},
  doi={10.1109/LRA.2018.2810887}}

@ARTICLE{ieeeGlovePostureDetection,
  author={DelPreto, Joseph and Hughes, Josie and D'Aria, Matteo and de Fazio, Marco and Rus, Daniela},
  journal={IEEE Robotics and Automation Letters}, 
  title={A Wearable Smart Glove and Its Application of Pose and Gesture Detection to Sign Language Classification}, 
  year={2022},
  volume={7},
  number={4},
  pages={10589-10596},
  doi={10.1109/LRA.2022.3191232}}

@INPROCEEDINGS{ieeeSkinStretchTesting,
  author={Zook, Zane A. and Fleck, Joshua J. and Tjandra, Tiffani W. and O’Malley, Marcia K.},
  booktitle={2019 IEEE World Haptics Conference (WHC)}, 
  title={Effect of Interference on Multi-Sensory Haptic Perception of Stretch and Squeeze}, 
  year={2019},
  volume={},
  number={},
  pages={371-376},
  doi={10.1109/WHC.2019.8816139}}

@ARTICLE{ieeeForceFeedbackInMRI,
  author={Frishman, Samuel and Kight, Ali and Pirozzi, Ileana and Coffey, Mela C. and Daniel, Bruce L. and Cutkosky, Mark R.},
  journal={IEEE Transactions on Haptics}, 
  title={Enabling In-Bore MRI-Guided Biopsies With Force Feedback}, 
  year={2020},
  volume={13},
  number={1},
  pages={159-166},
  doi={10.1109/TOH.2020.2967375}}

@ARTICLE{ieeeSkinStrechForXRFeeling,
  author={Salazar, Steeven Villa and Pacchierotti, Claudio and de Tinguy, Xavier and Maciel, Anderson and Marchal, Maud},
  journal={IEEE Transactions on Haptics}, 
  title={Altering the Stiffness, Friction, and Shape Perception of Tangible Objects in Virtual Reality Using Wearable Haptics}, 
  year={2020},
  volume={13},
  number={1},
  pages={167-174},
  doi={10.1109/TOH.2020.2967389}}

@ARTICLE{FW-Touch,
  author={Chen, Dapeng and Song, Aiguo and Tian, Lei and Fu, Liyue and Zeng, Hong},
  journal={IEEE Transactions on Haptics}, 
  title={FW-Touch: A Finger Wearable Haptic Interface With an MR Foam Actuator for Displaying Surface Material Properties on a Touch Screen}, 
  year={2019},
  volume={12},
  number={3},
  pages={281-294},
  doi={10.1109/TOH.2019.2920349}}

@INPROCEEDINGS{7591147,
  author={Husman, M. A. B. and Maqbool, H. F. and Awad, M. I. and Abouhossein, A. and Dehghani-Sanij, A. A.},
  booktitle={2016 38th Annual International Conference of the IEEE Engineering in Medicine and Biology Society (EMBC)}, 
  title={A wearable skin stretch haptic feedback device: Towards improving balance control in lower limb amputees}, 
  year={2016},
  volume={},
  number={},
  pages={2120-2123},
  doi={10.1109/EMBC.2016.7591147}}

@ARTICLE{ieeeEMGtbd,
  author={Iovene, Elisa and Monaco, Riccardo and Fu, Junling and Costa, Francesco and Ferrigno, Giancarlo and Momi, Elena De},
  journal={IEEE Transactions on Haptics}, 
  title={EMG-Based Variable Impedance Control for Enhanced Haptic Feedback in Real-Time Material Recognition}, 
  year={2025},
  volume={18},
  number={1},
  pages={220-231},
  doi={10.1109/TOH.2024.3524023}}

@ARTICLE{ieeeExoStairClimbing,
  author={Li, Zhijun and Deng, Chuanjie and Zhao, Kuankuan},
  journal={IEEE Transactions on Industrial Electronics}, 
  title={Human-Cooperative Control of a Wearable Walking Exoskeleton for Enhancing Climbing Stair Activities}, 
  year={2020},
  volume={67},
  number={4},
  pages={3086-3095},
  doi={10.1109/TIE.2019.2914573}}

@ARTICLE{ieeeExoSupportMaybe,
  author={Cao, Jinghui and Xie, Sheng Quan and Das, Raj},
  journal={IEEE Transactions on Control Systems Technology}, 
  title={MIMO Sliding Mode Controller for Gait Exoskeleton Driven by Pneumatic Muscles}, 
  year={2018},
  volume={26},
  number={1},
  pages={274-281},
  doi={10.1109/TCST.2017.2654424}}

@ARTICLE{ieeeCripHapticJam,
  author={Banerjee, Premankur and Sweidan, Selwa and Yunis, Catherine and Hern´andez Cruz, Vanessa and Culbertson, Heather},
  journal={IEEE Transactions on Haptics}, 
  title={Crip Haptic Jam: Vibrotactile Composition for Disability-Centered Dance}, 
  year={2026},
  volume={},
  number={},
  pages={1-13},
  doi={10.1109/TOH.2026.3709095}}

\appendix

\section{INITIAL STUDY WITH A PROOF-OF-CONCEPT}
\label{app:initial-study-with-poc}

\begin{figure}[!b]
    \centering
    \includegraphics[width=1\linewidth]{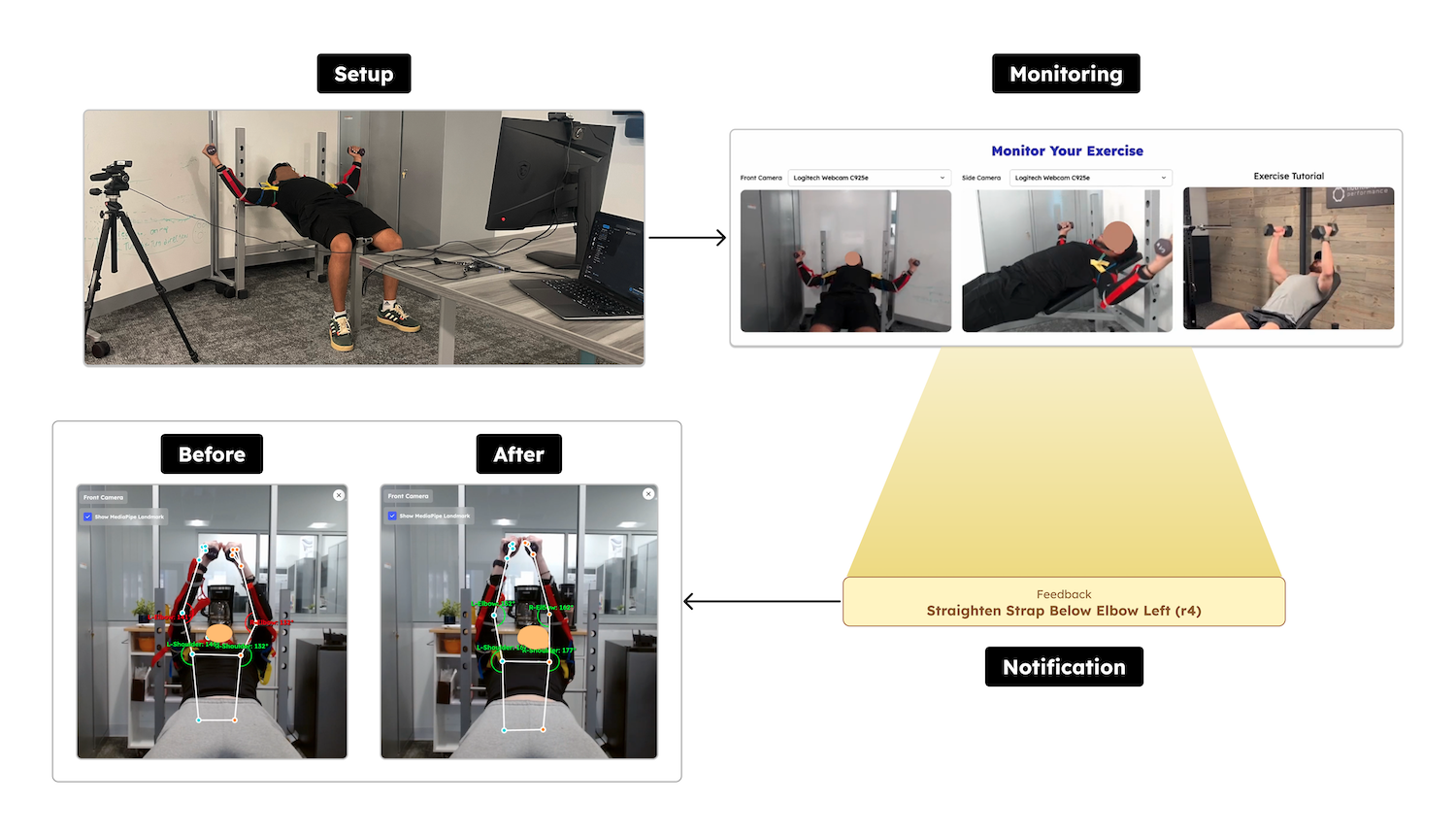}
    \caption{Our PoC approximates a closed-loop feedback system. A participant wears the PoC suit and is monitored through a system interface displaying \texttt{Front} and \texttt{Side} camera views. After each set, the interface issues notifications when deviations are detected (yellow = low, red = high). Researchers then tighten the corresponding straps based on the system's recommendations to correct the user's form.}
    \label{fig:System Operation}
\end{figure}

To take an initial step toward instantiating \sys{}, we developed a low-cost, strap-based proof-of-concept (PoC) wearable that implemented selected dimensions of the design space. The prototype consisted of a fitted shirt equipped with adjustable rigid straps across the upper body and elastic straps around the shoulders. The rigid straps provided opposing feedback by constraining movement beyond selected ranges, while the elastic straps provided guiding feedback through progressive resistance. We designed the prototype to explore how participants perceive different forms of embodied feedback during movement.

As illustrated in Fig.~\ref{fig:System Operation}, the PoC approximated a closed-loop feedback system through sensing, decision-making, and manually configured actuation. Two external cameras captured front and side views of participants' movements, which were processed using MediaPipe to estimate joint angles and identify deviations from reference postures. After each set of five repetitions, the system assessed recurring deviations and recommended appropriate feedback intensity levels. Researchers then manually adjusted the corresponding straps between sets to simulate adaptive feedback. 

We conducted workshops with six beginners in strength training (the study was approved by our institution's IRB). Each participant first underwent an individual calibration process to establish low and high feedback intensity levels. Participants then performed five repetitions of each exercise, followed by another five repetitions with adjusted strap configurations when corrective feedback was recommended. After completing the exercises, participants provided feedback through a questionnaire and semi-structured interview addressing perceived corrective support, comfort, and feedback appropriateness.

The preliminary observations suggested that participants could interpret and respond to different forms of embodied feedback. Rigid straps were generally perceived as clear movement boundaries, whereas elastic straps offered more gradual guidance that allowed greater movement continuity. Participants also highlighted the importance of phase-specific feedback and personalized resistance levels. However, stronger constraints sometimes restricted natural movement or caused discomfort, while elastic feedback could be too subtle without appropriate calibration. These observations reinforced several considerations identified in our design-space studies, particularly the trade-off between corrective guidance and movement continuity. Future implementations should investigate automated adaptation, improved calibration, and more comfortable wearable configurations.

\newcolumntype{Y}{>{\centering\arraybackslash}X}

\begin{table}[h]
\centering
\caption{Overview of co-design workshops (W1–W9) with matched trainer–trainee pairs. Ages in years; gender coded as F and M. Expert trainers (E1–E9) were recruited for their prior coaching experience, whereas novice trainees (N1–N9) were recruited with little to no strength training experience (< 3 months).}
\label{tab:codesbackground}
\rowcolors{2}{gray!12}{white}
\begin{tabularx}{\linewidth}{c c c c Y|c}
\hline
\textbf{Session} & \makecell{\textbf{Trainer} \\ \textbf{(Age, Gender)}} & \makecell{\textbf{Strength} \\ \textbf{Training} \\ \textbf{Experience}} & \makecell{\textbf{People} \\ \textbf{Coached}} & \textbf{Other Sports \& Fitness Background} & \makecell{\textbf{Trainee} \\ \textbf{(Age, Gender)}} \\
\hline
W1 & E1 (48, M) & 2 yrs & 5+ & -- & N1 (23, M) \\
W2 & E2 (26, M) & 1.5 yrs & 5--10 & -- & N2 (26, M) \\
W3 & E3 (25, F) & 1 yr & 2 & -- & N3 (33, F) \\
W4 & E4 (28, M) & 1 yr & 2 & -- & N4 (24, M) \\
W5 & E5 (24, F) & 8 yrs & 5+ & 
\begin{itemize}[leftmargin=1em]\vspace{-0.7\baselineskip}
    \item Licensed Zumba \& Aerobics instructor
    \item Experience teaching classes with handheld weights
\end{itemize}
& N5 (26, F) \\
W6 & E6 (23, F) & 9 yrs & 5+ &
\begin{itemize}[leftmargin=1em]\vspace{-0.7\baselineskip}
    \item Division I track/cross-country athlete
    \item Provided instruction on form and spotting to peers and juniors
\end{itemize}
& N6 (23, F) \\
W7 & E7 (24, F) & 5-7 yrs & 20+ & 
\begin{itemize}[leftmargin=1em]\vspace{-0.7\baselineskip}
    \item National-level rowing athlete
    \item Instructor of rowing camps
\end{itemize} 
& N7 (33, F) \\
W8 & E8 (22, M) & 8 yrs & 100+ & 
\begin{itemize}[leftmargin=1em]\vspace{-0.7\baselineskip}
    \item Fitness instructor (3 years)
    \item Track/cross-country athlete: PAC-12 Honor Roll, National U-20 champion
\end{itemize} 
& N8 (25, M) \\
W9 & E9 (25, M) & 8 yrs & 30+ & 
\begin{itemize}[leftmargin=1em]\vspace{-0.7\baselineskip}
    \item Gym trainer (5 months)
    \item Trained peers in calisthenics club
\end{itemize} 
 & N9 (31, M) \\
\hline
\end{tabularx}
\end{table}

\section{PARTICIPANT BACKGROUND}
\label{app:participant-background}

This section summarizes the demographic and professional backgrounds of participants in the two studies.

\subsection{CO-DESIGN WORKSHOPS}

Table~\ref{tab:codesbackground} summarizes the participants in the nine co-design workshops. Each session paired one novice trainee with one experienced trainer to collaboratively explore and prototype on-body feedback ideas.

\subsection{SPEED DATING}

Table~\ref{tab:sd-beginners} summarizes the beginner trainees who participated in the multi-party speed dating workshops. Participants were recruited with limited prior strength training experience and evaluated storyboard scenarios after briefly performing representative exercises. Table~\ref{tab:sd-trainers} summarizes the personal trainers who participated in the expert workshops, including their professional certifications and coaching backgrounds.

\begin{table}[h]
\centering
\caption{Overview of participants in the beginner trainee speed dating workshops. Participants (B1--B7) were recruited with fewer than three months of consistent strength training experience.}
\label{tab:sd-beginners}
\rowcolors{2}{gray!12}{white}
\begin{tabular}{c c c c}
\hline
\rowcolor{gray!12}
\textbf{Participant} &
\textbf{Age} &
\textbf{Gender} &
\textbf{Workshop} \\
\hline
B1 & 28 & M & Workshop 1 \\
B2 & 24 & M & Workshop 1 \\
B3 & 27 & F & Workshop 1 \\
B4 & 26 & M & Workshop 2 \\
B5 & 34 & M & Workshop 2 \\
B6 & 24 & M & Workshop 3 \\
B7 & 25 & F & Workshop 3 \\
\hline
\end{tabular}
\end{table}

\begin{table}[h]
\centering
\caption{Overview of recruited personal trainers (C1--C6) participating in the expert speed dating workshops.}
\label{tab:sd-trainers}

\rowcolors{2}{gray!12}{white}
\begin{tabularx}{\linewidth}{c c c c c X}
\hline
\textbf{Trainer} &
\textbf{Age} &
\textbf{Gender} &
\makecell{\textbf{Strength} \\ \textbf{Training} \\ \textbf{Experience}} &
\makecell{\textbf{People} \\ \textbf{Coached}} &
\makecell{\textbf{Sports \& Fitness} \\ \textbf{Background}} \\
\hline

C1 & 21 & M & 6 yrs & 20 &
\vspace{-0.7\baselineskip}
\begin{itemize}[leftmargin=1em]
    \item NASM Certified Personal Trainer
    \item F45 Certified Coach
\end{itemize} \\

C2 & 35 & M & 20+ yrs & 500+ &
\vspace{-0.7\baselineskip}
\begin{itemize}[leftmargin=1em]
    \item ACSM and NSCA Certified
\end{itemize} \\

C3 & 20 & F & 8 yrs & 15 &
\vspace{-0.7\baselineskip}
\begin{itemize}[leftmargin=1em]
    \item ACE Certified Personal Trainer
\end{itemize} \\

C4 & 56 & F & 20 yrs & 100+ &
\vspace{-0.7\baselineskip}
\begin{itemize}[leftmargin=1em]
    \item ACE and NASM Certified Personal Trainer
    \item Group fitness and yoga instructor
\end{itemize} \\

C5 & 19 & M & 5 yrs & 5 &
\vspace{-0.7\baselineskip}
\begin{itemize}[leftmargin=1em]
    \item ACE Certified Personal Trainer
    \item Lacrosse camp instructor
\end{itemize} \\

C6 & 49 & F & 20+ yrs & 20 &
\vspace{-0.7\baselineskip}
\begin{itemize}[leftmargin=1em]
    \item NASM Certified Personal Trainer
\end{itemize} \\

\hline
\end{tabularx}
\end{table}

\section{WORKSHEET DESIGN}
\label{app:worksheet-design} 

\begin{figure}[!b]
    \centering
    \includegraphics[height=0.32\textheight]{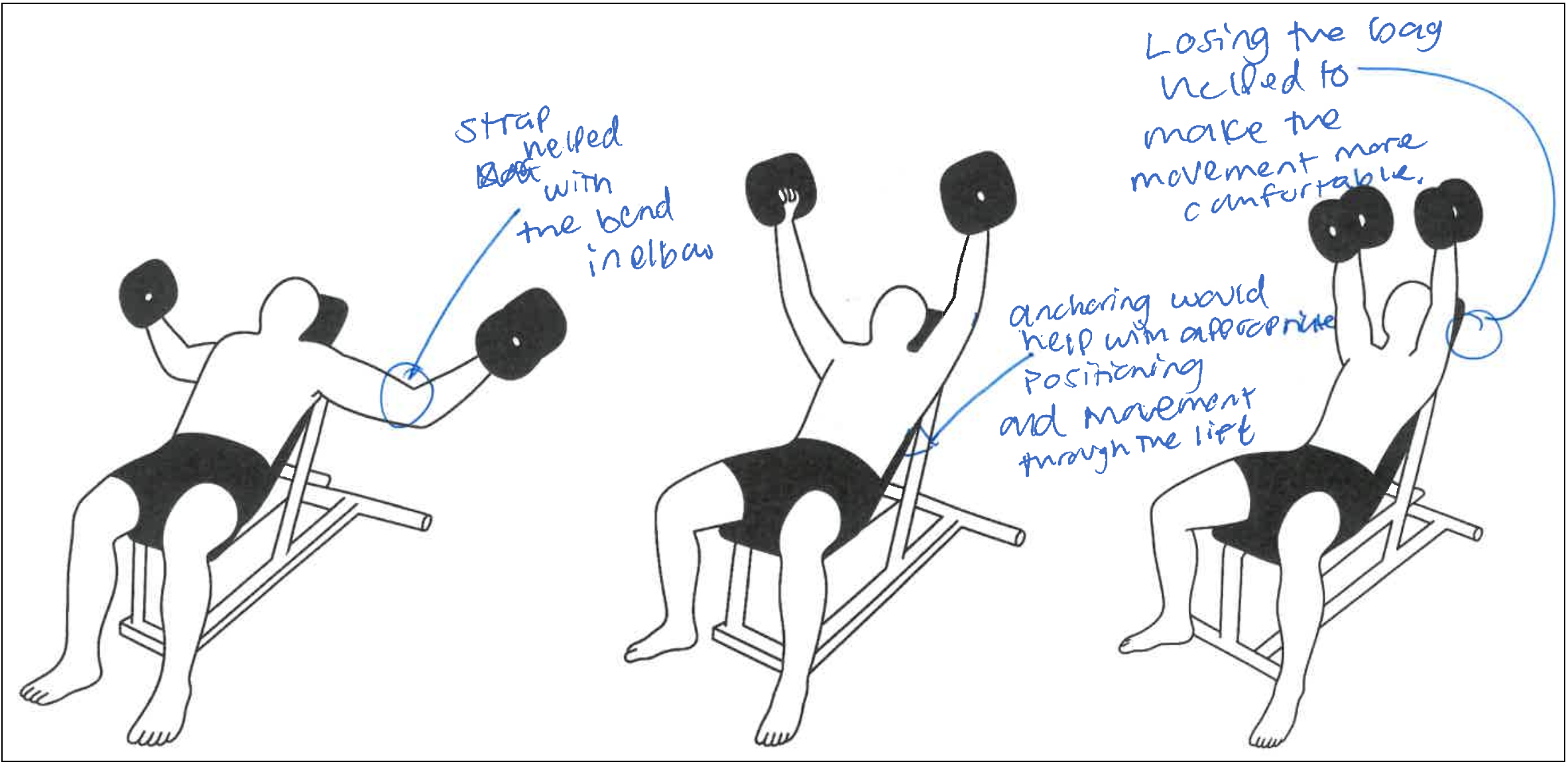}
    \caption{A sample worksheet for the dumbbell incline fly exercise, annotated during co-design workshops. The illustration captures multiple phases of the movement, with participant feedback highlighting design considerations such as strap support, anchoring for improved positioning, and adjustments to increase comfort throughout the lift.}
    \label{fig:fullworksheet}
\end{figure}

We prepared a worksheet for each exercise. Each worksheet included body maps \cite{bodymapsPaper} with figures of a person performing the exercise at representative moments, such as the starting position, mid-movement, and the maximum extension point. For the dumbbell front hold, an additional figure depicted an isometric hold at 15 seconds. These figures served as reference frames for participants to annotate, with the option of requesting extra sheets to sketch alternative designs or capture other parts of the movement cycle. As body maps are adaptable to diverse research contexts \cite{bodymapsPaper}, in our study they functioned both as a tool to guide the session and as an open-ended space for dialogue between trainer and trainee. They were designed for participants to sketch low-fidelity prototype designs, align on preferred configurations, and highlight specific body regions requiring attention. An example worksheet can be seen in Figure~\ref{fig:fullworksheet}.

\section{BODY ZONE MAPS OF EXTRINSIC AND INTRINSIC FEEDBACK} 
\label{app:body-zone-maps} 

\begin{figure}
    \centering
    \includegraphics[width=1\linewidth]{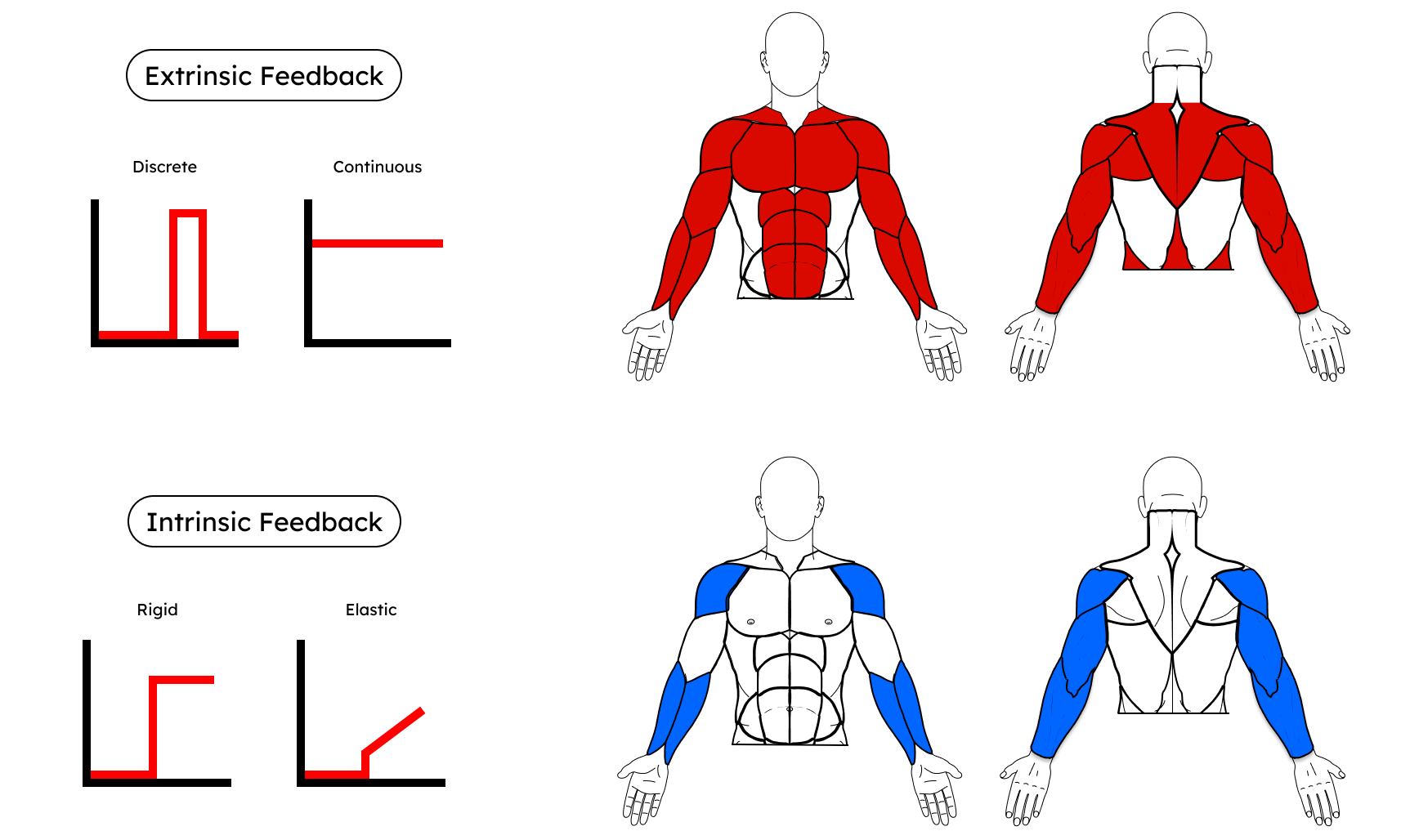}
    \caption{Body zone maps illustrating the regions where participants associated \FD{extrinsic} and \FD{intrinsic} feedback during the co-design workshops.}
    \label{fig:feedbackzonemap}
\end{figure}

Figure~\ref{fig:feedbackzonemap} summarizes the body locations discussed during the co-design workshops for \FD{extrinsic} and \FD{intrinsic} feedback. The maps aggregate participants' annotations across all sessions and provide an overview of where different feedback mechanisms were envisioned to support strength training. 

\section{CONNECTING SENSING \& FEEDBACK}
\label{app:connecting-sensing-and-feedback}

\begin{sidewaysfigure}
    \centering
    \raisebox{-25cm}{\includegraphics[width=\textheight]{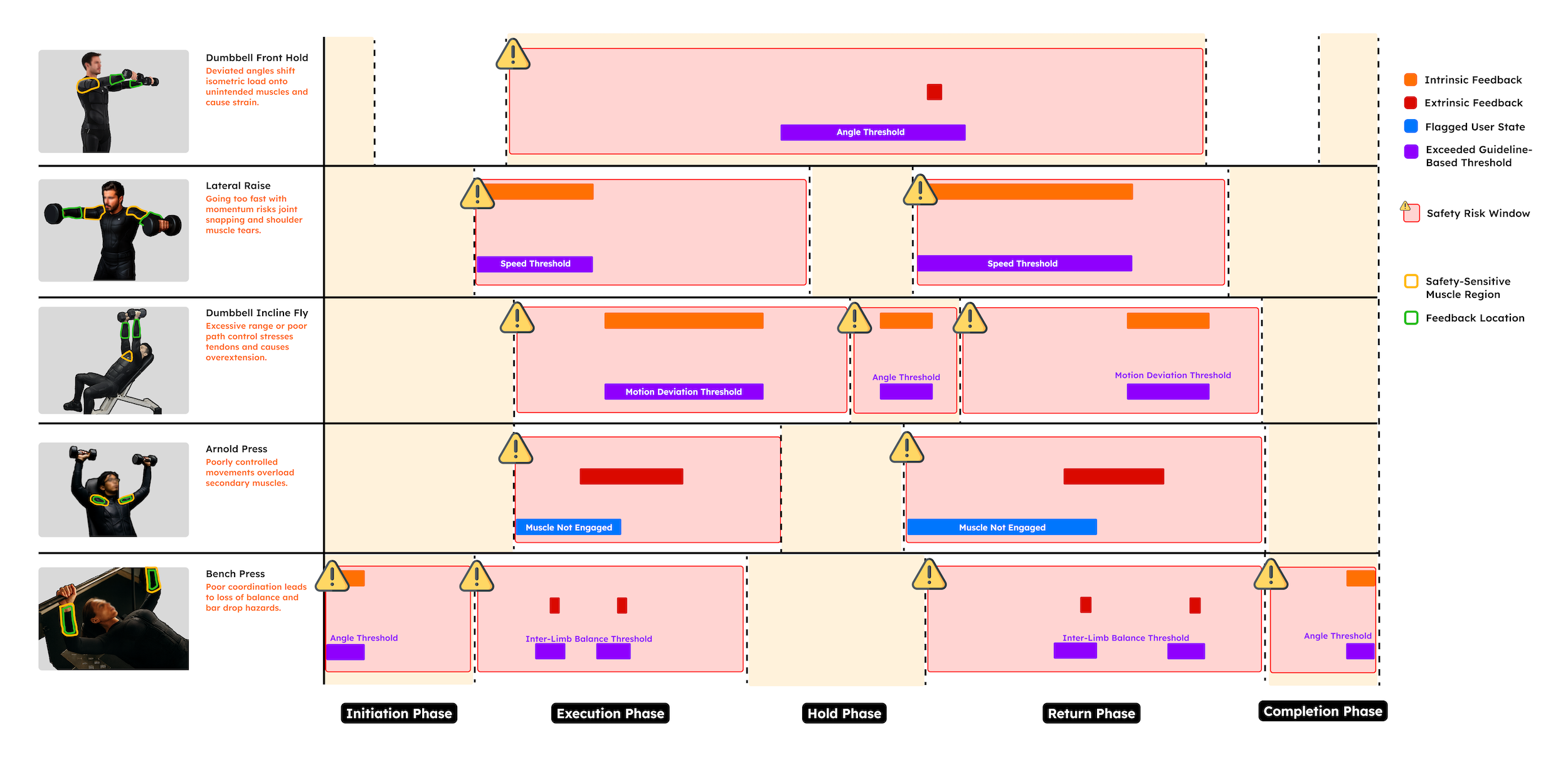}}
    \caption{Connecting \textit{Sensing \& Triggering} with \textit{Feedback Delivery}. The diagram illustrates five representative exercises, each segmented by exercise phase. Trigger metrics and policies determine which cues are flagged, while the corresponding feedback show how corrective cues are delivered.}
    \label{fig:graphdesignspace}
\end{sidewaysfigure}

This section illustrates how \textit{Sensing \& Triggering} and \textit{Feedback Delivery} operate together. Figure~\ref{fig:graphdesignspace} illustrates this integration across five representative exercises surfaced during co-design workshops, with example approaches for addressing issues at specific \textit{Trigger Locations} to scaffold safety during movement. Each row depicts an exercise across its phases of motion, highlighting the \textit{Trigger Metrics} and \textit{Trigger Policies} that determine when feedback is activated, alongside the corresponding feedback strategies.

This mapping underscores how \ST{expert-based} policies can guide the monitoring of metrics (e.g., \ST{joint angle}, \ST{motion path}) at locations where safety risk is elevated. It also extends to \ST{personalized} policies, such as monitoring \ST{muscle activity}, which can vary across body types and help detect cases where the intended muscle is not engaged. Once a cue is flagged, \textit{Feedback Delivery} defines the type and form of response. For example, in a lateral raise, exceeding a speed threshold of \ST{movement dynamics} triggered \FD{intrinsic} resistance to slow momentum, which could escalate with \FD{progressive} intensity, from \FD{low} to \FD{high} intensity, if the deviation worsens.  

By situating these elements side by side, the integrated design space highlights how different policies and feedback strategies can be tailored to the unique risks of each exercise. This connection makes explicit the logic of the system: cues sensed through body state and motion lead to feedback that constrains, guides, or withdraws in order to promote safe and effective strength training.

\section{LITERATURE REVIEW SEARCH TERMS}
\label{app:literature-search-terms}

The primary search terms are summarized below.

\begin{itemize}
    \item \textit{Wearable Form Factors.}

    \texttt{wearable},
    \texttt{on-body},
    \texttt{body-worn},
    \texttt{garment},
    \texttt{textile},
    \texttt{fabric},
    \texttt{smart clothing},
    \texttt{orthosis},
    \texttt{brace}, \\
    \texttt{exoskeleton}.

    \item \textit{Physical Movement Contexts.}

    \texttt{sport},
    \texttt{fitness},
    \texttt{exercise},
    \texttt{strength training},
    \texttt{weightlifting},
    \texttt{physical training},
    \texttt{movement},
    \texttt{motor learning},
    \texttt{rehabilitation},
    \texttt{posture}.

    \item \textit{Sensing and Feedback Mechanisms.}

    \texttt{sensing},
    \texttt{feedback},
    \texttt{haptic},
    \texttt{vibrotactile},
    \texttt{vibration},
    \texttt{electrical stimulation},
    \texttt{EMS},
    \texttt{TENS},
    \texttt{pneumatic},
    \texttt{pressure},
    \texttt{resistance},
    \texttt{variable stiffness},
    \texttt{soft robotics},
    \texttt{wearable robotics}.
    
\end{itemize}

Searches were performed by combining terms across these categories (e.g., wearable form factor + movement context + sensing or feedback mechanism). The resulting corpus was iteratively refined through title and abstract screening, full-text review, and backward/forward reference searching. The objective was to identify representative systems and recurring design decisions that informed the design landscape presented in Section~\ref{sec:design-landscape}.


\end{document}